\documentclass[11pt,letterpaper]{article}

\usepackage{amsmath,amsfonts,amsthm,mathabx,bbm}

\DeclareMathOperator*{\argmax}{arg\,max}

\usepackage{mathtools}   

\usepackage{enumitem}   

\usepackage{booktabs}

\usepackage{authblk} 

\usepackage{threeparttable}

\usepackage[titletoc,title]{appendix}  

\usepackage[normalem]{ulem}

\usepackage{amsfonts}
\usepackage[T1]{fontenc}

\usepackage[table,xcdraw,dvipsnames]{xcolor}   
\usepackage[colorlinks,urlcolor=blue,linkcolor=blue,citecolor=blue,urlcolor=blue]{hyperref}

\usepackage{caption}
\usepackage{subcaption}

\usepackage[linesnumbered,ruled,vlined]{algorithm2e}

\SetCommentSty{mycommfont}

\SetKwInput{KwInput}{Input}                
\SetKwInput{KwOutput}{Output}              

\renewcommand{\hat}{\widehat}

\newcommand{\bfm}[1]{\ensuremath{\boldsymbol{#1}}} 

   \def\bA{\bfm A}

     \def\EE{\mathbb{E}}
\def\bff{\bfm f}

     \def\PP{\mathbb{P}}
     
     \def\RR{\mathbb{R}}

\def\bz{\bfm z}

 \def\cA{{\cal  A}}
 \def\cB{{\cal  B}}

\def\calE{{\cal  E}} \def\cE{{\cal  E}}
 \def\cF{{\cal  F}}

 \def\cR{{\cal  R}}
 \def\cS{{\cal  S}}
\def\calT{{\cal  T}} 
 
 \def\cV{{\cal  V}}

\def\calY{{\cal  Y}} \def\cY{{\cal  Y}}
 \def\cZ{{\cal  Z}}

\def\mat{\hbox{\rm mat}}
\def\vec{\hbox{\rm vec}}

\usepackage{mathtools}
\DeclarePairedDelimiterX{\infdivx}[2]{(}{)}{%
  #1 \; \delimsize\| \; #2%
}

\DeclareMathOperator{\diag}{diag}
\newcommand{\E}[1]{{\mathbb{E}} \left[ #1 \right]}

\def\ideal{\text{\footnotesize ideal}}
\def\iso{{\rm iso}}

\newtheorem{assumption}{Assumption}[section]
\newtheorem{lemma}{Lemma}[section]

\newtheorem{theorem}{Theorem}[section]

\newtheorem{remark}{Remark}[section]

\definecolor{royalpurple}{rgb}{0.47, 0.32, 0.66}
\definecolor{greenfresh}{HTML}{00897B}
\definecolor{bluefresh}{HTML}{1E88E5}
\definecolor{redfresh}{HTML}{E53935}

\definecolor{royalpurple}{rgb}{0.47, 0.32, 0.66}

\def\beq{\begin{equation}}
\def\eeq{\end{equation}}

\def\bet{\begin{theorem}}
\def\eet{\end{theorem}}

\def\bel{\begin{lemma}}
\def\eel{\end{lemma}}

\def\vec{\mbox{vec}}

\def\tr{\mbox{tr}}

\def\lam {\lambda}

\newcommand{\R}{\mathbb{R}}

\usepackage[top=1in, bottom=1in, left=1in, right=1in]{geometry}

\usepackage{setspace}
\usepackage[authoryear]{natbib}  

\usepackage{graphicx}

\begin{document}
%
%
\def\TITLE{Estimation and Inference for CP Tensor Factor Models}
\newcommand{\blind}{0}

\if0\blind
{ \title{\bf \TITLE\footnote{We thank Yoosoon Chang, Rong Chen, Frank Diebold, Tae-Hwy Lee, Joon Park, Mahrad Sharifvaghefi and seminar participants at Duke University, Indiana University, University of Pittsburgh, University of California, Riverside, the 2024 Econometric Society Summer Meeting, and the workshop on Analysis of Complex Data: Tensors, Networks and Dynamic Systems for their useful comments and discussions. Any remaining errors are solely ours.}}
\author[1]{Bin Chen\thanks{ binchen@rochester.edu}}
\author[2]{Yuefeng Han \thanks{yuefeng.han@nd.edu}} 
\author[1]{Qiyang Yu\thanks{qyu13@ur.rochester.edu}}

\affil[1]{University of Rochester}
\affil[2]{University of Notre Dame}

	\maketitle
} \fi

\if1\blind
{
	\bigskip
	\bigskip
	\bigskip
	\title{\bf ...}
	\date{\vspace{-5ex}}
	\maketitle
	\medskip
} \fi

\begin{abstract}
\noindent
High-dimensional tensor-valued data have recently gained attention from researchers in economics and finance. We consider the estimation and inference of high-dimensional tensor factor models, where each dimension of the tensor diverges. Our focus is on a factor model that admits CP-type tensor decomposition, which allows for non-orthogonal loading vectors. Based on the contemporary covariance matrix, we propose an iterative simultaneous projection estimation method. Our estimator is robust to weak dependence among factors and weak correlation across different dimensions in the idiosyncratic shocks. We establish an inferential theory, demonstrating both consistency and asymptotic normality under relaxed assumptions. Within a unified framework, we consider two eigenvalue ratio-based estimators for the number of factors in a tensor factor model and justify their consistency. Simulation studies confirm the theoretical results and an empirical application to sorted portfolios reveals three important factors: a market factor, a long-short factor, and a volatility factor.

\bigskip

\bigskip

\end{abstract}

%
%
\bigskip


\section{Introduction}  \label{sec:intro}

Factor models have become one of the most popular tools for summarizing and extracting information from high-dimensional data in economics and finance (\cite{fan2021review}, \cite{Bai2016review}, \cite{Stock2016review}). Traditional factor models are designed to manage large panel data, where both cross-sectional and time series dimensions increase. These models admit a low-rank structure and have a common-idiosyncratic decomposition, allowing for the identification of significant variations within the panel of economic data.

In modern economics, researchers increasingly encounter vast, multi-dimensional datasets, or tensor. For example, monthly import-export volume time series spanning various product categories among countries can be represented as a three-dimensional tensor, with unavailable diagonal elements for each product category. Similarly, in portfolio selection, data often involve stock prices and various firm characteristics over time across different firms, forming a two-dimensional tensor. Additionally, macroeconomic studies on growth and productivity analyze multiple macro variables at the country-industry level, enabling cross-country comparative analyses, which are challenging with traditional panel data.

Statistical methods and economic applications for high-dimensional tensor factor analysis are still in their early stages of development. As in the classical panel setting, tensor factor models typically assume low-rank structures, with Canonical Polyadic (CP) and Tucker structures being the most common choices (see, e.g., \cite{KB2009review}). Recent studies have explored various estimation approaches and extensions. For example, working with Tucker decomposition, \cite{chen2022factor} consider two estimators based on the autocovariance matrices, while \cite{Han2021iterative} extend these methods using an iterative procedure with the matrix unfolding mechanism. \cite{chen2023statistical} propose an $\alpha$-PCA method that preserves the matrix structure and aggregates mean and contemporary covariance through a hyper-parameter $\alpha$. \cite{chen2024rank} introduce a pre-averaging technique for the Tucker tensor
factor model that significantly enhances the model's inherent signal strength under certain conditions. In parallel, \citet{lettau20243d} provides a pioneering empirical application of Tucker decomposition to a three-dimensional panel of double-sorted portfolios, showing substantial improvements over Fama–French factors and traditional PCA.
In the context of CP decomposition, \cite{han2024cp} propose an iterative simultaneous orthogonalization algorithm with warm-start initialization, while \cite{babii2022tensor} employ tensor principal component analysis (TPCA), assuming orthogonal factor loadings. \cite{chang2023modelling} develop an estimation procedure based on a generalized eigenanalysis (GE) constructed from the serial dependence structure of the underlying process.

In this paper, we focus on a tensor factor model with a CP low-rank structure due to its economic relevance and parsimonious features. We propose an iterative projection estimation based on the contemporary covariance and develop new inferential theories. Our contributions advance the existing literature in several ways:

First, unlike \cite{han2024cp}, which rely on lagged autocovariances, we use the contemporary covariance, which captures the full variance structure of the data. This is crucial in applications where serial dependence is weak, such as asset returns, where the efficient market hypothesis implies little autocorrelation. Our method significantly broadens the applicability of CP tensor factor models.

Second, inspired by techniques in \citet{anandkumar2014guaranteed} and \citet{sun2017provable} with noisy CP decomposition, we introduce an initialization strategy via randomized composite PCA (RC-PCA), which accommodates closely spaced eigenvalues. While those works consider CP decomposition of a single tensor, we establish performance bounds in the presence of noise and latent factor randomness---substantially complicating the theoretical analysis.

Third, we provide the first asymptotic normality results for estimated loading vectors in CP tensor factor models. This result enables valid statistical inference, such as confidence interval construction and hypothesis testing, tools that are essential for empirical applications\footnote{The importance of developing inferential theory in factor models is well established in the econometric literature; for instance, \cite{bai2003} laid the foundation for inference in high-dimensional factor models and has had a lasting impact. Our work brings similar inferential ideas to the tensor setting, filling an important gap in the existing literature.}. Our inferential framework is based on the asymptotic distribution of singular vectors and differs fundamentally from the technical tools used in both vector factor models (\citet{bai2003}) and Tucker tensor models (\citet{chen2023statistical}, \citet{yu2022projected}). Our approach can be extended to inference in many other matrix and tensor problems, such as the inference of factor loading vectors using lagged autocovariance matrices, as considered in \cite{lam2012,chang2023modelling}.

Fourth, we extend the eigenvalue ratio criterion---previously used in Tucker models (e.g., \citet{han2022rank})---to the CP tensor factor setting and establish its consistency. To our knowledge, this is the first work to do so.

Finally, our empirical analysis using characteristic-sorted portfolios demonstrates the practical value of our method. The proposed CP factor model outperforms benchmark approaches in out-of-sample forecasts. The factors extracted from the CP low-rank structure yield smaller cross-sectional pricing errors than classical factor models in the literature. Moreover, the estimated loadings and factors are interpretable without post-hoc rotation, revealing economically meaningful factors: a market factor, a long-short factor, and a volatility factor.


The remaining sections of this paper are organized as follows. Section \ref{sec:model} introduces the high-dimensional tensor factor model with a CP low rank structure allowing for non-orthogonal loading vectors. In Section \ref{sec:estimation}, we present an iterative projection estimation procedure and two generalized eigenvalue ratio-based estimators for the number of latent factors. Section \ref{sec-theory} establishes inferential theory. We assess the finite sample performance through simulation in Section \ref{sec-simul} and apply our method to characteristic portfolios in Section \ref{sec-applit}. Finally, Section \ref{sec-conclu} concludes the paper with all mathematical proofs and additional simulations included in the Appendix.
\subsection{Notations and preliminaries}
In this subsection, we introduce essential notations and basic tensor operations. For an in-depth review, readers may refer to \cite{KB2009review}.

Let $[n]$ denote the set $\{1, 2, \ldots, n\}$. Let $\|x\|_q = (x_1^q+...+x_p^q)^{1/q}$, $q\ge 1$, for any vector $x=(x_1,...,x_p)^\top$. We employ the following matrix norms: matrix spectral norm $\|M\|_{2} = \underset{\|x\|_2=1,\|y\|_2=1}{\max}  \|x^\top M y\|_2 = \sigma_1 (M)$, where $\sigma_1(M)$ is the largest singular value of $M$. For two sequences of real numbers $\{a_n\}$ and $\{b_n\}$, we write $a_n\asymp b_n$ if there exists a constant $C$ such that $|a_n|\leq C |b_n|$ and $|a_n|\geq C |b_n|$ hold for all sufficiently large $n$, and $a_n\lesssim b_n$ if there exists a constant $C$ such that $a_n\le Cb_n$.

Consider two tensors $\cA\in\RR^{d_1\times d_2\times \cdots \times d_K}, \cB\in \RR^{r_1\times r_2\times \cdots \times r_N}$. The tensor product $\otimes$ is defined as $\cA\otimes \cB\in \RR^{d_1\times \cdots \times d_K \times r_1\times \cdots \times r_N}$, where
$$(\cA\otimes\cB)_{i_1,...,i_K,j_1,...,j_N}=(\cA)_{i_1,...,i_K}(\cB)_{j_1,...,j_N} .$$
The $k$-mode product of $\cA\in\RR^{d_1\times d_2\times \cdots \times d_K}$ with a matrix $U\in\RR^{m_k\times d_k}$ is an order $K$ tensor of dimension $d_1\times \cdots \times d_{k-1} \times m_k\times d_{k+1} \times \cdots \times d_K$, denoted as $\cA\times_k U$, where
$$ (\cA\times_k U)_{i_1,...,i_{k-1},j,i_{k+1},...,i_K}=\sum_{i_k=1}^{d_k} \cA_{i_1,i_2,...,i_K} U_{j,i_k}.  $$
Similarly, for a matrix $V\in \RR^{d_k\times d_\ell}$, define $\cA\times_\ell\times_k V\in\RR^{d_1\times\cdots \times d_{\ell-1}\times d_{\ell+1} \times\cdots \times d_{k-1}\times d_{k+1}\times\cdots \times d_K}$ as
$$ (\cA\times_\ell\times_k V)_{i_1,...,i_{\ell-1},i_{\ell+1},...,i_{k-1},i_{k+1},...,i_K}=\sum_{i_{\ell}=1}^{d_{\ell}}\sum_{i_k=1}^{d_k} \cA_{i_1,i_2,...,i_K} V_{i_\ell,i_k}.$$
The mode-$k$ matricization of a tensor $\cA\in \R^{d_1\times\cdots\times d_K}$ is denoted as $\mat_k(\cA)\in \R^{d_k\times d_{-k}}$, where $d=\prod_{j=1}^K d_j$ and $d_{-k}=d/d_k=\prod_{j=1,j\neq k}^K d_j$. It is obtained by setting the $k$-th tensor mode as its rows and collapsing all the others into its columns. And the vectorization of the matrix/tensor $\cA$ is denoted as ${\rm{vec}}(\cA)\in \R^d$. With a slight abuse of notation, we still define $\mat_k(\vec(\cA))=\mat_k(\cA)$. For nonempty $J\subseteq [K]$, $\text{mat}_J(\cA)$ is the mode $J$ matrix unfolding which maps $\cA$ to $d_J\times d_{-J}$ matrix with $d_J=\prod_{j\in J}d_j$ and $d_{-J}=d/d_J$, e.g. $\text{mat}_{\{1,2\}}(\cA)=\text{mat}_3^\top(\cA)$ for $K=3$.


\section{Model} \label{sec:model}
We consider a tensor-valued time series $\mathcal{Y}_t \in \mathbb{R}^{d_1 \times d_2 \times \cdots \times d_K}$, where $ K \ge 2$ and $1 \le t \le T$\footnote{When $K=1$, $\mathcal{Y}_t$ reduces to a vector, and model (\ref{eqn:fm}) becomes the classical factor model extensively studied in the literature (\cite{bai2002} and \cite{stock2002}). The identification condition discussed in Remark \ref{identification} does not apply in this case, and hence we assume $K\ge 2$.
}. Our focus is on a tensor factor model with a CP low-rank structure:
\begin{equation}\label{eqn:fm}
\mathcal{Y}_t = \sum_{i=1}^r f_{it}  \left(\Gamma_{i1} \otimes  \Gamma_{i2} \cdots \otimes  \Gamma_{iK}\right) + \mathcal{E}_t
    = \sum_{i=1}^r w_i f_{it}  \left(a_{i1} \otimes  a_{i2} \cdots \otimes  a_{iK}\right) + \mathcal{E}_t,
    \quad  t \le T,
\end{equation}
where $\otimes$ denotes the tensor product, $f_{it}$ is a one-dimensional latent factor, $\Gamma_{ik}$ denotes the $d_k$-dimensional loading vector, which needs not to be orthogonal\footnote{Our framework naturally accommodates orthogonal loadings or factors as a special case, although our identification condition does not require orthogonal loadings or factors \citep{KB2009review}.}. Without loss of generality and to ensure identifiability, we assume that $\EE f_{it}^2 =1$ and normalize the factor loadings $\Gamma_{ik}$ so that $ a_{ik}= \Gamma_{ik} /\|\Gamma_{ik}\|_2$, for all $1\le i\le r$ and $1\le k\le K$. Consequently, all factor strengths are captured by $w_i$ with $w_i=\prod_{k=1}^K \|\Gamma_{ik}\|_2$. In the strong factor model case, $\|\Gamma_{ik}\|_2\asymp d_k^{1/2}$, which implies that $w_i \asymp \sqrt{d_1d_2\cdots d_K}$.
Unlike the uncorrelated factors assumed in \cite{han2024cp}, we allow for moderately strong correlation structures among the individual factors. 
The noise tensor $\mathcal{E}_t$ is assumed to be uncorrelated with the latent factors but may exhibit weak correlations across different dimensions. The rank $r$ may either be fixed or divergent.

A natural alternative approach to extracting common factors is to vectorize the data:
\begin{equation} \label{eqn:vec}
\vec(\cY_t) = \Xi F_t + \vec(\cE_t),
\end{equation}
where $\vec(\cY_t) \in \mathbb{R}^{d}$ with $d=d_1d_2\cdots d_K$ and $F_t=\left(f_{1t},f_{2t},\cdots,f_{rt}\right)^\top \in \mathbb{R}^{r}$. However, this method ignores the tensor structure of the data and hence substantially increases the number of parameters in the loading matrices from $(d_1+d_2+\cdots+d_K)r$ in the tensor case to $(d_1d_2\cdots d_K)r$ in the stacked vector version. In contrast, our proposed model preserves the tensor structure by modeling $\vec(\cY_t)$ as $A W F_t + \vec(\cE_t)$, $W=\text{diag}(w_1,...,w_r), A=(a_1,\cdots,a_r)$ and $a_i=\vec( a_{i1}\otimes a_{i2}\otimes\cdots\otimes a_{iK})$. This structure not only reduces the parameter dimension but also leads to improved convergence rates due to its parsimonious structure.

\begin{figure}[htbp!]
\begin{center}
\caption{Time series plots of characteristic decile portfolio returns}
\begin{threeparttable}
 \includegraphics[width=1\linewidth,height=\textheight,keepaspectratio=true]{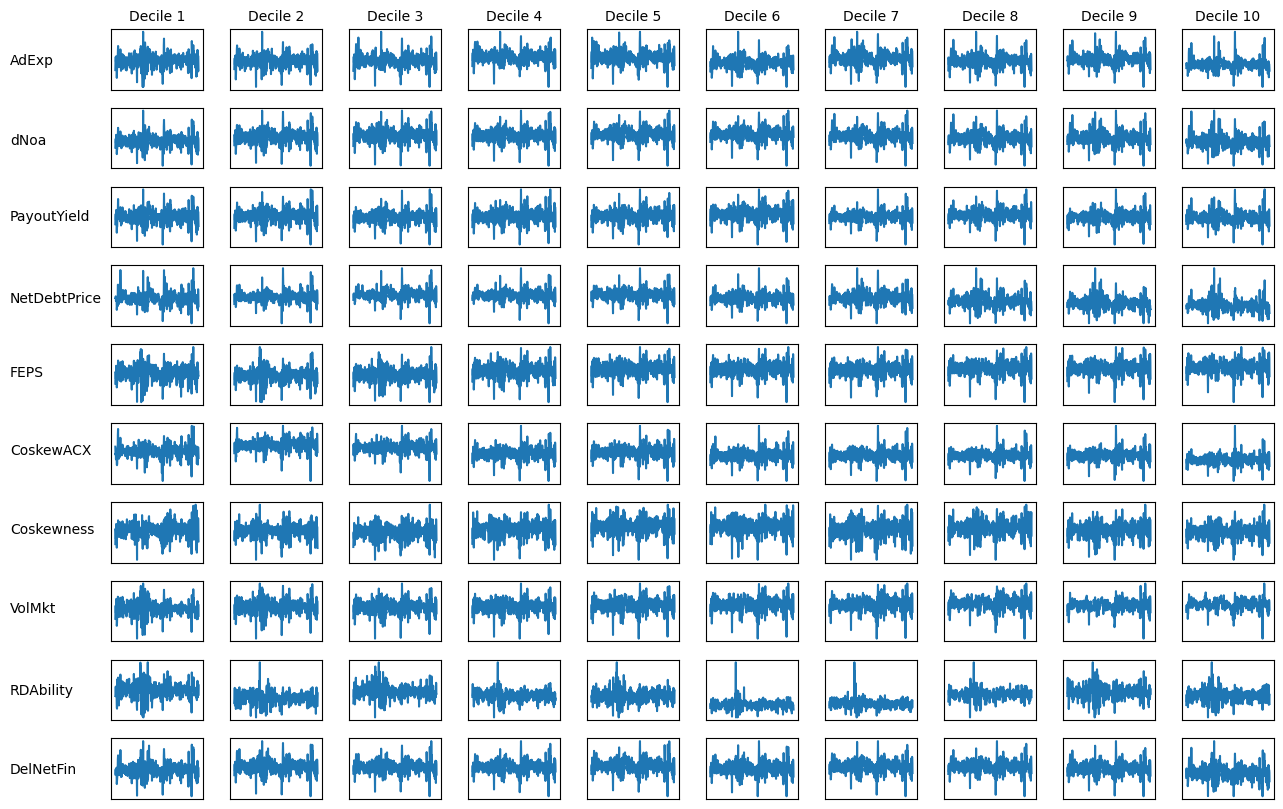}
\begin{tablenotes}
\small
\item Note: sample period from January 1990 to December 2022.
\end{tablenotes}
\end{threeparttable}
\label{fig:tfden}
\end{center}
\end{figure}

Consider an illustrative example of sorted portfolios, detailed in Section \ref{sec-applit}. The observed $\mathcal{Y}_t$ is represented as a matrix, where $d_1$ is the number of characteristic-sorted portfolios, $d_2=10$, and $K=2$. Each entry $\mathcal{Y}_{t,jl}$ corresponds to the excess return of the $l^{\text{th}}$-decile of the $j^{\text{th}}$ characteristic at time $t = 1, \ldots, T$. Figure \ref{fig:tfden} shows a time series plot of $\mathcal{Y}_t$ for ten characteristics from January 1990 to December 2022. Model (\ref{eqn:fm}) identifies $r$ latent factors, which can be interpreted as systematic risk factors. The element $ a_{i1,j}$ of the loading vector $ a_{i1}$, where $i = 1,\cdots,r$ and $j = 1,\cdots,d_1$, captures the heterogeneous exposure of the $j^{\text{th}}$ characteristic to the $i^{\text{th}}$ risk factor. Similarly, the entry $ a_{i2,l}$ of the loading vector $ a_{i2}$, where $i = 1,\cdots,r$ and $l = 1,\cdots,10$, determines the exposure of the $l^{\text{th}}$ decile to the $i^{\text{th}}$ risk factor. We allow the number of risk factors $r$ to increase with the dimensions $d_1$, $d_2$ and the sample size $T$.

In the literature, an alternative tensor factor model based on Tucker decomposition has been explored (see, e.g., \cite{Han2021iterative}, \cite{wang2017tensor}, \cite{lettau20243d}):
\begin{equation} \label{eq:tucker-model}
\cY_t= \cF_{t}\times_1 A_{1}\times\cdots\times A_{K}+\cE_t,
\end{equation}
where the core tensor $\cF_t\in \R^{r_1\times\cdots \times r_K}$ is the latent factor process in a tensor form, and $ A_i$'s are $d_i\times r_i$ loading matrices. As discussed in \cite{babii2022tensor} and \cite{han2024cp}, unlike the CP decomposition, the Tucker decomposition is generally non-unique, leading to identification difficulties: even with the usual identification
restrictions, only the column spaces of loading matrices are identified. Consequently, estimation results from model (\ref{eq:tucker-model}) may exhibit ambiguity, undermining meaningful discussions of individual factors\footnote{For classical factor models, \cite{stock2002} point out, ``because the factors are identified only up to a $k\times k$ matrix, detailed discussion of the individual factors is unwarranted''. The same comment applies to Tucker tensor factor models.}. In contrast, the CP  tensor factor model (\ref{eqn:fm}) yields $r$ scalar factors, aligning with the conventional wisdom in economic applications regarding common factors. This set of one-dimensional latent factors serves as natural inputs for diffusion index forecasting and factor-augmented regressions\footnote{A prominent example is the modeling of global yield curve dynamics. \citet{diebold2008} show that the generalized Nelson-Siegel model accurately captures the dynamics of yield curves and delivers strong predictive performance. Their model assumes that global yields follow a linear factor structure with interpretable loadings. This specification fits naturally within our CP factor model framework, where $f_t$ captures global factors, and $a_{i1}$ and $a_{i2}$ represent country-specific and maturity-specific loadings, respectively.}. We regard the CP tensor factor as a more parsimonious yet flexible and economically relevant alternative. Further comparison of the performance of these two tensor factor models will be presented in Section \ref{sec-applit}.

\begin{remark}[Identifiability Condition]\label{identification}
By incorporating time, we may stack $\cY_t$ into an order-$(K + 1)$ tensor $\cY\in\R^{d_1\times\cdots\times d_K\times T}$, with time $t$ as the $(K + 1)$-th mode, referred to as the time-mode. Subsequently, model \eqref{eqn:fm} can be reformulated as
\begin{equation*}
\cY=\sum_{i=1}^r w_i a_{i1}\otimes a_{i2}\otimes\cdots\otimes a_{iK}\otimes \bff_i+\cE,
\end{equation*}
where $\bff_i=(f_{i1},...,f_{iT})^\top$.
Ignoring the random noise $\cE$, the CP decomposition above is unique up to scaling and permutation indeterminacy if $\sum_{k=1}^K\cR( A_k)+\cR( F)\ge 2r+K$, where $ A_k=( a_{1k},..., a_{rk}),  F=(\bff_1,...,\bff_r)$ and $\cR( A) = \max\{s:$ any $s$ columns of the matrix $ A$ are linearly independent$\}$. 
Such a requirement provides a sufficient condition for uniqueness as per \cite{KB2009review}. In the subsequent estimation procedure, we analyze the estimation of the covariance tensor $\Sigma$ in \eqref{eq:sigma} below. The sufficient identifiability condition for the CP decomposition of the covariance tensor becomes $2\sum_{k=1}^K\cR( A_k)\ge 2r+2K-1$. This condition is significantly milder compared to the condition necessary to ensure statistical convergence. Note that the vector factor model with $K=1$ always violates this identifiability condition\footnote{When $K=1$, any invertible matrix yields the same fit and hence this ``rotation'' freedom cannot be ruled out without extra constraints (e.g., orthogonality, sparsity, sign/scale conventions). In contrast, when $K > 1$ we have a tensor CP model where each rank-1 component is a $(K+1)$-fold outer product (e.g., $a_{i1} \otimes a_{i2} \otimes f_i$ for $K = 2$). The key is that a general rotation that mixes components in one mode would have to be accompanied by \emph{the same mixing} in every other mode to preserve the rank-1 outer-product structure across all modes. Except for trivial rescalings and permutations, such coupled rotations are impossible once the factor columns across modes are ``independent enough''. That is why CP admits intrinsic identifiability that the classical factor model does not.}.

The identifiability condition is \textit{sufficient but not necessary}---it is conservative. Identification can still hold even when it fails, but the guarantee is lost. It will definitely fail (and identification will typically be weak or lost) when a mode has very low rank, e.g., if some loading vectors are constant or proportional so that many columns are nearly collinear. For instance, if one mode effectively has rank $= 1$ (e.g., several components share the same constant loading vector), then the sum of ranks can't clear the threshold unless the rank is trivial; in practice, this allows components to be mixed within the low-rank subspace, recreating a rotation-like ambiguity. More broadly, near-violations (high collinearity, duplicate columns, extremely unbalanced ranks/dimensions) lead to weak identification and numerical instability even if the identifiability condition barely holds.
\end{remark}

\section{Estimation}  \label{sec:estimation}
We consider a two-step estimation procedure to derive the loading vectors and latent factors. This approach begins with initialization via RC-PCA, followed by an iterative refinement step utilizing an iterative simultaneous orthogonalization (ISO) procedure.

We start by defining the contemporary covariance as the expected value of the outer product of $\cY_t$:
\begin{equation} \label{eq:sigma}
\begin{aligned}
   \Sigma & =\E {\mathcal{Y}_t\otimes\mathcal{Y}_t} \\
   & = \sum_{i,j=1}^r \Theta_{ij} \otimes_{l=1}^K a_{il} \otimes_{l=1}^K a_{jl}  + \E{ \cE_t \otimes \cE_t},
  \end{aligned}
\end{equation}
where $\Theta_{ij}=w_i w_j \E {f_{it}f_{jt}}$. Its sample analogue, denoted as $\hat{\Sigma}$, is computed as the average outer product over $T$ observations:
\begin{equation} \label{eq:hatsigma}
\begin{aligned}
   \hat{\Sigma} & =\sum_{t=1}^T \frac{\mathcal{Y}_t\otimes\mathcal{Y}_t}{T}.
  \end{aligned}
\end{equation}

We aim to estimate the loading vectors by minimizing the empirical quadratic loss, formulated as:
\begin{align} \label{problem:ls}
\min_{\substack{a_{i1},a_{i2},...,a_{iK},1\le i\le r, \\ \|a_{i1}\|_2=\ldots=\|a_{iK}\|_2 =1 } } \left\| \hat{\Sigma} -\sum_{i,j=1}^r \Theta_{ij} \otimes_{l=1}^K a_{il} \otimes_{l=1}^K a_{jl} \right\|_{\rm F}^2,
\end{align}
where $\|\cA\|_{F}$ denotes the Frobenius norm of a tensor $\cA$. However, this optimization problem is non-convex and prone to multiple local optima. To counter this problem, we employ a two-step approach. The first step focuses on obtaining a suitable initialization close to the global optimum.


The contemporary covariance $\Sigma$ in (\ref{eq:sigma}) can be unfolded to a $d\times d$ matrix
\begin{equation} \label{eq:sigma*}
\begin{aligned}
   \Sigma_0 & = A \Theta  A^\top ,
  \end{aligned}
\end{equation}
where $\Theta= W (\EE F_t F_t^\top) W$, $F_t=(f_{1t},\cdots,f_{rt})^\top$, $W=\diag(w_1,...,w_r)$. This unfolding enables classical PCA estimation if the columns of the loading matrix $ A$ are orthogonal. Our framework accommodates general non-orthogonal $a_i$'s and hence the PCA procedure introduces a bias component, which motivates the second stage refinement. The accuracy of the PCA estimator hinges on the maximum correlation among the loading vectors. When the additional orthogonality condition is imposed as in \cite{babii2022tensor}, the maximum correlation reduces to $0$ and hence bias disappears. The first step, termed initialization via RC-PCA, is detailed in Algorithm \ref{alg:initialize-cp}.

To further relax the eigengap assumption imposed in \cite{babii2022tensor} and \cite{han2024cp}, we incorporate randomized projection (Procedure \ref{alg:initialize-random}) into RC-PCA approach\footnote{Closely spaced eigenvalues do not imply that the covariance matrix is singular. A covariance matrix can remain full rank even when some of its eigenvalues are identical or nearly identical. In our theoretical analysis, ``closely spaced eigenvalues'' specifically refers to identical or nearly identical eigenvalues of the population covariance matrix of the noiseless data, defined as $\Theta= W (\EE F_t F_t^\top) W$, where $F_t=(f_{1t},\cdots,f_{rt})^\top$, $W=\diag(w_1,...,w_r)$. The estimation error for an individual eigenvector depends on the corresponding eigen-gap \citep{yu2015useful}. When this gap is small, or when the factor number $r$ grows and $\lambda_1 \asymp \lambda_r$, the performance of standard PCA deteriorates, whereas our RC-PCA remains effective. Moreover, the theoretical investigation of RC-PCA under relaxed assumptions is non-trivial and holds independent value.}. Random projection, also known as random slicing \citep{anandkumar2014guaranteed,sun2017provable} is a well-recognized initialization method in noiseless tensor CP decomposition, which accommodates closely spaced eigenvalues. We extend this approach to the tensor CP factor model. In Algorithm \ref{alg:initialize-cp} and Procedure \ref{alg:initialize-random}, the tuning parameters $c_0$ and $\nu$ are not highly sensitive to their specific values. In both the simulation and empirical studies presented in the main text, we use the default settings $c_0 = 0.1$ and $\nu = 0.8$, and we find that small or moderate deviations from these values do not affect the results\footnote{The parameter $c_0$ is related to the noise magnitude. Since the randomized projection step guarantees consistency under orthogonal loadings regardless of whether the eigen-gap condition holds, using a larger $c_0$ is a safe choice. The parameter $\nu$ is a pre-specified threshold used to filter out tuples that are too similar to the tuple already selected. Each tuple consists of vectors that estimate factor loadings for an unknown rank $i \in \{ 1, \ldots, r\}$. Therefore, the role of $\nu$ is to ensure that redundant tuples corresponding to the same factor are removed. Further simulation studies regarding $c_0$ and $\nu$ can be found in Appendix \ref{appendix:simulation}.}. Note that the condition $\min\{|\widehat\lambda_i-\widehat\lambda_{i-1}|,|\widehat\lambda_i-\widehat\lambda_{i+1}| \} > c_0 \widehat\lambda_r $ is checked within Algorithm 1 and automatically determines the appropriate regime based on information from the data.

\begin{algorithm}[htpb!]
\caption{Initialization via Randomized Composite PCA}\label{alg:initialize-cp}
    \SetKwInOut{Input}{Input}
    \SetKwInOut{Output}{Output}
    \Input{The observations $\cY_t\in\RR^{d_1\times\cdots\times d_K}$, $t=1,...,T$, the number of factors $r$, small constant $0<c_0<1$.}

Evaluate $\widehat\Sigma$ in \eqref{eq:hatsigma}, and unfold it to $d\times d$ matrix $\widetilde\Sigma=\mat_{[K]}(\widehat\Sigma)$.

Obtain $\widehat\lambda_i,\widehat u_i, 1\le i\le r$, the top $r$ eigenvalues and eigenvectors of $\widetilde\Sigma$. Set $\widehat \lambda_0=\infty$ and $\widehat\lambda_{r+1}=0$.

\If{$\min\{|\widehat\lambda_i-\widehat\lambda_{i-1}|,|\widehat\lambda_i-\widehat\lambda_{i+1}| \} > c_0 \widehat\lambda_r $  }{
 Compute $\widehat a_{ik}^{\rm rcpca}$ as the top left singular vector of ${\rm mat}_k(\widehat  u_i)\in\R^{d_k\times (d/d_k)}$, for all $1\le k\le K$.
}
\Else{
Form disjoint index sets $I_1,...,I_N$ from all contiguous indices $1\le i\le r$ that do not satisfy the above criteria of the eigengap.

For each $I_j$, form $d\times d$ matrix $\widetilde \Sigma_{j}=\sum_{\ell\in I_j} \widehat \lambda_{\ell} \widehat u_{\ell}\widehat  u_{\ell}^\top$, and formulate it into a tensor $\widehat\Sigma_j\in \R^{d_1\times\cdots\times d_K\times d_1\times\cdots\times d_K}$. Then run Procedure \ref{alg:initialize-random} on $\widehat\Sigma_j$ to obtain $\widehat a_{ik}^{\rm rcpca}$ for all $i\in I_j,1\le k\le K$.
}
    \Output{Warm initialization $\hat  a_{ik}^{\rm rcpca}, 1\le i\le r, 1\le k\le K$}
\end{algorithm}

\SetAlgorithmName{Procedure}{procedure}{List of Procedures}
\begin{algorithm}[htpb!]
\caption{Randomized Projection}\label{alg:initialize-random}
    \SetKwInOut{Input}{Input}
    \SetKwInOut{Output}{Output}
    \Input{Noisy tensor $\Xi\in\RR^{d_1\times\cdots\times d_K\times d_1\times\cdots\times d_K}$, rank $s$, number of random projections $L$, mode $h$ for random projection, tuning parameter $\nu$.}

If $h=\emptyset$, pick $h=\arg\max_{1\le k\le K} (d_k)$

   \For{$\ell = 1$ to $L$}{
        Randomly draw a $d_h\times d_h$ Gaussian matrix $\theta$ whose entries are i.i.d. $N(0,1)$.

        Compute $\Xi\times_{h}\times_{K+h}\theta$ and compute its leading singular value and left singular vector $\eta_\ell,\widetilde u_{\ell}$.

        Compute $\widetilde a_{\ell k}$ as the top left singular vector of ${\rm mat}_k(\widetilde  u_{\ell})\in\R^{d_k\times (d/(d_kd_h))}$, for all $k\neq h$.

        Compute $\widetilde a_{\ell h}$ as the top left singular vector of $\Xi\times_{k\neq h,K+h} \widetilde  a_{\ell k}$, with $\widetilde a_{\ell, k}=\widetilde a_{\ell, k-K}$ for $k>K$.

        Add the tuple $(\widetilde  a_{\ell k},1\le k\le K)$ to $\cS_L$.
    }


    \For{$i = 1$ to $s$}{
        Among the remaining tuples in $\cS_L$, choose one tuple $(\widetilde  a_{\ell k},1\le k\le K)$ that correspond to the largest $\|\Xi\times_{k=1}^K \widetilde  a_{\ell k}\times_{k=K+1}^{2K}\widetilde  a_{\ell, k-K}\|_2$. Set it to be $\widehat  a_{ik}^{\rm rcpca}=\widetilde  a_{\ell k}$.

        Remove all the tuples with $\max_{1\le k\le K} |\widetilde  a_{\ell' k}^\top \widehat a_{ik}^{\rm rcpca}|>\nu$.
    }

    \Output{Warm initialization $\hat  a_{ik}^{\rm rcpca}, 1\le i\le s, 1\le k\le K$}
\end{algorithm}

\begin{remark}\label{rmk:random_projection}
To illustrate the idea behind Procedure 2, consider a matrix time series case with $K=2$. If the collinearity or coherence among the CP loading vectors $(a_{1k},...,a_{rk})$ is low, selecting a random projection vector $b$ for the first mode that is close to $a_{11}$ will result in $|a_{i1}^\top b|$ being small for all $i>1$ and $|a_{11}^\top b|\approx1$. Consequently, the projected data $\cY_t b=\sum_{i=1}^r w_i f_{it} (a_{i1}^\top b) a_{i2}+\cE_t b$ retains the signal corresponding to the first CP factor almost unchanged while significantly reducing the signals from the other factors.
\end{remark}

Following initialization, we refine the estimation using an ISO procedure (Algorithm \ref{algorithm:projection}). This step aims to enhance estimation accuracy and extract latent factors. The procedure is motivated by the vector factor structure of the denoised $\cY_t$:
\begin{align} \label{eq:cp-ideal}
\cZ_{t,ik}=w_i f_{it} a_{ik} + \cV_{t,ik},
\end{align}
where \begin{align}
\cZ_{t,ik}=&\; \cY_t \times_1 b_{i1}^\top \times_2 \cdots \times_{k-1} b_{i,k-1}^\top \times_{k+1} b_{i,k+1}^\top \times_{k+2}\cdots\times_K b_{iK}^\top , \label{eq:z}\\
\cV_{t,ik}=&\; \cE_t \times_1 b_{i1}^\top \times_2 \cdots \times_{k-1} b_{i,k-1}^\top \times_{k+1} b_{i,k+1}^\top \times_{k+2}\cdots\times_K b_{iK}^\top,  \label{eq:e*}
\end{align}
$B_k = A_k(A_k^{\top} A_k)^{-1} = (b_{1k},...,b_{rk}) \in\R^{d_k\times r}$, $A_k=(a_{1k},\ldots,a_{rk})\in \R^{d_k\times r}$, and we have used the fact that $b_{ik}$ is orthogonal to all
$a_{jk}, j\neq i$ by construction.
The orthogonalization projection, which takes place in all except the $k$th mode simultaneously in each computational iteration, transforms the tensor $\cY_t$ to a $d_k\times 1$ vector, reducing dimensions and noise substantially. This transformation enables easy and accurate estimation of the classical vector factor model in equation ($\ref{eq:cp-ideal}$).

Consider a simple illustrative example with $K=2$, $r=2$: \begin{align*}
  \mathcal{Y}_t = w_1f_{1t} a_{11} \otimes  a_{12}+w_2 f_{2t} a_{21}\otimes  a_{22} + \mathcal{E}_t  = w_1f_{1t} a_{11}  a_{12}^\top +w_2 f_{2t} a_{21}  a_{22}^\top + \mathcal{E}_t\in \R^{d_1\times{d_2}}.
\end{align*}
The ideal projection with true $b_{12}$ yields
\begin{align*}
  \cZ_{t,11}&=\cY_t \times_2 b_{12}=\cY_t\  b_{12} \\
 &= w_1 \ f_{1t}\ a_{11} (\LaTeXunderbrace{a_{12}^\top b_{12}}_{1})
+ w_2 \ f_{2t}\ a_{21} (\LaTeXunderbrace{a_{22}^\top b_{12}}_{0})+\cE_{t,11}^* \\
 & =w_1\ f_{1t} a_{11}+\cE_{t,11}^* ,
\end{align*}
where $\cE_{t,11}^*=\cE_t\times_2b_{12}\in \R^{d_1}$. This projection reduces the dimension from $\R^{d_1\times{d_2} }$ to $\R^{d_1}$, while simultaneously  increasing the signal-to-noise ratio.

In practice, we do not observe $b_{ik}$ and iterations can be applied to update the estimations. Given the previous estimates $\widehat a_{ik}^{(m-1)}$, where $m$ is the iteration number, $\cY_t$ can be denoised via
\begin{align*}
&\cZ_{t,ik}^{(m)}=\cY_t \times_1 \widehat  b_{i1}^{(m)\top} \times_2 \cdots \times_{k-1} \widehat  b_{i,k-1}^{(m)\top} \times_{k+1} \widehat  b_{i,k+1}^{(m-1)\top} \times_{k+2}\cdots\times_K \widehat  b_{iK}^{(m-1)\top} ,
\end{align*}
for $t=1,...,T$, and consequently, updated loading vectors $\widehat a_{ik}^{(m)}$ are obtained through eigenanalysis based on the contemporary covariance $\widehat\Sigma( \cZ_{1:T,ik}^{(m)} )=\frac{1}{T}\sum_{t=1}^T \cZ_{t,ik}^{(m)} \cZ_{t,ik}^{(m)\top}$. The iteration continues until convergence or the maximum number of iterations is reached. The default accuracy is set to $\epsilon = 10^{-5}$; a smaller value indicates more precise estimation, which typically results in a larger number of iterations. When our proposed RC-PCA is used as the first-stage (initial) estimator, Algorithm \ref{algorithm:projection} converges very rapidly. In our simulation and empirical applications, we set the maximum number of iterations to $M=100$, but convergence is typically achieved in fewer than 5 iterations.

\SetAlgorithmName{Algorithm}{algorithm}{List of Algorithms}
\begin{algorithm}[htbp]
\caption{Iterative Simultaneous Orthogonalization} \label{algorithm:projection}
\SetKwInOut{Input}{Input}
\SetKwInOut{Output}{Output}
\Input{The observations $\cY_t\in\RR^{d_1\times\cdots\times d_K}$, $t=1,...,T$, the number of factors $r$, the warm-start initial estimates $\widehat a_{ik}^{(0)}$, $1\le i\le r$ and $1\le k\le K$, the tolerance parameter $\epsilon>0$, and the maximum number of iterations $M$.}

Compute $\widehat B_k^{(0)} = \widehat A_k^{(0)}(\widehat A_k^{(0)\top}\widehat A_k^{(0)})^{-1} = (\widehat b_{1k}^{(0)},...,\widehat b_{rk}^{(0)})$ with $\widehat A_k^{(0)}=(\widehat a_{1k}^{(0)},\ldots,\widehat a_{rk}^{(0)})\in\R^{d_k\times r}$ for $k=1,\ldots, K$.
Set $m=0$.

\Repeat{$m=M$ or $\max_{1\le i\le r}\max_{1\le k\le K}\| \widehat a_{ik}^{(m)} \widehat a_{ik}^{(m)\top} - \widehat a_{ik}^{(m-1)} \widehat a_{ik}^{(m-1)\top} \|_{2}\le \epsilon$}{

Let $m=m+1$.

\For{$k=1$ to $K$}{

\For{$i=1$ to $r$}{

Given previous estimates $\widehat a_{ik}^{(m-1)}$, calculate
\begin{align*}
&\cZ_{t,ik}^{(m)}=\cY_t \times_1 \widehat  b_{i1}^{(m)\top} \times_2 \cdots \times_{k-1} \widehat  b_{i,k-1}^{(m)\top} \times_{k+1} \widehat  b_{i,k+1}^{(m-1)\top} \times_{k+2}\cdots\times_K \widehat  b_{iK}^{(m-1)\top} ,
\end{align*}
for $t=1,...,T$. Let $\widehat\Sigma \left( \cZ_{1:T,ik}^{(m)} \right)=\frac{1}{T}\sum_{t=1}^T \cZ_{t,ik}^{(m)} \cZ_{t,ik}^{(m)\top}$.

Compute $\widehat a_{ik}^{(m)}$ as the top eigenvector of $\widehat\Sigma( \cZ_{1:T,ik}^{(m)} )$.
}

Compute $\widehat B_k^{(m)} = \widehat A_k^{(m)}(\widehat A_k^{(m)\top}\widehat A_k^{(m)})^{-1} = (\widehat b_{1k}^{(m)},...,\widehat b_{rk}^{(m)})$ with $\widehat A_k^{(m)}=(\widehat a_{1k}^{(m)},\ldots,\widehat a_{rk}^{(m)})$.
}
}

\Output{Estimates
\begin{align*}
\widehat  a_{ik}^{\iso}&=\widehat  a_{ik}^{(m)},\quad i=1,...,r,\ \ k=1,...,K, \\
\widehat w_{i} &= \left( \frac1T\sum_{t=1}^T \left(\cY_t\times_{k=1}^K \widehat b_{ik}^{(m)\top}\right)^2 \right)^{1/2}, \quad \quad i=1,...,r, \\
\widehat f_{it}&= \widehat w_{i}^{-1} \cdot \cY_t\times_{k=1}^K \widehat  b_{ik}^{(m)\top}, \quad  i=1,...,r, \ \ t=1,...,T, \\
\widehat\cY_t &= \sum_{i=1}^r \widehat f_{it} \otimes_{k=1}^K \widehat  a_{ik}^{(m)} , \quad  t=1,...,T.
\end{align*}
}
\end{algorithm}

The above estimation procedure assumes that the rank $r$ is known. However, we need to estimate $r$ in practice. We consider two estimation procedures based on the eigenvalue ratio method proposed by \cite{ahn2013}.

For the first procedure, we unfold the sample contemporary covariance $\widehat\Sigma$ in \eqref{eq:hatsigma} to a $d\times d$ matrix $\widetilde\Sigma$. Let $\hat{\lambda}_{1} \geq \hat{\lambda}_{2} \geq \cdots \geq \hat{\lambda}_{r} \geq 0$ be the ordered eigenvalues of $\widetilde\Sigma$. The CP tensor factor model \eqref{eqn:fm} can also be adapted to a vector factor model \eqref{eqn:vec} with the same number of factors $r$. Thus, the eigenvalue ratio-based estimator derived from the unfolded covariance matrix $\widetilde\Sigma$ can be defined as
\begin{equation} \label{eqn:eigen_ratio}
\hat{r}^{\rm uer} = \argmax_{1 \leq i \leq r_{\max}} \frac{\hat{\lambda}_{i}}{\hat{\lambda}_{i+1}},
\end{equation}
and $r_{\max}$ is a selected upper bound.

Alternatively, we can define the mode-$k$ covariance with the inner product:
\begin{align*}
\widehat\Sigma_k = \sum_{t=1}^T \frac{\operatorname{mat}_k(\cY_t) \operatorname{mat}^\top_k(\cY_t)}{T} \in \R^{d_k \times d_k}.
\end{align*}
Let $\hat{\lambda}_{1k} \geq \hat{\lambda}_{2k} \geq \cdots \geq \hat{\lambda}_{rk} \geq 0$ be the ordered eigenvalues of $\widehat\Sigma_k$. The eigenvalue ratio-based estimator using the inner product can be defined as
\begin{equation}\label{eqn:eigen_ratio_ip}
  \hat{r}^{\rm ip} = \operatorname{max}(\hat{r}_1, \hat{r}_2, \ldots, \hat{r}_K),
\end{equation}
where $\hat{r}_k = \argmax_{1 \leq i \leq r_{\max}} \frac{\hat{\lambda}_{ik}}{\hat{\lambda}_{i+1,k}}$. We have adopted the setup in the CP tensor factor model \eqref{eqn:fm} where the number of spiked eigenvalues $r_k$ remains constant across different mode-$k$ covariance. Further details of these two procedures can be found in Algorithms \ref{alg:uer} and \ref{alg:tipuper}.

\begin{algorithm}[htpb!]
\caption{Unfolded Eigenvalue Ratio Method}\label{alg:uer}
    \SetKwInOut{Input}{Input}
    \SetKwInOut{Output}{Output}
    \Input{The observations $\cY_t\in\RR^{d_1\times\cdots\times d_K}$, $t=1,...,T$, the upper bound of the number of factors $r_{\max}$. }

Evaluate $\widehat\Sigma$ in \eqref{eq:hatsigma}, and unfold it to $d\times d$ matrix $\widetilde\Sigma$, i.e. $\widetilde\Sigma = \operatorname{\mat}_{[K]}(\widehat\Sigma)$.

Obtain $\widehat\lambda_i, 1 \leq i \leq r_{\max}+1$, the top $r_{\max}+1$ eigenvalues of $\widetilde\Sigma$.

Obtain $\hat{r}^{\rm uer}$ by
\begin{align*}
\hat{r}^{\rm uer} = \argmax_{1 \leq i \leq r_{\max}} \frac{\hat{\lambda}_{i}}{\hat{\lambda}_{i+1}} .
\end{align*}

    \Output{Estimate of the number of factors $\hat{r}^{\rm uer}$.}
\end{algorithm}

\begin{algorithm}[htpb!]
\caption{Eigenvalue Ratio Method through Inner Product}\label{alg:tipuper}
    \SetKwInOut{Input}{Input}
    \SetKwInOut{Output}{Output}
    \Input{The observations $\cY_t\in\RR^{d_1\times\cdots\times d_K}$, $t=1,...,T$, the upper bound of the number of factors $r_{\max}$. }

\For{$k = 1$ to $K$}{
Evaluate
\begin{align*}
\widehat\Sigma_k = \sum_{t=1}^T \frac{\operatorname{mat}_k(\cY_t) \operatorname{mat}^\top_k(\cY_t)}{T} \in \R^{d_k \times d_k}     .
\end{align*}

Obtain $\hat{\lambda}_{ik}, 1 \leq i \leq r_{\max}+1$, the top $r_{\max} + 1$ eigenvalues of $\widehat\Sigma_k$.

Obtain $\hat{r}_k$ by
\begin{align*}
\hat{r}_k = \argmax_{1 \leq i \leq r_{\max}} \frac{\hat{\lambda}_{ik}}{\hat{\lambda}_{i+1,k}}  .
\end{align*}
}

Calculate $\hat{r}^{\rm ip}$ by
\begin{align*}
\hat{r}^{\rm ip} = \max(\hat{r}_1, \hat{r}_2, \ldots, \hat{r}_K) .
\end{align*}

    \Output{Estimate of the number of factors $\hat{r}^{\rm ip}$.}
\end{algorithm}

It is noteworthy that \cite{han2024cp} explore a similar tensor CP factor model as (\ref{eqn:fm}), albeit within a distinct setting where latent factors are assumed uncorrelated and noise follows a white noise process. Methodologically, their approach rely on the autocovariance between $\cY_{t-h}$ and $\cY_{t}$, where $h\geq 1$, whereas our method employs contemporaneous covariance. The autocovariance-based method may not be ideal for datasets with low temporal dependence, such as asset return data, which often exhibit minimal serial correlation possibly due to market efficiency.

Another closely related approach is tensor PCA (TPCA) proposed in \cite{babii2022tensor}. They consider a CP tensor factor model with orthogonal loading vectors. Unlike Tucker factor models, the identification of CP factor models does not necessarily require orthogonality. Applying TPCA to models with non-orthogonal loadings introduces a bias component of higher order than our first-stage RC-PCA. Even when the loadings are orthogonal, our contemporary variance-based iterative estimation exhibits a faster convergence rate than TPCA due to dimension and noise reduction. A comparison of our estimator with the autocovariance-based estimator and TPCA through simulation will be presented in Section \ref{sec-simul}.


\section{Theory} \label{sec-theory}

In this section, we study the statistical properties of the algorithms introduced previously. Our theoretical framework offers guarantees for consistency and outlines the statistical error rates for estimating the factor loading vectors $ a_{ik}$, where $1\le i\le r, 1\le k\le K$, given certain regularity conditions. Considering that the loading vector $ a_{ik}$ can only be identified with a change in sign, we employ
\begin{align*}
\|\widehat a_{ik}\widehat a_{ik}^\top  - a_{ik} a_{ik}^\top \|_{2}=\sqrt{1-(\widehat a_{ik}^\top  a_{ik})^2 } = \sup_{\bz\perp  a_{ik}}|\bz^\top \widehat  a_{ik}| = \| \sin\angle(\widehat a_{ik}, a_{ik}) \|_2
\end{align*}
to quantify the discrepancy between $\widehat a_{ik}$ and $ a_{ik}$. This measure provides a meaningful and practical way to assess the discrepancy between the estimated and true loadings, and we shall apply it in our simulation as well.

To present theoretical properties of the proposed procedures, we impose the following assumptions.

\begin{assumption}\label{asmp:error}
Let $\xi_t = (\xi_{1t}, \xi_{2t}, \ldots, \xi_{pt})$ be independent $p$-dimensional random vector with each entry $\xi_i$ independent and satisfying $\EE(\xi_{it}) = 0$, $\EE(\xi_{it}^2 ) = 1$ and for $0<\vartheta\le 2$
\begin{align}
\max_i\PP\left( \left| \xi_{it} \right| \ge x \right) \le c_1 \exp\left( -c_2x^{\vartheta} \right).
\end{align}
Let $\vec(\cE_t)=H \xi_t$, where $H$ is a deterministic matrix and $p\ge d$. The eigenvalues of the covariance matrix of $\vec(\cE_t)$ satisfies $C_0^{-1}\le \lambda_d(\Sigma_e)\le \cdots \le \lambda_1(\Sigma_e)\le C_0$ where $\Sigma_e=\EE \vec(\cE_t)\vec(\cE_t)^\top$ and $C_0$ is a constant.
\end{assumption}

\begin{assumption}\label{asmp:eigenvalue}
Let $F_t=(f_{1t},...,f_{rt})^\top, W=\diag(w_1,...,w_r)$, and $\Theta= W (\EE F_t F_t^\top) W$. Assume that the eigenvalues of $\EE (F_t F_t^\top)$ satisfy $c_0^{-1}\le \lambda_r(\EE F_t F_t^\top)\le\cdots \le \lambda_1(\EE F_t F_t^\top)\le c_0$ for a constant $c_0$. Define $\lambda_i=\lambda_i(\Theta)$ and, without loss generality, let $\lambda_1\ge\lambda_2\ge\cdots\ge \lambda_r>0$. For any $v\in\R^{r}$ with $\|v\|_2=1$,
\begin{align}\label{cond2}
\PP\left( \left| v^\top F_{t} \right| \ge x \right) \le c_1 \exp\left( -c_2x^{\gamma_1} \right),
\end{align}
where $c_1,c_2$ are some positive constants and $0<\gamma_1\le 2$.
\end{assumption}

\begin{assumption}\label{asmp:mixing}
Assume the factor process $f_{it}, 1\le i\le r$, is stationary and $\alpha$-mixing in $t$. The mixing coefficient satisfies
\begin{align}\label{cond1}
\alpha(m) \le \exp\left( - c_0 m^{\gamma_2} \right)
\end{align}
for some constant $c_0>0$ and $\gamma_2\ge 0$, where
\begin{align*}
\alpha(m) = \sup_t\Big\{\Big|\PP(A\cap B) - \PP(A)\PP(B)\Big|:
A\in \sigma(f_{is}, 1\le i\le r, s\le t), B\in \sigma(f_{is}, 1\le i\le r, s\ge t+m)\Big\}.
\end{align*}
\end{assumption}

Assumption \ref{asmp:error} allows for cross-sectional dependence in errors and is closely aligned with the noise conditions presented in seminal works such as \cite{bai2002}, \cite{bai2003}, \cite{lam2011}, \cite{lam2012}, and others within the factor model literature. It requires that $\xi_{it}$ has exponential-type tails, allowing us to apply large deviation theory. This is a standard condition in the literature on ultrahigh-dimensional data analysis (see, e.g., \cite{chang2023modelling,fan2013}). When $\vartheta=2$, the condition corresponds to a sub-Gaussian tail. Although it might be possible to replace this assumption with finite moment conditions through a significantly more involved theoretical analysis, we have chosen to retain it to focus on the essentials.
For simplicity, we assume that the noise tensor remains independent across time $t$, allowing for weak cross-sectional dependence. While incorporating weak temporal correlation among the noise, as suggested by \cite{bai2002}, is plausible, it substantially complicates our theoretical analysis. Therefore, we defer this exploration to future research. Nonetheless, our simulation studies show that the proposed methods remain robust even when the noise exhibits weak temporal dependence.

Assumption \ref{asmp:eigenvalue} ensures the unique identification of all factor loading vectors $ a_{ik}$ up to sign changes. Unlike the eigen decomposition of a matrix, if some $\lambda_i$ are equal, the estimation of the loading vectors $ a_{ik}$ isn't subject to rotational ambiguity but only to the signed permutation of loading vectors. Consistent with the classical literature on factor models (e.g., Assumption A in \cite{bai2002}, Assumption M(d) in \cite{stockwatson1998}), Assumption \ref{asmp:eigenvalue} allows factors to be arbitrarily correlated. Furthermore, Assumption \ref{asmp:eigenvalue} specifies that the tail probability of $f_{it}$ must exhibit exponential decay. Specifically, when $\gamma_1 = 2$, it implies that $f_{it}$ follows a sub-Gaussian distribution. We exclude the case of identical factors. In extreme cases, such as when $f_{1t}=f_{2t}$, our composite PCA method (steps 3 and 4 in Algorithm \ref{alg:initialize-cp}) will fail. However, based on Remark \ref{rmk:random_projection}, the random projection method (Procedure \ref{alg:initialize-random}) should still yield reasonable initializations. We do not pursue this direction further in our theoretical analysis, as it is beyond the scope of our current paper.

Assumption \ref{asmp:mixing} is a widely recognized standard condition that accommodates a broad range of time series models, including causal ARMA processes with continuously distributed innovations, as further detailed in works such as \cite{tong1990non, bradley2005, tsay2005analysis, fan2008nonlinear, rosenblatt2012markov, tsay2018nonlinear}, among others. 

While Assumptions \ref{asmp:error} and \ref{asmp:eigenvalue} currently assume exponential tails for both noise and factor processes, these conditions can be extended to accommodate polynomial-type tails (under bounded moment conditions) when the number of factors $r$ is fixed, albeit at the cost of a more complex theoretical analysis.

Recall $ A_k$ defined in equation (\ref{eq:e*}) with $ a_{ik}$ as its columns. 
As $ \| a_{ik}\|_2^2=1$, the correlation among columns of $ A_k$ can be measured by
\begin{align}\label{corr-k}
\delta_k = \| A_k^\top  A_k - I_{r}\|_{2}.
\end{align}
Similarly we use
\begin{align}\label{corr-all}
\delta = \| A^\top A - I_{r}\|_{2}
\end{align}
to measure the correlation of the matrix $ A = ( a_1,\ldots, a_r)\in \R^{d\times r}$ with $ a_i = \vec(\otimes_{k=1}^K  a_{ik})$ and
$d=\prod_{k=1}^K d_k$. Let $\delta_{\max} =\max\{\delta_1,\cdots \delta_K \}$. Using properties of the Kronecker product, we can show that $\delta\leq \prod_{k=1}^K \delta_k < \min_{1\leq k\leq K} \delta_k\le \delta_{\max}$.

Theorem \ref{thm:initial} below presents the performance bounds, which depends on the coherence (the degree of non-orthogonality) of the factor loading vectors.

\begin{theorem}\label{thm:initial}
Suppose Assumptions \ref{asmp:error}, \ref{asmp:eigenvalue}, \ref{asmp:mixing} hold. Let $1/\gamma=2/\gamma_1+1/\gamma_2$, and $\delta<1$ with $\delta$ defined in \eqref{corr-all}. 

\noindent (i). The eigengaps satisfy $\min\{\lam_i-\lam_{i+1},\lam_{i-1}-\lam_{i}\} \ge c \lam_r$ for all $1\le i\le r$, with $\lambda_0=\infty, \lambda_{r+1}=0$, and $c$ is sufficiently small constant. With probability at least $1-T^{-C_1}-d^{-C_1}$, the following error bound holds for the estimation of the loading vectors $ a_{ik}$ using Algorithm \ref{alg:initialize-cp},
\begin{align}\label{thm:initial:eq1}
\|\widehat a_{ik}^{\rm rcpca}\widehat a_{ik}^{\rm rcpca\top}  - a_{ik} a_{ik}^\top \|_{2} &\le \left(1+\frac{2\lambda_1}{\lambda_r}\right)\delta+\frac{C_2 \phi^{(0)} }{\lambda_r},
\end{align}
for all $1\le i\le r$, $1\le k\le K$, where $C_1,C_2$ are some positive constants, and
\begin{align}\label{thm:initial:eq2}
\phi^{(0)} &= \lambda_1 \left(\sqrt{\frac{r+ \log T}{T}} + \frac{(r+ \log T)^{1/\gamma}}{T}  \right)+ \sqrt{\frac{\lambda_1 d\log d}{T}}+\frac{\sqrt{\lambda_1 dr} \log(d)(\log T)^{1+\frac{2}{\vartheta}+\frac{1}{\gamma_1}} + d \log(d)(\log T)^{\frac{2\vartheta+4}{\vartheta}}}{T}  +1.
\end{align}

\noindent (ii). The eigengaps condition in (i) is not satisfied. Assume $\lam_1\asymp \lam_r$ and the number of random projections $L\ge Cd^2 \vee Cdr^{2(\lam_1/\lam_r)^2}$. With probability at least $1-T^{-C_1}-d^{-C_1}$, the following error bound holds for the estimation of the loading vectors $ a_{ik}$ using Algorithm \ref{alg:initialize-cp},
\begin{align}\label{thm:initial:eq1*}
\|\widehat a_{ik}^{\rm rcpca}\widehat a_{ik}^{\rm rcpca\top}  - a_{ik} a_{ik}^\top \|_{2} &\le C_3 \sqrt{ \delta_{\max} }+ C_3 \sqrt{\frac{\phi^{(0)} }{\lambda_r} }.
\end{align}
\end{theorem}

The first term of the upper limit in \eqref{thm:initial:eq1} and \eqref{thm:initial:eq1*} arises from the non-orthogonality of the loading vectors $ a_{ik}$, which can be interpreted as bias. Meanwhile, the second term in \eqref{thm:initial:eq1} stems from a concentration inequality for random noise and thus reflects a form of stochastic error.

When the eigengap condition is not met, we employ randomized projection to determine the statistical convergence rate as shown in \eqref{thm:initial:eq1*}, which is slower than the rate in \eqref{thm:initial:eq1}.
A broader result than \eqref{thm:initial:eq1*}, permitting a more general eigen ratio $\lam_1/\lam_r$ for part (ii), is detailed in the appendix.
In practice, since the sample covariance tensor includes both the average of signal-by-noise cross-products and the average of noise-by-noise cross-products, it is uncommon to encounter nearly identical sample spiked eigenvalues. Our simulation study demonstrates that while the original composite PCA provides viable initializations when $\lambda_1 = \lambda_r$, its performance is not as good as that of RC-PCA using Procedure \ref{alg:initialize-random}.

\begin{remark}\label{rmk:init}
With minor modifications to the proof of Theorem \ref{thm:initial}(i), we are able to show
\begin{align}\label{thm:initial:eq3}
\|\widehat a_{ik}^{\rm rcpca}\widehat a_{ik}^{\rm rcpca\top}  - a_{ik} a_{ik}^\top \|_{2} =O_{\PP}\left( (\lambda_1/\lambda_r)\delta+\frac{\lambda_1}{\lambda_r} \left(\sqrt{\frac{r}{T}} + \frac{r^{1/\gamma}}{T}  \right)+ \frac{\sqrt{\lambda_1 d} }{\lambda_r\sqrt{T}} + \frac{1}{\lambda_r} \right) .
\end{align}
In the typical strong factor models where $\lambda_1\asymp \lambda_r \asymp d$ (i.e. $w_i\asymp \sqrt{d}$) and $r$ fixed, the rate becomes $O_{\PP}(\delta+\sqrt{1/T}+1/d)$, aligning with the convergence rate for the vector factor model when $\delta=0$.
\end{remark}

\begin{remark}
Procedure \ref{alg:initialize-random} provides a theoretically guaranteed strategy for resolving the repeated eigenvalue issue and can, intuitively, also be applied uniformly in Algorithm \ref{alg:initialize-cp}. However, due to the random projection step, the enhanced incoherence parameter $\delta$ is compromised, resulting in slower convergence rates as indicated in \eqref{thm:initial:eq1*} compared to \eqref{thm:initial:eq1}. In contrast, \cite{auddy2023large} assumes an orthogonal decomposable tensor, thereby avoiding bias issues; for their purpose, they only provide preliminary results with a convergence rate slower than $1/4$ under the same error metric. Similarly, \cite{anandkumar2014guaranteed} considers the random projection method, but their analysis is limited to tensor CP decomposition with the eigen-ratio assumed to be of constant order.

Moreover, when the eigen-ratio is large, the number of required initializations increases exponentially with the eigen-ratio (see Theorem \ref{thm:initial}(ii)). From a computational cost perspective, the composite PCA remains valuable.

\end{remark}

Let the statistical error bound of the initialization used in Algorithm \ref{algorithm:projection} be $\psi_0$ (for example, the right hand side of \eqref{thm:initial:eq1}), and also let
\begin{align}\label{eq:final_rate}
\psi^{\ideal}=\max_{1\le k\le K} \left(  \frac{1}{\lambda_r} \sqrt{\frac{d_k\log d}{T}} + \sqrt{\frac{d_k\log d}{\lambda_r T}}+ \frac{1}{\lambda_r} \right) .
\end{align}

\begin{theorem}\label{thm:projection}
Suppose Assumptions \ref{asmp:error}, \ref{asmp:eigenvalue}, \ref{asmp:mixing} hold. Assume that $\delta_{\max}=\max_{k\le K}\delta_k<1$ with $\delta_k$ defined in \eqref{corr-k}, and $r=O(T)$. Let $1/\gamma=2/\gamma_1+1/\gamma_2$. 
Assume
\begin{align}
&C_{1,K}\sqrt{r}\psi_0+C_{1,K}\left(\frac{\lambda_1}{\lambda_r} \right) \psi_0^{2K-3}
+ C_{1,K}\sqrt{\frac{\lambda_1}{\lambda_r} }\left(\sqrt{\frac{r+\log T }{T}} + \frac{(r+\log T)^{1/\gamma}}{T}  \right) \psi_0^{K-2} \le \rho <1 .  \label{thm-projection:eq1a}
\end{align}
Then, after at most $M=O(\log (\psi_0/\psi^{\ideal}))$ iterations of Algorithm \ref{algorithm:projection}, with probability at least $1-T^{-C}- d^{-C}$, the final estimator satisfies
\begin{align}\label{thm-projection:eq2}
\|\widehat a_{ik}^{\iso} \widehat a_{ik}^{\iso\top}  - a_{ik} a_{ik}^\top \|_{2} &\le C_{0,K} \psi^{\ideal},
\end{align}
for all $1\le i\le r$, $1\le k\le K$, where $C_{0,K}$ and $C_{1,K}$ are some constants depending on $K$ only and $C$ is a positive numeric constant.
\end{theorem}

It is important to note that the error bound $\psi_0$ for initialization is intended for each individual loading vector $ a_{ik}$. When applying Algorithm \ref{algorithm:projection}, which requires the inverse of $\widehat A_k^\top \widehat A_k$, the condition $\sqrt{r}\psi_0\lesssim 1$ ensures a reliable initial estimate of the loading matrix $\widehat A_k$. Although the other components in \eqref{thm-projection:eq1a} may seem complex, they are designed to ensure the error contraction effect in each iteration. This ensures that as iterations progress, the error bound will approach the desired statistical upper bound. A more detailed discussion of initial estimates can be found in Appendix \ref{appendix:initial}.

\begin{remark}\label{rmk:thm2}
With slightly modifications to the proof of Theorem \ref{thm:projection}, we can show
\begin{align}\label{thm-projection:eq3}
\max_i\|\widehat a_{ik}^{\iso} \widehat a_{ik}^{\iso\top}  - a_{ik} a_{ik}^\top \|_{2} =O_{\PP}\left( \sqrt{\frac{d_{k} }{\lambda_r T}} + \frac{1}{\lambda_r} \right) .
\end{align}
In the typical strong factor models where $\lambda_1\asymp \lambda_r \asymp d$ (i.e. $w_i\asymp \sqrt{d}$), the rate simplifies to $O_{\PP}(\sqrt{d_{k}/(dT)}+1/d)$. This rate is significantly faster than that found in the vector factor model.
Moreover, in comparison with the initial estimator discussed in Remark~\ref{rmk:init}, the term $1/d$ arises from the cross-sectional dependence among the noise tensor $\cE_t$ in \eqref{eqn:fm}, and is therefore irreducible. Notably, the bias $\delta$ is eliminated by the iterative refinement process. More importantly, the main source of estimation error is reduced from $O_{\PP}(\sqrt{1/T})$ to $O_{\PP}(\sqrt{d_{k}/(dT)})$.
\end{remark}

\begin{remark}
To further examine the constant in the error contraction condition \eqref{thm-projection:eq1a}, we assume that the constant in the convergence rate of the final estimator \eqref{thm-projection:eq2} satisfies $C_{0,K}=C_0 \alpha^{2-2K}$, for some absolute constant $C_0>0$, and define $\alpha = \sqrt{1-\delta_{\max}}-(r^{1/2}+1)\psi_0/\sqrt{1-1/(4r)}$. Then, condition \eqref{thm-projection:eq1a} can be detailed as follows:
\begin{align*}
&\alpha>0,\ C_0 \alpha^{2-2K}\left(\frac{\lambda_1}{\lambda_r} \right) \psi_0^{2K-3} <1,\\
& C_0 \alpha^{2-2K}\sqrt{\frac{\lambda_1}{\lambda_r} }\left(\sqrt{\frac{r+\log T }{T}} + \frac{(r+\log T)^{1/\gamma}}{T}  \right) \psi_0^{K-2}  <1   .
\end{align*}
Assuming $\lambda_1\asymp \lambda_2\cdots\asymp\lambda_r$, condition \eqref{thm-projection:eq1a} requires that $r\psi_0\lesssim 1$, which in turn implies $r\delta \lesssim1$ under the setting of Theorem~\ref{thm:initial}(i). This condition is automatically satisfied under orthogonal loadings, but it also accommodates the case with mild non-orthogonal loadings.
\end{remark}

The following Theorem \ref{thm:factors} specifies the convergence rate for the estimated factors $f_{it}$\footnote{Asymptotic normality results for estimated factors are developed in our companion paper on diffusion index forecasting: \url{https://papers.ssrn.com/sol3/papers.cfm?abstract_id=5213594}, which builds directly on the theoretical framework established here.}.
\begin{theorem}\label{thm:factors}
Suppose Assumptions \ref{asmp:error}, \ref{asmp:eigenvalue}, \ref{asmp:mixing} hold. Assume that $\delta_{\max}=\max_{k\le K}\delta_k<1$ with $\delta_k$ defined in \eqref{corr-k}, $r=O(T)$, $1\lesssim \lambda_r$ and condition \eqref{thm-projection:eq1a} holds. Let $d_{\max}=\max_k d_k$.
Then, after at most $M=O(\log (\psi_0/\psi^{\ideal}))$ iterations of Algorithm \ref{algorithm:projection}, with probability at least $1-T^{-C}- d^{-C}$, the final estimator satisfies
\begin{align}
w_i^{-1} \left| \widehat w_i \widehat f_{it}- w_i f_{it} \right| \le C \left( \sqrt{\frac{d_{\max} \log d}{\lambda_r T}} + \sqrt{\frac{1}{\lambda_r}} \right)  \label{thm-factors:eq1}
\end{align}
for $1\le i\le r, 1\le t\le T$, and some constant $C>0$.
\end{theorem}

Note that $w_i$ is a scalar representing the factor strength. In a strong factor model, consistent with \cite{bai2003,stock2002,chen2023statistical}, it is typically set as $w_i=\sqrt{d}=\sqrt{d_1d_2\cdots d_K}$, and we use $\widehat w_i \widehat f_{it}/w_i$ as the estimated factors for further interpretation and analysis.

We now demonstrate the feasibility of obtaining a more precise bound by closely examining the leading order term. This process allows us to ascertain the asymptotic behavior of the estimator $ a_{ik}$. Specifically, we will establish that
\begin{align*}
\widehat a_{ik}^{\iso} -\text{sign}(a_{ik}^{\top} \widehat a_{ik}^{\iso} ) a_{ik} =P_{ a_{ik},\perp} \left[\frac{1}{ \Theta_{ii}T} \sum_{t=1}^T f_{it} \left( \cE_t \times_{\ell\in [ K] \backslash{k} } b_{i\ell}^{\top} \right) \right] + O_{\PP}\left(\frac{1}{\Theta_{ii}}\left( \frac{d_k}{T} + \sqrt{\frac{d_k}{T}} + 1\right) \right),
\end{align*}
where $P_{ a_{ik},\perp}=I_{d_k}- a_{ik} a_{ik}^\top$ and $\Theta=(\Theta_{ij})_{r\times r}$, with $\Theta$ defined in Assumption \ref{asmp:eigenvalue}.
This enables the determination of asymptotic distributions for linear forms of $a_{ik}$.

The following theorem shows the asymptotic distribution of a linear form of the factor loading vector $u^\top  a_{ik}$ for some fixed vector $u$. Note that in the strong factor model, we have $\Theta_{ii}\asymp w_i^2 \asymp d$ for all $1\le i\le r$.

\begin{theorem}\label{thm:clt}
Suppose the conditions in Theorem \ref{thm:projection} are satisfied. Let $\lambda_1\asymp \lambda_r \asymp \Theta_{ii}$. Assume that $\lim\inf_{d_k\to\infty}$ $\|P_{ a_{ik},\perp} u\|_2>0$, for each $1\le i\le r, 1\le k\le K$, we have:
\begin{enumerate}
\item[(i)] If $T/(d_k \lambda_r)\to 0$, then
\begin{equation}\label{eqn:clt1}
\sqrt{T\Theta_{ii}}\sigma_{u,ik}^{-1} u^\top\left(\widehat a_{ik}^{\iso} -\text{sign}(\widehat a_{ik}^{\iso\top}  a_{ik})\cdot a_{ik} \right)\xrightarrow{d} N(0,1),
\end{equation}
where $\sigma_{u,ik}^2= h_{ik}^\top\Sigma_e h_{ik}$, $h_{ik}= b_{iK}\odot \cdots \odot b_{i,k+1} \odot P_{ a_{ik},\perp}u \odot b_{i,k-1}\odot \cdots \odot b_{i1} \in \R^d$ and $\odot$ represents Kronecker product.

\item[(ii)] If $d_k \lambda_r=O(T)$, then
\begin{equation}\label{eqn:clt2}
\Theta_{ii} u^\top\left(\widehat a_{ik}^{\iso} -\text{sign}(\widehat a_{ik}^{\iso\top}  a_{ik} )\cdot a_{ik} \right)=O_{\PP}(1).
\end{equation}
\end{enumerate}
\end{theorem}


\begin{remark}
Theorem \ref{thm:clt} is analogous to Theorem 2 of \cite{bai2003}. The dominant case is (i), which exhibits asymptotic normality, while case (ii) is of theoretical interest when a specific convergence rate is required. In the typical strong factor model,  we have $\Theta_{ii}\asymp w_i^2\asymp d$ for all $1\le i\le r$. Thus if $T/(d_kd)\rightarrow0$, case (i) applies; conversely, if $d_kd=O(T)$, the error from the noise covariance dominates, leading to case (ii). In practice, we can estimate $\Theta$ by $\hat \Theta=\frac{1}{T}\sum_{t=1}^T\hat W \hat{F}_t\hat{F}_t^\top \hat{W}$ and then compare $d_k\hat \Theta_{ii}$ with $T$ to determine the appropriate case. In the special case where the noise tensor $\cE_t$ has i.i.d. entries, only case (i) is applicable.
\end{remark}

\begin{remark}
    The asymptotic normality results provide theoretical guarantees that allow us to construct confidence intervals and conduct hypothesis tests. For example, with consistent estimators $\hat \Theta$ and $\hat\sigma_{u,ik}^2= \hat h_{ik}^\top\hat \Sigma_e \hat h_{ik}$, we can form a $(1-\alpha)$-level confidence interval for $\widehat a_{ijk}^{\iso}$ as $\left(\widehat a_{ijk}^{\iso}-q_{1-\alpha/2}\sqrt{T\hat \Theta_{ii}}\hat \sigma_{u,ik}^{-1},\right.$
    $\left.\widehat a_{ijk}^{\iso}+q_{1-\alpha/2}\sqrt{T\hat \Theta_{ii}}\hat \sigma_{u,ik}^{-1}\right)$, $1\le i\le r, 1\le j \le d_k, 1\le k\le K$, where $q_{1-\alpha/2}$ is the $1-\alpha/2$ quantile of the standard normal distribution. Figures \ref{fig:char_loading_ci} and \ref{fig:dec_loading_ci} in Section \ref{sec-applit} illustrate this by reporting estimated loadings for each characteristic and decile with 95\% confidence intervals. Beyond this, asymptotic normality opens the door to a range of testing strategies that are central in empirical asset pricing. As \citet{bai2003} emphasizes, estimated factor loadings can be used to construct portfolios in the spirit of \citet{lehmann1988}. Similarly, \citet{lettau20243d} interpret Tucker factor loadings as portfolio weights. Our asymptotic distribution theory for CP loadings makes analogous portfolio-based inference feasible: for example, comparing the relative importance of portfolio weights associated with different characteristics\footnote{A further illustration comes from alpha testing. In empirical asset pricing, an important question is whether portfolio alphas vanish after controlling for observed and latent factors, as in \citet{giglio2021thousands}. Our CP factor model could potentially be extended to this context. The asymptotic normality results in Theorem \ref{thm:clt} provide the foundation for constructing valid \textit{t}-statistics for such tests. Exploring this extension would require substantial additional work and therefore lies beyond the scope of the present paper, but it illustrates how our inferential framework can, in principle, be applied to important empirical questions in asset pricing.}.
\end{remark}

Since $\sigma_{u,ik}^2\asymp d_k$, Theorem \ref{thm:clt}(i) implies that $u^\top\left(\widehat a_{ik}^{\iso} -\text{sign}(\widehat a_{ik}^{\iso\top}  a_{ik})\cdot a_{ik} \right)=O_{\PP}(\sqrt{d_k/(\lambda_r T})$. In contrast, Theorem \ref{thm:clt}(ii) yields $u^\top\left(\widehat a_{ik}^{\iso} -\text{sign}(\widehat a_{ik}^{\iso\top}  a_{ik})\cdot a_{ik} \right)=O_{\PP}(1/\lambda_r) $.
In Theorem \ref{thm:clt}, we focus on vectors $u$ with the property that $\|P_{ a_{ik},\perp} u\|_2>0$ when $d_k$ is large, which effectively assumes $\sin\angle(u, a_{ik})> 0$. Conversely, when $\sin\angle(u, a_{ik}) =0$, the convergence rate of the estimated linear form is faster, and its asymptotic distribution is a mixture of $\chi_1^2$ distributions.\footnote{Theorem 4.5 complements Theorem 4.4 by addressing the case where $\sin\angle(u, a_{ik}) =0$, i.e., when the direction of interest $u$ is aligned with the loading vector $a_{ik}$. In most economic applications, however, we are interested in inference for a particular component of the loading, in which case $u$ is a unit vector aligned with a coordinate axis. Hence, the condition $\sin \angle(u, a_{ik}) > 0$ is typically satisfied, and Theorem 4.4 remains applicable. As a safeguard, one can check whether $\|P_{\widehat a_{ik}, \perp} u\|_2$ is close to zero using the estimated $\widehat a_{ik}$ and the user-specified $u$. }
Similarly, we will establish that
\begin{align*}
1-\left( \widehat a_{ik}^{\top}  a_{ik} \right)^2 =&  \frac{1}{\Theta_{ii}^2 } \left[\frac{1}{ T} \sum_{t=1}^T w_i f_{it} \left( \cE_t \times_{\ell\in [ K] \backslash\{k\} }b b_{i\ell}^{\top} \right) \right]^\top  P_{ a_{ik},\perp}  \left[\frac{1}{ T} \sum_{t=1}^T w_i f_{it} \left( \cE_t \times_{\ell\in [ K] \backslash\{k\} } b_{i\ell}^{\top} \right) \right]    \notag\\
&+O_{\PP}\left( \frac{1}{\Theta_{ii}^2}\left( \frac{d_k\log(d)}{T} + \sqrt{\frac{d_k \log(d)}{T}} + 1\right)^2 \right) .
\end{align*}

\begin{theorem}\label{thm:clt2}
Suppose the conditions in Theorem \ref{thm:projection} are satisfied. Let $\lambda_1\asymp \lambda_r \asymp \Theta_{ii}$. For each $1\le i\le r, 1\le k\le K$, we have:
\begin{enumerate}
\item[(i)] With probability at least $1-T^{-C}- d^{-C}$,
\begin{equation}\label{eqn2:clt1}
1-(\widehat a_{ik}^{\iso\top} a_{ik})^2 \le C_{0,K} (\psi^{\ideal})^2,
\end{equation}
where $\psi^{\ideal}$ is defined in \eqref{eq:final_rate}.

\item[(ii)] If $T/(d_k \lambda_r)\to 0$, then
\begin{equation}\label{eqn2:clt2}
T\Theta_{ii} d_k^{-1}\left( 1-(\widehat a_{ik}^{\iso\top} a_{ik})^2 \right)\xrightarrow{d} \sum_{j=1}^{d_k} \varpi_j \chi_1^2,
\end{equation}
where $\varpi_j,1\le j\le d_k$ are the eigenvalues of $\Sigma_{ik}^{1/2} P_{ a_{ik},\perp}\Sigma_{ik}^{1/2}$, with $\Sigma_{ik}=d_k^{-1} \EE[ (\cE_t\times_{\ell\neq k}^K  b_{i\ell})(\cE_t\times_{\ell\neq k}^K  b_{i\ell})^\top ]$.

\item[(iii)] If $d_k \lambda_r=O(T)$, then
\begin{equation}\label{eqn2:clt3}
\Theta_{ii}^2 \left( 1-(\widehat a_{ik}^{\iso\top} a_{ik})^2 \right)=O_{\PP}(1).
\end{equation}
\end{enumerate}
\end{theorem}

Drawing parallels with traditional PCA is insightful; in PCA, a debiasing process is often necessary to achieve asymptotic normality in linear combinations of the principal components, as discussed in \cite{koltchinskii2016asymptotics, koltchinskii2017concentration, koltchinskii2020efficient}. For the CP tensor factor model, however, merely meeting the signal strength requirement $T/(d_k \Theta_{ii})\to 0$ is enough to render the bias inconsequential. This observation aligns with findings by \cite{bai2003} regarding vector factor models.

The estimators are constructed with a specified rank $r$, although in the theoretical analysis, $r$ is allowed to increase. Practically, $\hat r$ can be estimated using the generalized eigenvalue ratio-based estimators detailed in Algorithms \ref{alg:uer} or \ref{alg:tipuper}. The asymptotic validity of $\hat r^{\rm uer}$ and $\hat r^{\rm ip}$ are established in Theorem \ref{thm:rank} below.

\begin{theorem}\label{thm:rank}
Suppose Assumptions \ref{asmp:error}, \ref{asmp:eigenvalue}, \ref{asmp:mixing} hold and $r_{\max}$ is a predetermined constant no smaller than $r$. Assume $r=O(T)$ and $\lambda_r^{-1/2}d^{1/2}T^{-1/2} + \lambda_r^{-1}=o(1)$. Then
\begin{align*}
&\PP ( \hat{r}^{\rm uer}=r) \to 1,    \\
&\PP ( \hat{r}^{\rm ip}=r) \to 1,
\end{align*}
as $d_k\to \infty$ and $T\to \infty$.
\end{theorem}
Theorem \ref{thm:rank} derives the consistency of rank estimators $\hat r^{\rm uer}$ and $\hat r^{\rm ip}$ based on the eigenvalue ratios. This can be viewed as a generalization of Theorem 1 of \cite{ahn2013} from vector factor models to CP tensor factor models.


\section{Simulation} \label{sec-simul}

In this section, we conduct empirical comparisons among different methods for estimating loading vectors across various simulation scenarios, and verify the limiting distribution of the estimated loading vectors. We assess the performance of the contemporary covariance-based iterative simultaneous orthogonalization procedure (CC-ISO) proposed in this paper, auto-covariance-based iterative simultaneous orthogonalization procedure by \cite{han2024cp} (AC-ISO), and TPCA by \cite{babii2022tensor}. The auto-covariance considered by \cite{han2024cp} is defined by the following lagged-cross product operator:
$$
\boldsymbol{\Sigma}_h = \mathbb{E}\left[ \frac{\mathcal{Y}_{t-h} \otimes \mathcal{Y}_t}{T-h} \right] \in \R^{d_1 \times \cdots d_K \times d_1 \times \cdots \times d_K}.
$$
In this section, we fix $h = 1$. The estimation error measures the angle between the estimated loading vector and the true loading vector, computed as:
$$
\max_{i,k} \| \hat a_{ik} \hat a_{ik}^\top - a_{ik} a_{ik}^\top  \|_2 = \max_{i,k} \sqrt{1-(\widehat a_{ik}^\top  a_{ik})^2} .
$$



Throughout our analysis, the observations $\cY_t$'s are simulated according to model \eqref{eqn:fm} with $K=2$ and $r=3$. As mentioned in Section \ref{sec:estimation}, we fix $c_0 = 0.1$, $\nu = 0.8$, $L = 2r^2$, $\epsilon = 10^{-5}$ and $M = 100$. The true loading vectors are generated as follows. The elements of matrices $\widetilde{A}_k=\left(\widetilde{a}_{1 k}, \ldots, \widetilde{a}_{r k}\right) \in$ $\mathbb{R}^{d_k \times r}, 1 \leqslant k \leqslant K$, are drawn from i.i.d. $N(0,1)$ distributions and then orthonormalized via QR decomposition. To conveniently control nonorthogonality, we introduce the parameter $\eta=\max_{1\le i<j\le r}|a_i^\top a_j|$, analogous to $\delta$ defined in \eqref{corr-all}. It can be shown that $\delta \leq (r-1)\eta$.
If $\eta=0$, we simply set $A_k=\widetilde{A}_k$; otherwise, we set $a_{1 k}=\widetilde{a}_{1 k}$ and $a_{i k}=\left(\widetilde{a}_{1 k}+\theta \widetilde{a}_{i k}\right) /\left\|\widetilde{a}_{1 k}+\theta \widetilde{a}_{i k}\right\|_2$ for all $i \geqslant 2$ and $1 \leqslant k \leqslant K$, with $\theta=\left(\eta^{-2 / K}-1\right)^{1 / 2}$. By varying $\eta$, we control the correlations between loading vectors. It is evident that an increase in $\delta$ (or equivalently $\eta$) leads to a higher degree of linear dependence among the loading vectors.

The factor processes $f_{it}$ exhibit weak temporal dependence and are generated as an independent AR(1) process, with the factor strength $w_i$ being a scalar depending on $d_1, d_2$ and $r$:
\begin{align}\label{eqn:factorDGP}
f_{i,t+1} &= \phi f_{it} + \epsilon_{it}   ,\\
w_i &= \frac{1}{5} (r-i+1) \sqrt{d_1 \times d_2},
\end{align}
where $\phi=0.1$, $\operatorname{Var}(\epsilon) = 1 - \phi^2 = 0.99$.
In this case, the tensor factor model in \eqref{eqn:fm} represents a typical strong factor model, where $\lambda_i = w_i^2 = (r-i+1)^2 d_1 d_2 / 25 = O(d)$ when $r$ is fixed. The results with $\phi=0.5$ are reported in Appendix \ref{appendix:simulation}.

The following three configurations are adapted from \cite{babii2022tensor} and \cite{han2024cp} with modifications made for comparative analysis of the empirical performances among TPCA, AC-ISO and CC-ISO.
In these configurations, $(d_1, d_2) \in \{(40,40),(40,60),(60,60)\}$, $T \in \{100,300,500\}$ and $r = 3$.

\begin{itemize}
    \item[I.] (Orthogonal loading matrix) Set $\eta = 0$ so that the columns of loading matrix $A_k$ are orthonormal. Each entry of error term $\cE_t$ is generated independently from $N(0,1)$.
    \item[II.] (Non-orthogonal loading matrix) Vary $\eta$ in the set $\{0.05,0.15,0.25\}$ so that the columns of loading matrix $A_k$ are not orthogonal. Each entry of error term $\cE_t$ is generated independently from $N(0,1)$.
    \item[III.](Serial correlation in $\cE_t$) Set $\eta = 0.1$. The errors $\cE_t$ are generated according to $\cE_t = \Psi_1^{1/2} Z_t \Psi_2^{1/2}$, where
    \begin{itemize}
        \item $\Psi_1 = \Psi_2 = \{ \sigma_{e,ij} \}$ with $\sigma_{e,ij} = 0.5^{|i-j|}$;
        \item $\vec(Z_t) = \Phi \vec(Z_{t-1}) + U_t$ where $U_t \sim i.i.d.\ N(0, I_{d_1 d_2})$ and $\Phi \in \R^{d_1 d_2 \times d_1 d_2}$ is a diagonal matrix with all diagonal elements equal to $\rho$.
    \end{itemize}
    We vary $\rho$ in the set $\{0.1, 0.3, 0.5\}$ to investigate the robustness of our algorithm under weak cross-sectional correlation and serial correlation in the error term.
\end{itemize}

The following configuration aims to assess the robustness of our proposed algorithm under weak factor structures.

\begin{itemize}
        \item[IV.](Weak factors) Set $r = 3$, and $\eta = 0.1$. The error terms are generated according to  $\cE_t = \Psi_1^{1/2} Z_t \Psi_2^{1/2}$, where
        \begin{itemize}
            \item $\Psi_1 = \Psi_2 = \{ \sigma_{e,ij} \}$ with $\sigma_{e,ij} = 0.5^{|i-j|}$;
            \item $Z_{ijt} \sim i.i.d.\ N(0,1)$.
        \end{itemize}
        The scaling multiplier in factor process $w_i = (r-i+1) (d_1 d_2)^{1/\alpha} $, where $\alpha$ varies in the set $\{2.5,3,3.5,4\}$. Note that when $\alpha = 2$, the factor structure is considered strong.  A larger $\alpha$ indicates a weaker factor structure.
\end{itemize}

\begin{figure} [htbp!]
    \centering
    \includegraphics[width = 0.6\columnwidth]{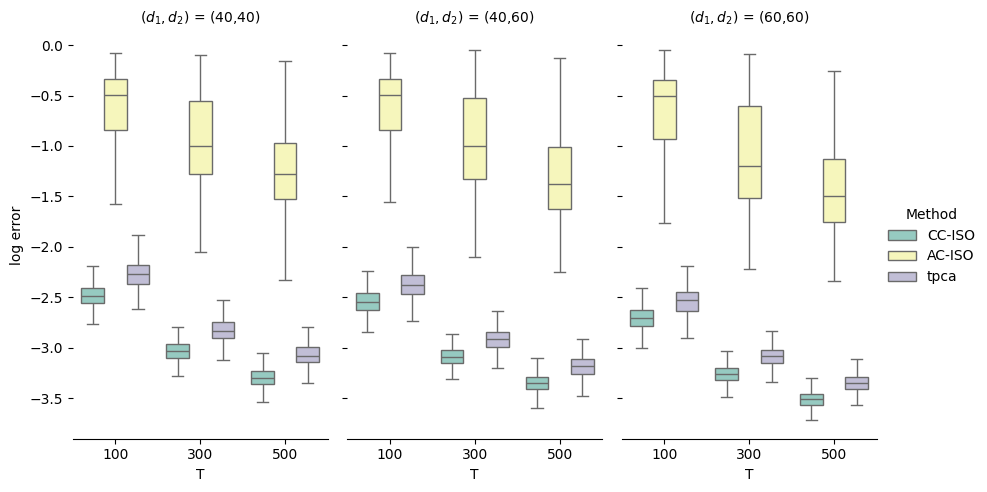}
    \caption{Boxplots of the estimation error over 500 replications under configuration I}
    \label{fig:config1}
\end{figure}

For each configuration, we conduct the experiment 500 times and present the box plots of the results. Figure \ref{fig:config1} shows the estimation errors for CC-ISO, AC-ISO and TPCA under configuration I. Notably, CC-ISO consistently outperforms the other two algorithms across various dimensions. The estimation by AC-ISO deviates significantly from the true value due to the weak signal in the auto-covariance matrix resulting from the weak temporal dependence in the factor process.

\begin{figure} [htbp!]
    \centering
    \includegraphics[width = 0.6\columnwidth]{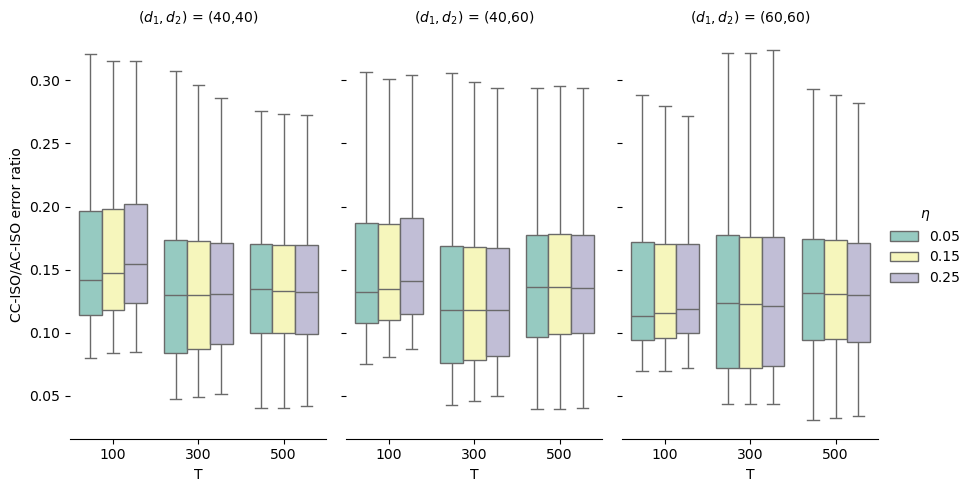}
    \includegraphics[width = 0.6\columnwidth]{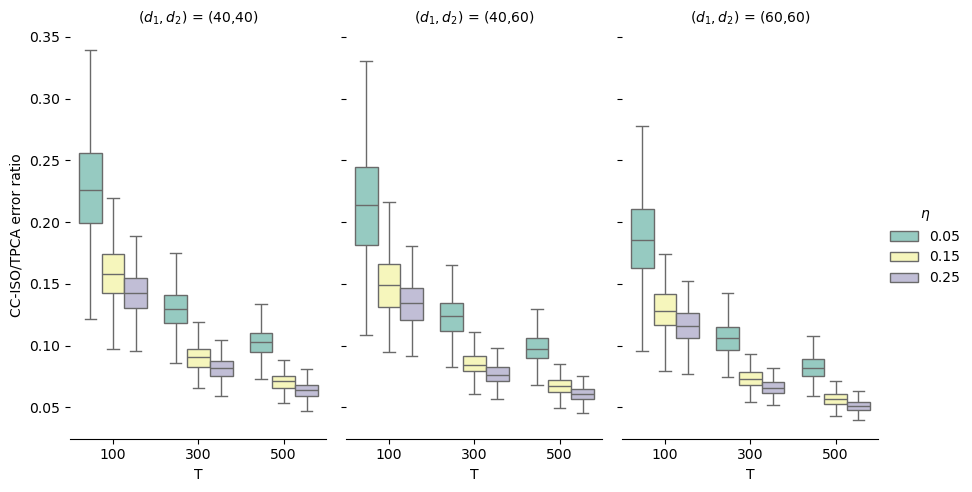}
    \caption{Boxplots of the estimation error over 500 replications under configuration II. Note: The first panel shows the ratio of the estimation error of CC-ISO on AC-ISO. The second panel shows the ratio of the estimation error of CC-ISO on TPCA.}
    \label{fig:config2}
\end{figure}

In Configuration II, we assess the impact of non-orthogonal factor loadings on estimation. Figure \ref{fig:config2} shows the ratio of the estimation errors of CC-ISO to AC-ISO (first panel) and to TPCA (second panel) across different values of $\eta$ and dimensions. In the first panel, the error ratio of CC-ISO to AC-ISO remains around 0.15, indicating the superior accuracy of CC-ISO. The ratio remains relatively stable because the signal part in AC-ISO, albeit small, also increases with dimensions, limiting its relative improvement. In contrast, in the second panel, the error ratio of CC-ISO to TPCA decreases and converges as dimensions grow. This is because TPCA is unable to recover non-orthogonal factor loading vectors, resulting in stable estimation errors regardless of dimension. Meanwhile, CC-ISO effectively identifies non-orthogonal loadings, leading to estimation errors that diminish and eventually converge to zero.

\begin{figure} [htbp!]
    \centering
    \includegraphics[width = 0.6\columnwidth]{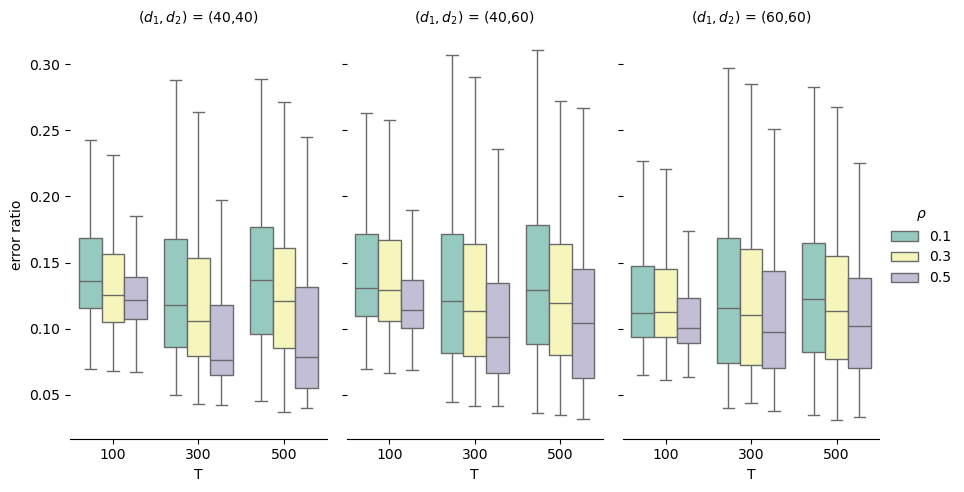}
    \caption{Boxplots of the estimation error over 500 replications under configuration III}
    \label{fig:config3}
\end{figure}

Figure \ref{fig:config3} shows the ratio of estimation errors of CC-ISO to AC-ISO under configuration III, designed to evaluate the robustness of the proposed CC-ISO algorithm against serial correlation in the error term. We observe that CC-ISO's performance improves monotonically as $T$ increases. In contrast, AC-ISO's performance deteriorates as the serial correlations in the error term strengthen. This decline is due to the contamination of signals in the auto-covariance by the serial correlations in the error terms. However, CC-ISO demonstrates robustness against such serial correlations.


\begin{figure} [htbp!]
    \centering
    \includegraphics[width = 0.6\columnwidth]{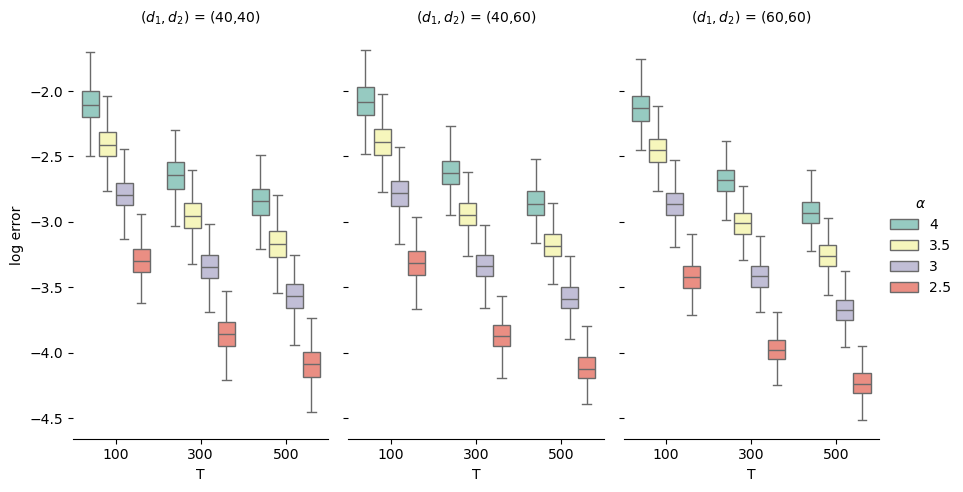}
    \caption{Boxplots of the estimation error over 500 replications under configuration IV}
    \label{fig:config4}
\end{figure}

In Figure \ref{fig:config4}, we show the box plots of the logarithm of the estimation errors of CC-ISO algorithm under a weak factor configuration. It is evident that the estimation errors decrease as $T$ increases. Additionally, the rate of decrease in estimation errors depends on the value of $\alpha$: a higher $\alpha$ leads to a faster decrease. These results validate the robustness of the CC-ISO algorithm against a certain degree of weak factor structure, in line with the conclusions drawn in Theorem \ref{thm:projection}.

We also examine the performance of Randomized Projection (RP) from Procedure \ref{alg:initialize-random} and compare it with composite PCA.
\begin{itemize}
    \item[V.](composite PCA vs. RC-PCA) $r = 5$. $d_1 = d_2 = \Bar{d}$ with $\Bar{d} \in \{20,40,80\}$ and $T \in \{100,200,500\}$. The columns of factor loadings $A_k$ are orthonormal and are generated as described in Configuration I. Furthermore, the factors $f_{it}$ are also orthonormal, generated using QR decomposition after deriving from AR(1) processes. In this setting, the singular values of the common components $\sum_{i=1}^r w_i f_{it} a_{i1} \otimes a_{i2}$ are solely determined by $w_i$. We set $w_i = w = 10$ to ensure identical eigenvalues of common components. Error terms are generated from i.i.d. $N(0,1)$. Though the top $r$ eigenvalues of the $\widetilde\Sigma$ are not identical due to noise, their differences are relatively small, allowing randomized projection algorithms to ensure the accuracy of initial estimations. For the remaining parameters, we set $\nu = 0.8$, $c_0 = 0.1$ and $L = 2 r^2$.
\end{itemize}

\begin{figure} [htbp!]
    \centering
    \includegraphics[width = 0.8\columnwidth]{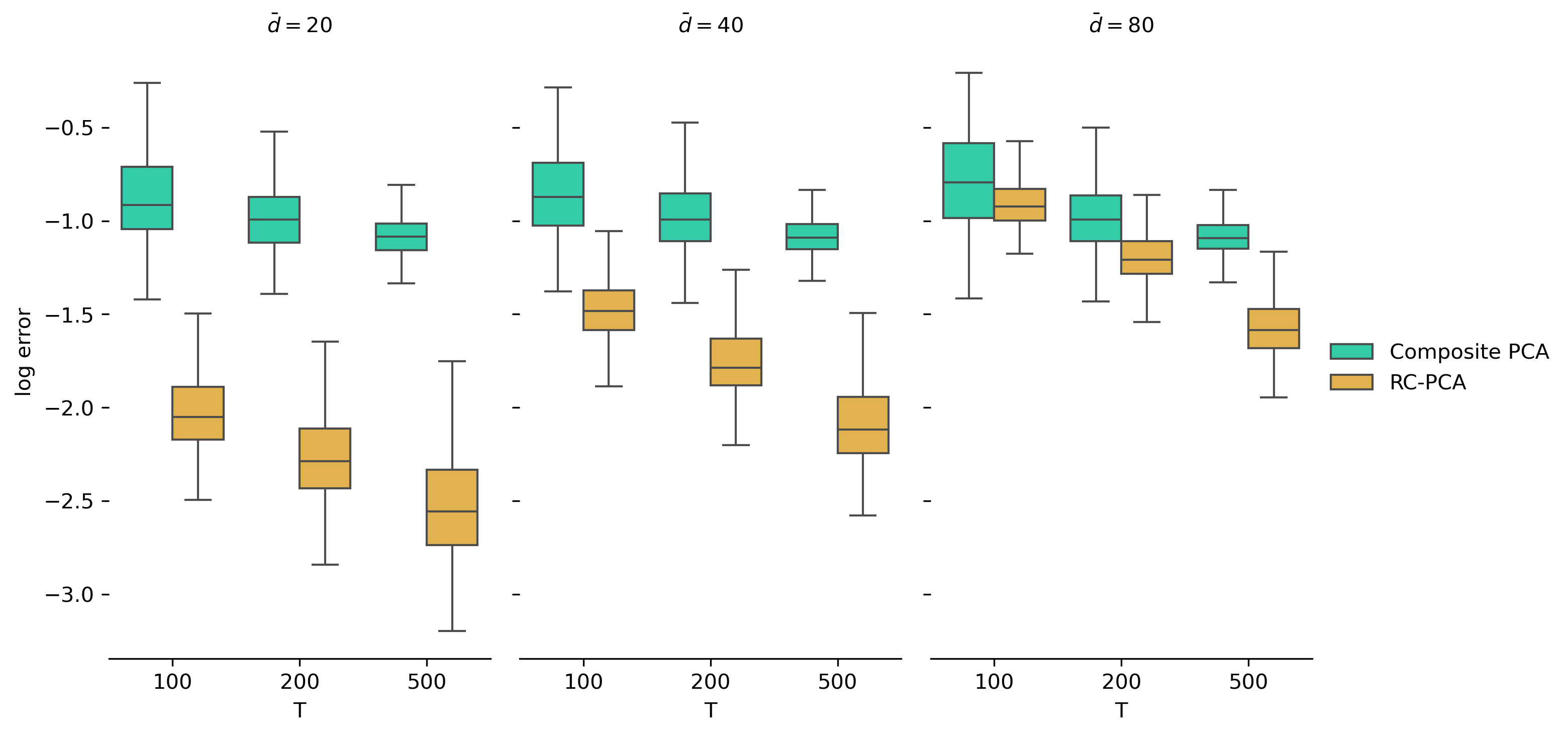}
    \caption{Boxplots of the estimation error over 500 replications under configuration V}
    \label{fig:config5}
\end{figure}

Given the close empirical performances of CC-ISO under both initialization methods under configuration V, our focus shifts to the estimation errors of the initial estimations, as illustrated in Figure \ref{fig:config5}. RC-PCA outperforms composite PCA in terms of the accuracy of initial estimations, particularly pronounced when $d$ is smaller and $T$ is larger. This occurs because the sample covariance of $\operatorname{Vec}(\cE_t)$ approaches the identity matrix when $d$ is small and as $T$ increases. Consequently, $\widetilde\Sigma$ is more likely to have eigenvalues that are closer together.

The subsequent simulation verifies the robustness of the CC-ISO algorithm against weak mis-specification of the model. The data is generated following the Tucker factor model with $K = 2$:

$$
\cY_t = \bA_1 \cF_t \bA_2^\top + \cE_t,
$$
where $\cF_t \in \R^{r \times r}$ is the factor process in the Tucker factor model. In the CP factor model, $\cF_t$ is diagonal with the $i^{th}$ diagonal element equal to $f_{it}$. In the mis-specification setting, we allow the off-diagonal entries to deviate from 0. Denote the $(i,j)^{th}$ entry of $\cF_t$ by $\cF_{t,ij}$. Let $\cF_{t,ij} = w_{ijt} f_{ijt}$, where $f_{ijt}$ is generated as specified in \eqref{eqn:factorDGP}. The configuration is as follows:

\begin{itemize}
    \item[VI.] (Mis-specification) $r = 3, (d_1, d_2) \in \{(40,40),(40,60),(60,60)\}$ and $T \in \{100,300,500\}$. The loading vectors and error terms are generated as in Configuration IV, allowing for correlation between loading vectors and weak cross-sectional correlation in the error term. Set $w_{iit} = \sqrt{d_1 d_2} / 5$ and $w_{ijt} = (d_1 d_2)^{1/\alpha} / 5$ if $i \neq j$ with $\alpha \in \{3,4,5\}$. A smaller $\alpha$ indicates a more severe mis-specification in the model.
\end{itemize}

\begin{figure}
    \centering
    \includegraphics[width = 0.8\columnwidth]{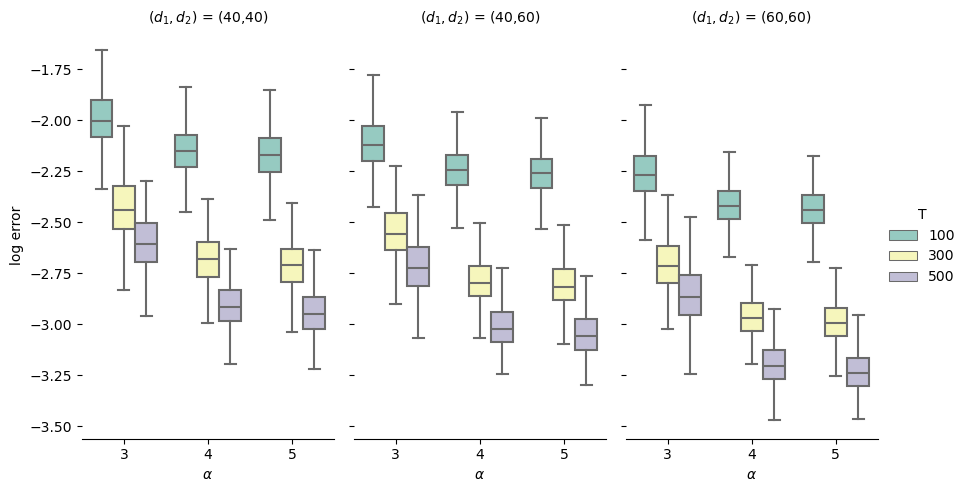}
    \caption{Boxplots of the estimation error over 500 replications under configuration VI}
    \label{fig:config6}
\end{figure}

Figure \ref{fig:config6} shows the results under configuration VI. Given $\alpha$, the estimation error decreases in $T$ or in $d$, which illustrates the robustness of CC-ISO against weak mis-specification.

Next simulation is conducted to verify the results in Theorem \ref{thm:clt}(i). The configuration is as follows:
\begin{itemize}
    \item[VII.] (CLT) $r = 3$. $d_1 = d_2 = \Bar{d} \in \{20,60,100\}$. For each $\Bar{d}$, we set $T = 200$ and $w_i = (r-i+1)\sqrt{d_1d_2}$. For factor loading vectors, we let $\eta = 0.1$ to allow for non-orthogonal loading vectors. The error $\cE_{i,j,t}$ are generated as in Configuration IV to allow for weak cross-sectional correlations. We simulate the distribution of $a_{ik}$ in \eqref{eqn:clt1} with $i = 1$, $k = 1$ under three choices of $u$: $u_1 = 1 / \sqrt{\Bar{d}}$, $u_{2} = [1,0,0, \ldots,0]^\top$ and $u_3 = [0,1,0,0,\ldots,0]^\top$.
\end{itemize}

\begin{figure} [htbp!]
    \centering
    \includegraphics[width = 0.95\columnwidth]{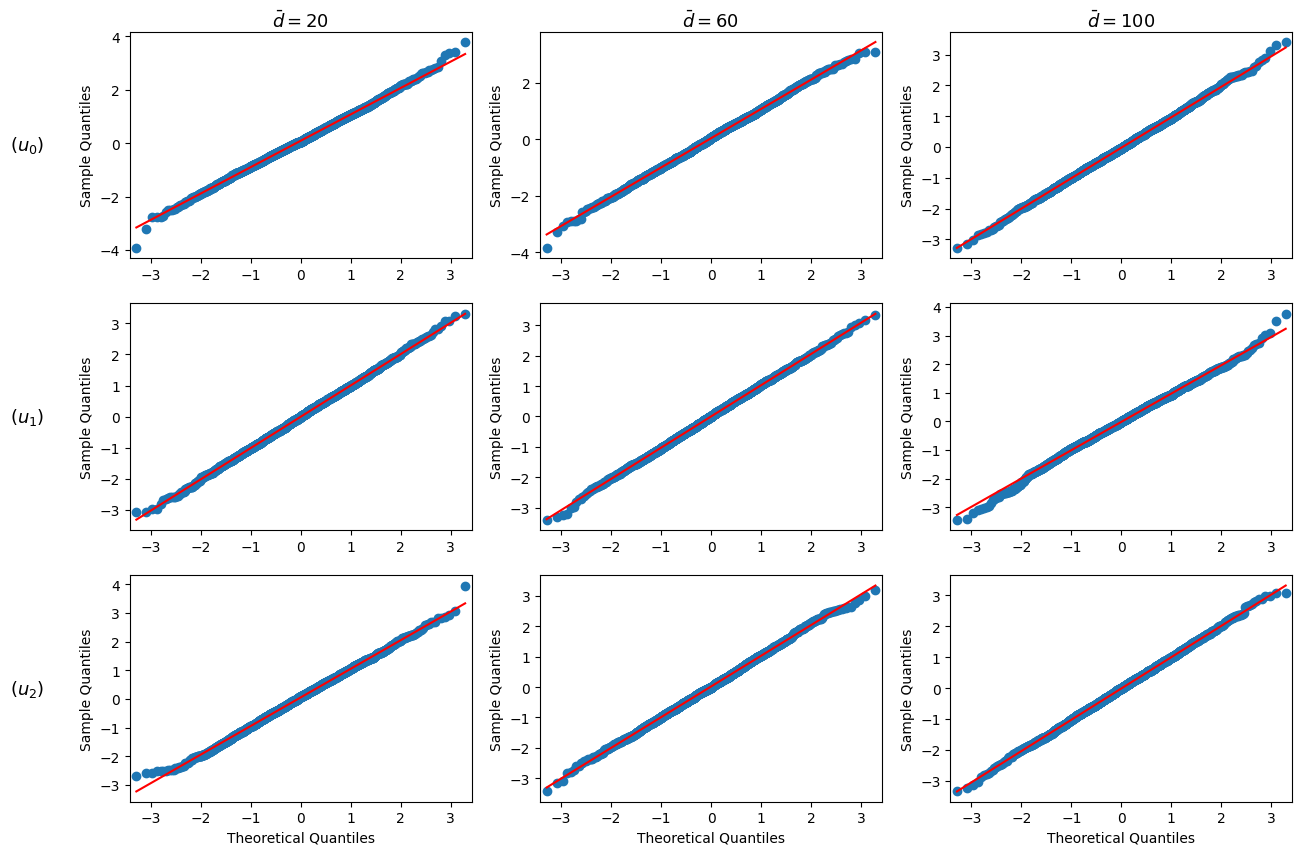}
    \includegraphics[width = 0.95\columnwidth]{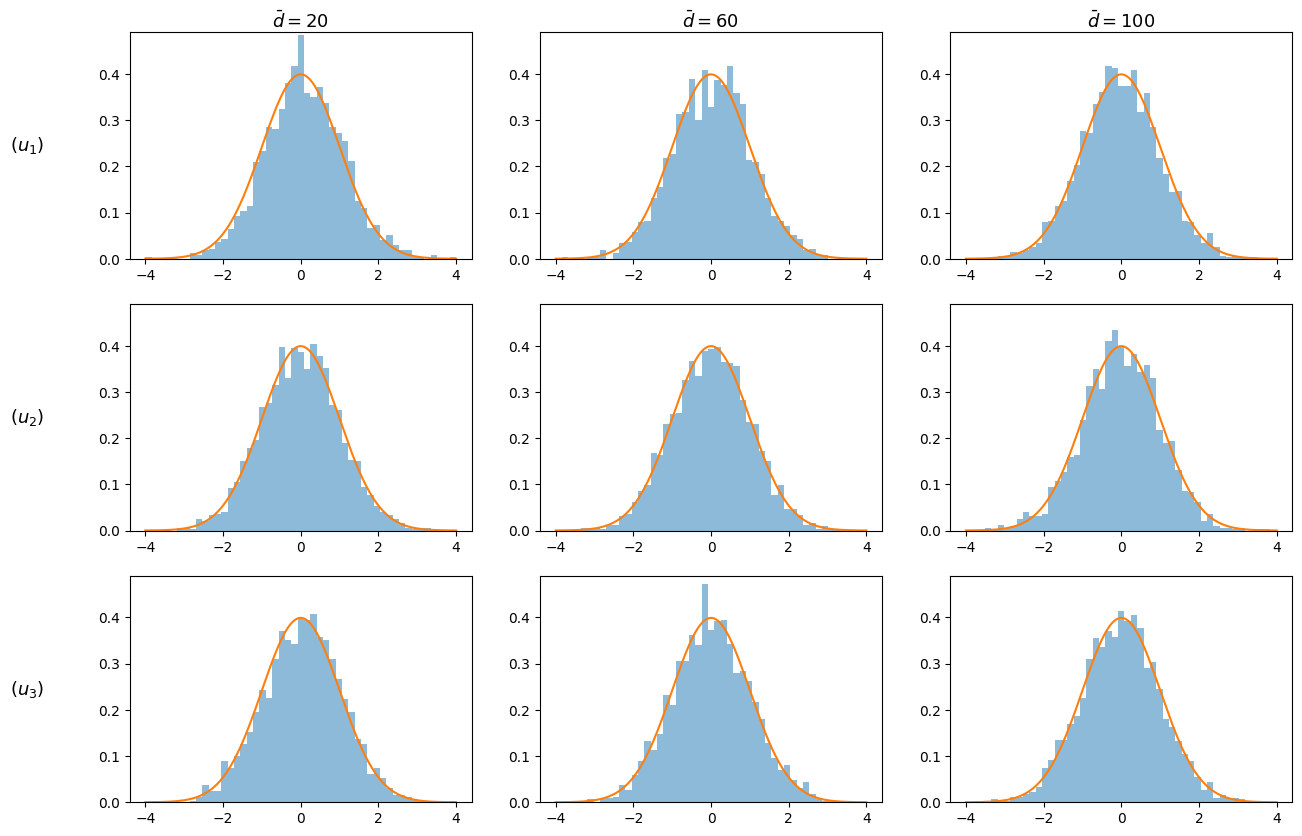}
    \caption{\small QQ plots and histograms of $ \sqrt{T \Theta_{ii} }u^\top\left(\widehat a_{ik}^{\iso} -\text{sign}(\widehat a_{ik}^{\iso\top}  a_{ik})\cdot a_{ik} \right) / \sigma_{u,ik}$ under Configuration VII. Notes: (1) The row displays the results for $u_i$. The column shows the results for different dimension $\Bar{d}$. (2) The red curve plots the distribution of standard normal distribution.}
    \label{fig:config7}
\end{figure}

Figure \ref{fig:config7} shows the QQ plots and histograms of $ \sqrt{T \Theta_{ii} }u^\top\left(\widehat a_{ik}^{\iso} -\text{sign}(\widehat a_{ik}^{\iso\top}  a_{ik})\cdot a_{ik} \right) / \sigma_{u,ik}$ derived from Theorem \ref{thm:clt} under Configuration VII. It is observed that the normalized empirical distribution closely approximates the standard normal distribution. Figure \ref{fig:sim_clt_wcs} in Appendix C shows the corresponding histograms when the variance $\sigma_{u,ik}^2$ is replaced by its estimate $\hat\sigma_{u,ik}^2$, using the thresholding estimator proposed in \cite{chy2025}. The approximation to the standard normal distribution remains accurate even with estimated variance, demonstrating the robustness of the inference.


Next, we evaluate the performance of two proposed rank estimation algorithms: the unfolded eigenvalue ratio method and the eigenvalue ratio method via inner product, over the following DGP configuration:

\begin{itemize}
    \item[VIII] (Rank Estimation) $r = 3$. $d_1 = d_2 = \Bar{d} \in \{ 20,40,60,80 \}$ and $T \in \{100,300,500\}$. Set $\eta = 0.1$ to allow for correlation among factor loading vectors. Error terms are generated as in Configuration IV to accommodate weak cross-sectional correlation. Factors are generated according to \eqref{eqn:factorDGP} with $w_i = (r-i+1)\Bar{d}$ and $\phi \in \{0.1,0.5\}$.
\end{itemize}

The results are presented in Table \ref{tab:sim_rankest}. The numbers in the table denote the relative frequency of correct rank estimation over 500 replications. Both methods perform very well with accuracy levels close to 1.

\begin{table}[htbp]
  \centering
   \caption{Rank estimation }
  \label{tab:sim_rankest}%
    \begin{tabular}{rcrrrr}
    \toprule
          & \multicolumn{1}{l}{$\rho$} & \multicolumn{2}{c}{0.1} & \multicolumn{2}{c}{0.5} \\
    \midrule
    \multicolumn{1}{c}{$(d_1,d_2)$} & $T$     & \multicolumn{1}{c}{$\hat{r}^{uer}$} & \multicolumn{1}{c}{$\hat{r}^{ip}$} & \multicolumn{1}{c}{$\hat{r}^{uer}$} & \multicolumn{1}{c}{$\hat{r}^{ip}$} \\
    \midrule
    \multicolumn{1}{r}{(20,20)} & 100   & 1     & 0.98 & 1     & 0.95 \\
          & 300   & 1     & 1 & 1     & 1 \\
          & 500   & 1     & 1     & 1     & 1 \\
    \midrule
    \multicolumn{1}{r}{(40,40)} & 100   & 1     & 1     & 1     & 1 \\
          & 300   & 1     & 1     & 1     & 1 \\
          & 500   & 1     & 1     & 1     & 1 \\
    \midrule
    \multicolumn{1}{r}{(60,60)} & 100   & 1     & 1     & 1     & 1 \\
          & 300   & 1     & 1     & 1     & 1 \\
          & 500   & 1     & 1     & 1     & 1 \\
    \midrule
    \multicolumn{1}{r}{(80,80)} & 100   & 1     & 1     & 1     & 1 \\
          & 300   & 1     & 1     & 1     & 1 \\
          & 500   & 1     & 1     & 1     & 1 \\
    \bottomrule
    \end{tabular}%

  \begin{tablenotes}
\small
\item Note: Relative frequency of correct rank estimation over 500 replications.
\end{tablenotes}
\end{table}%

Finally we evaluate the factor estimation of the proposed algorithm and compare it with AC-ISO.
\begin{itemize}
    \item[IX] (Factor estimation) Set $r=3$ and $\eta = 0.1$. The error $\calE_{i,j,t}$ is generated as in Configuration IV. We vary $(d_1,d_2)$ from the set $\{(40,40),(40,60),(60,60)\}$ and $T$ from $\{100,300,500\}$. The estimation error is evaluated by the mean square error:
    $$
    MSE = \frac{1}{rT} \sum_{t=1}^T \sum_{i=1}^r \left( \hat f_{it} - h_if_{it}\right)^2,
    $$
    where $h_i = w_i / \hat w_i.$
\end{itemize}

\begin{figure}[htbp!]
    \centering
    \includegraphics[width=0.8\linewidth]{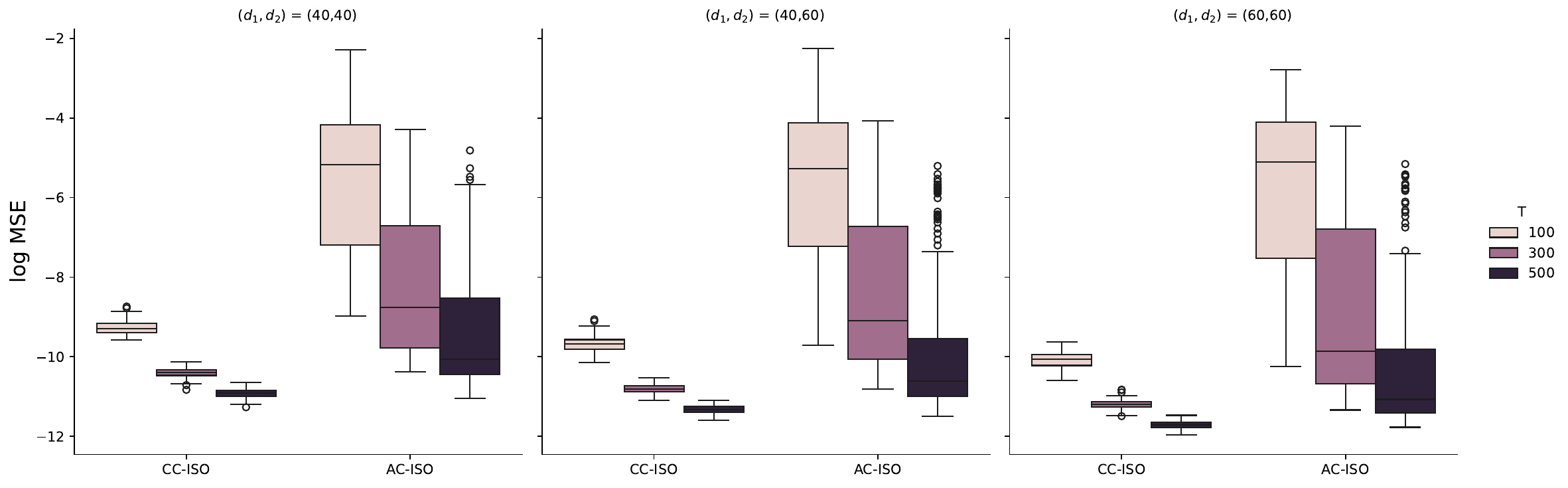}
    \caption{Boxplots of Factor Estimation Errors under Configuration IX.}
    \label{fig:sim_ferr}
\end{figure}

Figure \ref{fig:sim_ferr} shows the results under Configuration IX. The CC-ISO method outperforms AC-ISO in terms of both the accuracy and variability of the factor estimates. In addition, the estimation errors decrease as $d$ and $T$ increase, which is consistent with Theorem \ref{thm:factors}.


\section{Empirical Application: Characteristic decile portfolios} \label{sec-applit}

Latent factor models have been extensively employed in the asset pricing literature to identify risk factors and measure risk exposures. In this section, we conduct an empirical analysis using the dataset compiled by \cite{ChenZimmermann2021}, which consists of over 200 characteristic-sorted portfolios drawn from previous studies on stock market anomalies. This data has also been analyzed by \cite{babii2022tensor}. We utilize the August 2023 release of the database, which contains monthly portfolio excess returns sorted into 10 deciles based on firm-level characteristics, spanning from January 1990 to December 2022. To ensure a balanced panel of portfolios throughout the sample period, we restrict our analysis to 127 characteristics\footnote{The issue of missing values in tensor factor models is both important and challenging. To the best of our knowledge, the only existing study that addresses this issue is \cite{cen2025tensor}, which adopts a Tucker tensor factor structure;
    no prior work has considered missing values under the CP tensor factor framework. Tackling this problem entails substantial additional effort, which we plan to undertake in a separate project.}. Consequently, the dimension of the tensor-valued time series $\cY_t$ is $127 \times 10$ with a sample size of $T=396$. The risk-free rate, obtained from the Kenneth French data library, is used to compute the excess return of each portfolio.

As noted by \cite{babii2022tensor}, \cite{pelger2020RFS} also study a similar dataset, which includes 10 single-sorted decile portfolios for each of 37 characteristics. However, rather than utilizing the natural tensor structure, they flatten the data into a total of 370 portfolios and apply risk-premium PCA based on a vector factor model. In contrast, we exploit the inherent tensor structure by estimating a CP tenor factor model via the proposed CC-ISO method. The CP tensor factor model effectively identifies risk factors and factor exposures without rotation ambiguity. This key property ensures that the estimated factors and loadings are uniquely identified up to sign, independent of rotations or the choice of a varimax algorithm, therefore enhancing the reliability of their economic interpretation.

To implement this, we estimate the CP tensor factor model introduced in Equation \eqref{eqn:fm}. As discussed in Section \ref{sec:model}, the latent factors can be interpreted as systematic risk factors, while the factor loadings represent heterogeneous risk exposures.

\begin{figure}
    \centering
    \includegraphics[width=\linewidth]{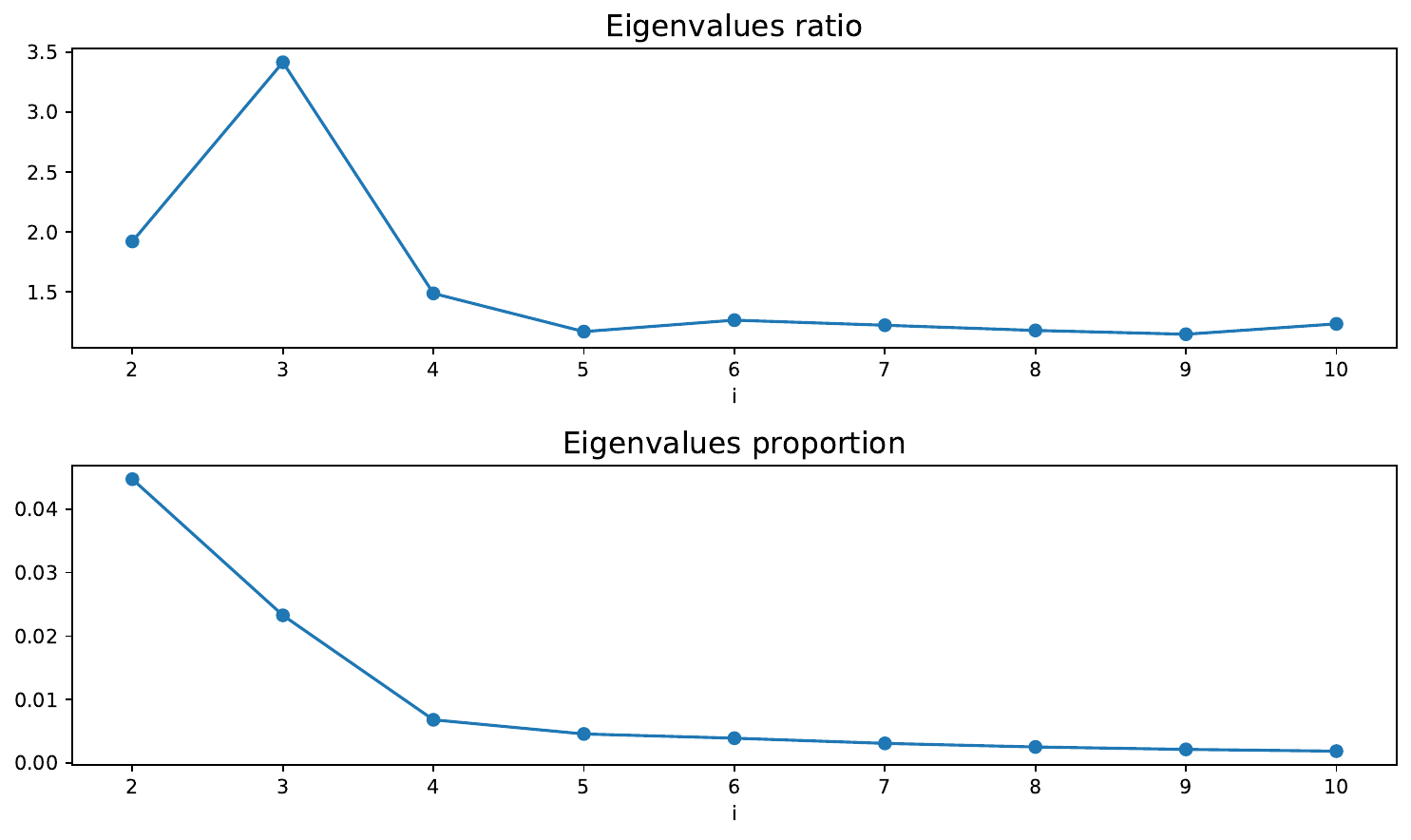}
    \caption{\small The plot of aggregate eigenvalue ratio (upper panel) and the aggregate scree plot (lower panel) of the excess return of characteristic decile portfolios, starting from the second largest eigenvalue. In the scree plot, the y-axis indicates the proportion of the $i^{th}$ largest eigenvalue to the sum of eigenvalues. In the eigenvalue ratio plot, the y-axis indicates the ratio of the $i^{th}$ eigenvalue to the $(i+1)^{th}$ eigenvalue. The plots suggest 3 factors for our analysis.}
    \label{fig:decile_agg_screeplot}
\end{figure}

To determine the number of factors, we examine the eigenvalues of the sample covariance matrix. The first factor is substantially stronger than the others, with its corresponding eigenvalue accounting for 86\% of the total sum of eigenvalues, which might suggest the presence of a dominant market factor. To ensure that our model captures effects beyond this dominant factor, we set the minimum number of factors to two. Guided by the eigenvalue ratio plot and the scree plot (Figure \ref{fig:decile_agg_screeplot}), we select three factors for our analysis. For the Tucker factor models, the mode-wise eigenvalue ratio and scree plots (Figure \ref{fig:decile_mode_screeplot} in Appendix D) suggest the rank pair $(2,3)$.

We estimate the CP factor model in Equation $\eqref{eqn:fm}$ using various algorithms: CC-ISO, TPCA, AC-ISO with $h=1$, GE proposed by \cite{chang2023modelling} with $K = 1$, the CC-based iterative projection methods based on Tucker decomposition (Tucker-CC-IP) proposed by \cite{Han2021iterative} with rank $(2,3)$, the HOOI algorithm based on Tucker decomposition applied by \cite{lettau20243d} (3D-PCA), and the observed factor model with Fama-French 3 factors (FF3). We compare the in-sample performance by $R^2$ defined as:
$$
R^2 = 1 - \frac{ \| \mathcal{Y} - \hat{\mathcal{Y}} \|_{F}^2 }{ \| \Tilde{\mathcal{Y}} \|_{F}^2 },
$$
where $\hat{\mathcal{Y}}$ is the fitted value of $\mathcal{Y}$ with various methods and $\Tilde{\mathcal{Y}}$ is the demeaned tensor of $\mathcal{Y}$ with the $t^{th}$ entry defined as $\Tilde{\mathcal{Y}}_{t} = \mathcal{Y}_{t} - \frac{1}{T} \sum_{t=1}^T \mathcal{Y}_{t}.$

To evaluate out-of-sample performance, we conduct one-step-ahead forecasts over the final two years of the sample, following the scheme proposed by \cite{chang2023modelling}. Specifically, for each $s \in \{1,2,\ldots,24\}$, we estimate the model using the sample $\{\calY_t\}_{t=1}^{371+s}$ and fit the estimated latent factors with a VAR(1) model. The one-period-ahead excess return $\calY_{372+s}$ is then predicted using the estimated loadings and the predicted one-period-ahead factors. The out-of-sample MSE is defined as:
$$
\text{Out-of-sample MSE} = \frac{1}{24} \sum_{s=1}^{24}\left\| \calY_{372+s} - \hat \calY_{372+s}  \right\|_F^2.
$$

Table \ref{tab:r2sumstat} displays the $R^2$ and out-of-sample MSE ratios (normalized by Tucker-CC-IP) for six algorithms along with $p$-values from the one-sided \cite{Diebold1995} (DM) Forecast comparison tests.  In the DM tests, row ``DM I'' evaluates whether each competing algorithm outperforms Tucker-AC-IP, while row ``DM II'' tests whether CC-ISO outperforms the respective competing method. Among the CP-based methods, CC-ISO achieves the best in-sample fit, with the highest $R^2$. Within the ISO methods, CC-ISO outperforms AC-ISO. In contrast, The GE algorithm yields the lowest $R^2$ value of 0.768.

Given that Tucker decomposition generally involves a much larger number of factors than CP decomposition ($6$ vs $3$ in our example), it is not surprising that Tucker-CC-IP achieves slightly higher $R^2$ than the CP-based methods. Nonetheless, the improvement is modest, with Tucker-CC-IP increasing $R^2$ by only $2$ to $3$ percentage points compared to CP models. This observation suggests that the characteristic decile portfolio data might exhibit a CP-like factor structure, where a low-rank CP model effectively captures the underlying risk factors.

In terms of forecast performance, smaller MSE ratios indicate better prediction accuracy. CC-ISO significantly outperforms Tucker-CC-IP, 3D-PCA, and GE, while performing comparably to AC-ISO and TPCA. Interestingly, although TPCA achieves the lowest MSE, its improvement over Tucker-CC-IP is not statistically significant. Overall, the Tucker-based methods tend to produce higher MSEs, likely due to overfitting from the increased number of factors. These findings highlight the advantage of CP-based methods in balancing model parsimony with predictive performance.


\begin{table}[htbp]
  \centering
  \fontsize{10}{15}\selectfont{
    \begin{tabular}{lccccccc}
    \toprule
          & CC-ISO & AC-ISO & TPCA  & GE    & Tucker-CC-IP & 3D-PCA  & FF3 \\
    \midrule
    $R^2$    & 0.8531 & 0.8445 & 0.8108 & 0.7682 & 0.8783 & 0.8783 & 0.839 \\
    OOS MSE Ratio & 0.9401 & 0.9471 & 0.9359 & 1.0084 & 1.0000 & 1.0002 & 0.9748 \\
    DM I  & 0.0153 & 0.0738 & 0.1733 & 0.6221 & -     & 0.6353 & 0.1403 \\
    DM II & -     & 0.3362 & 0.5235 & 0.0083 & 0.0153 & 0.0151 & 0.1165 \\
    \bottomrule
    \end{tabular}%
    }
    \caption{\small Comparison of different algorithms for characteristic decile portfolio excess returns. \\ Notes: (1) $R^2$: the in-sample $R^2$ across all algorithms. OOS MSE ratio: the out-of-sample mean square error ratios relative to Tucker-CC-IP, which has the largest MSE. (2) DM I: one-sided p-value from  \cite{Diebold1995} test, with the alternative that the algorithm in each column is more accurate than Tucker-CC-IP. DM II: one-sided p-value from  \cite{Diebold1995} test based on the pairwise comparison between the CC-ISO and the algorithm on the column.}
  \label{tab:r2sumstat}%
\end{table}%
 Nest, we evaluate cross-sectional pricing errors in the spirit of \cite{lettau20243d}. Specifically, for each decile portfolio, we estimate

    $$
    r_{ijt} = \alpha_{ij} + \beta_{ij}' f_t + u_{ijt}, \quad i = 1, \ldots, 127, j = 1, \ldots, 10,
    $$
    and compute the mean squared pricing errors (MSPE) of $f_t$ as
    $$
    MSPE_{f} = \frac{1}{127 \times 10} \sum_{i,j} \alpha_{ij}^2.
    $$

    For the Tucker factor model with factor matrix $F_t \in \R^{r_1 \times r_2}$ from Tucker factor model, we vectorize it prior to regression, i.e. $f_t = \Vec(F_t) \in \R^{r_1r_2}$.

    Table \ref{tab:mpe} reports MSPE and MSPE ratios of different methods relative to Tucker-CC-IP with rank (2,3), as suggested by the mode-wise scree plots. For completeness, we include the 3D-PCA approach of \cite{lettau20243d} and the Fama-French three-factor model. As shown, the factors obtained from CC-ISO and GE achieve the lowest MSPE, demonstrating that our model performs strongly not only in out-of-sample forecasting but also in cross-sectional pricing.

    \begin{table}[htbp]
      \centering
        \begin{tabular}{cccccccc}
        \toprule
        Method & CC-ISO & AC-ISO & TPCA  & GE & Tucker-CC-IP  & 3D-PCA & FF3 \\
        \midrule
        MSPE & 0.0603 &	0.0925 & 0.0632	& 0.0594 & 0.0799 & 0.0785 & 0.0883 \\
        MSPE Ratio  & 0.7547 & 1.1573 & 0.7913 & 0.7434 & 1.0000 & 0.9825 & 0.8744 \\
        \bottomrule
        \end{tabular}%
        \caption{\small Table of mean pricing error ratios for different methods. The MSPE row shows the mean squared pricing errors of each method. The MSPE Ratio row shows the mean pricing error ratios relative to Tucker-CC-IP with rank (2,3).}
      \label{tab:mpe}%
    \end{table}%

\begin{figure}[htbp!]
    \centering
    \includegraphics[width=\linewidth]{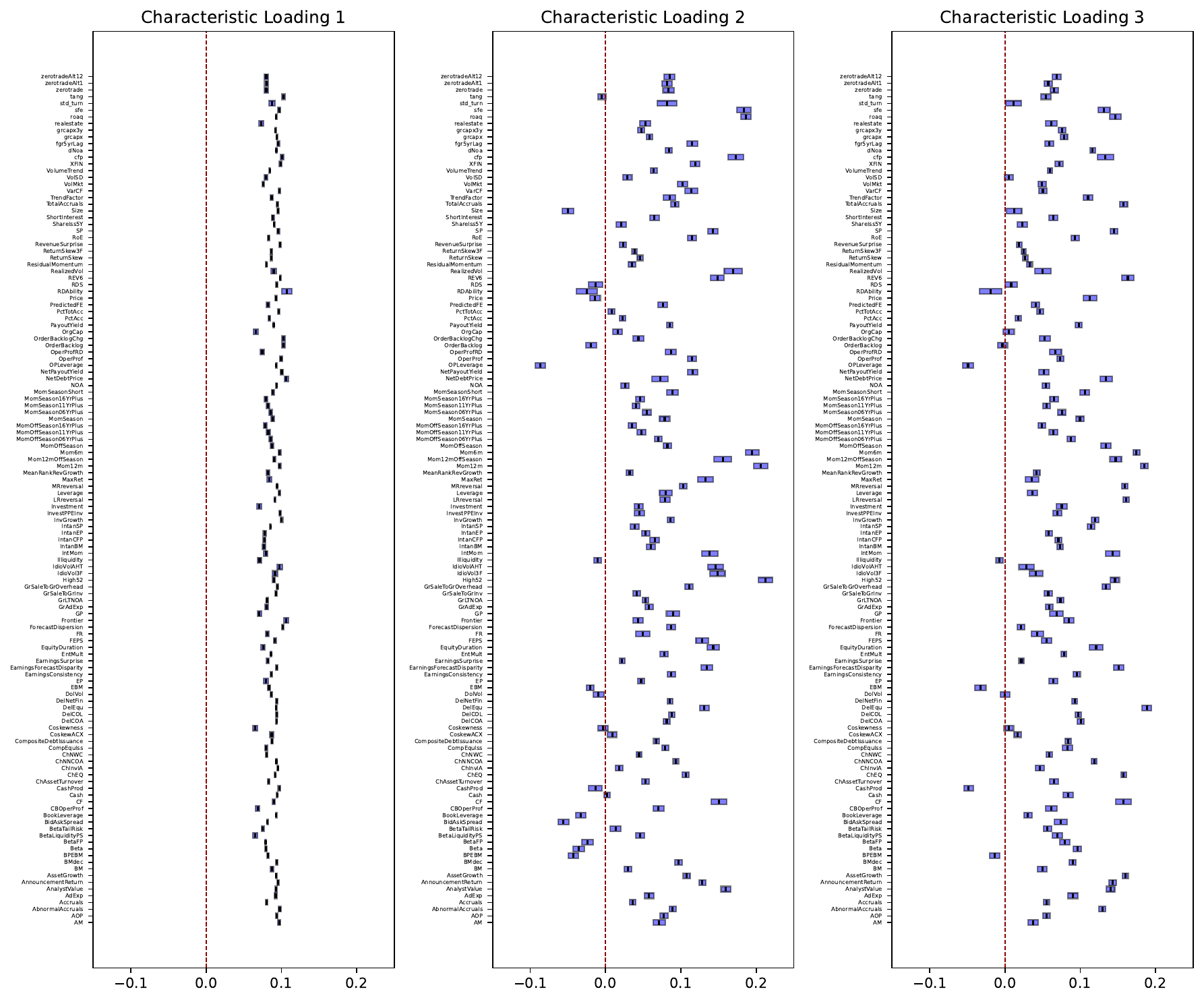}
    \caption{\small Estimated loadings on characteristics with 95\% confidence interval.}
    \label{fig:char_loading_ci}
\end{figure}

\begin{figure}[t]
    \centering
    \includegraphics[width= 0.5\linewidth]{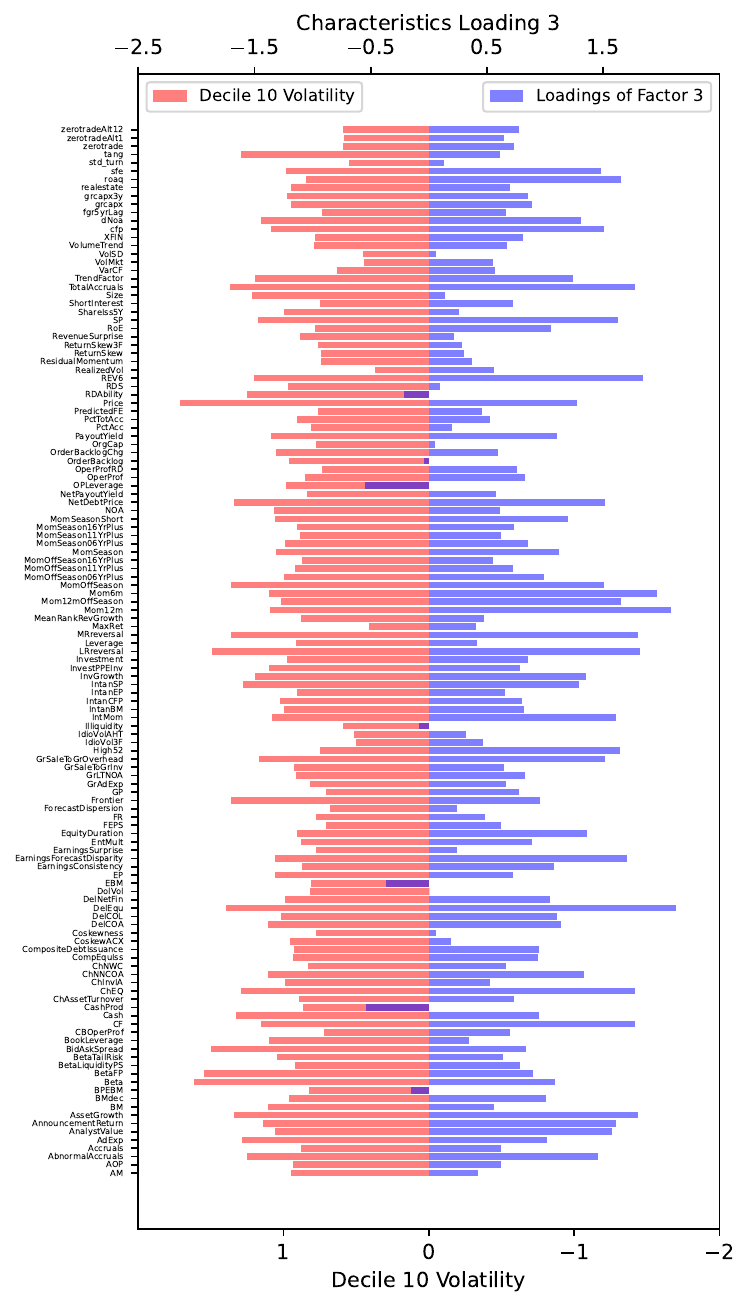}
    \caption{\small Characteristics loading 3 vs. Decile 10 portfolio volatility.}
    \label{fig:dec10vola_vs_loading3}
\end{figure}

Finally, we further examine the estimated factors and loadings in our CP factor model. The estimated loadings reveal several notable patterns. Figure \ref{fig:char_loading_ci} presents the estimated loadings along with their 95\% confidence intervals across characteristics. The loadings on the first factor are uniformly positive, resembling ``long-only'' portfolio weights. The loadings on the second and third factors display more variation, with predominantly positive elements. These loadings are closely linked to the volatilities of the decile portfolios. In particular, Figure \ref{fig:dec10vola_vs_loading3} compares the volatilities of decile 10 portfolios with the corresponding characteristic loadings on the third factor. The figure highlights a strong correlation (0.875) between the two: characteristics with higher decile 10 portfolio volatility tend to have larger third-factor loadings, while those with lower volatility have smaller or even negative loadings.

\begin{figure}[htbp!]
    \centering
    \includegraphics[width=\linewidth]{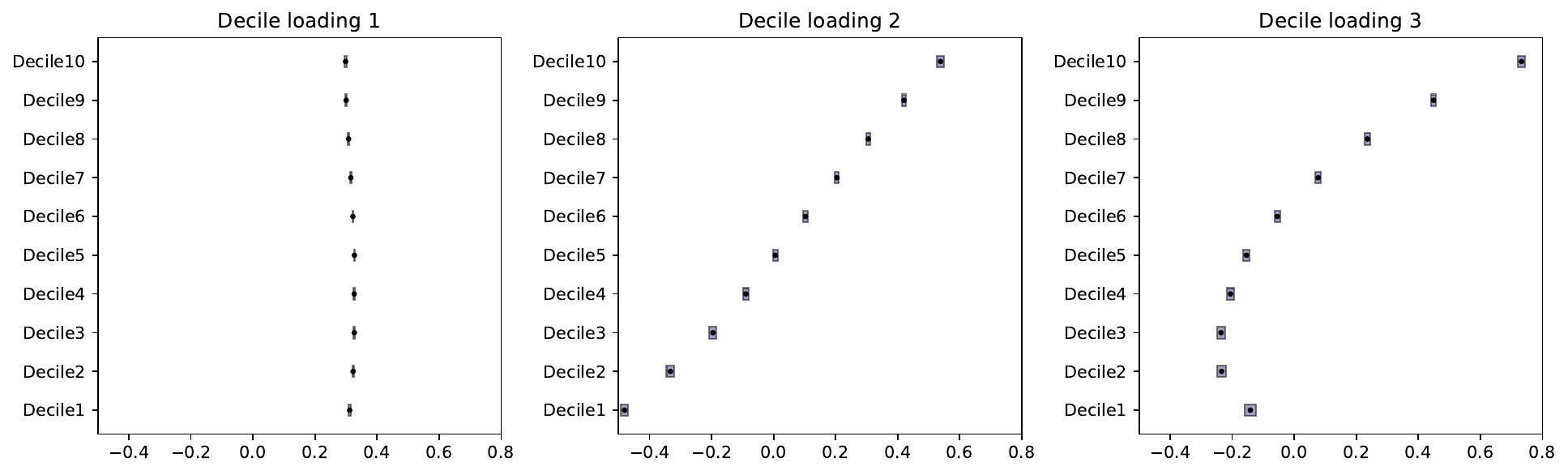}
    \caption{\small Estimated loadings on decile with 95\% confidence interval.}
    \label{fig:dec_loading_ci}
\end{figure}

The factor loadings along the decile mode follow a distinct level-slope-curvature pattern, as shown in Figure \ref{fig:dec_loading_ci}. The first factor follows a ``long-only'' structure, with strictly positive and nearly uniform weights across deciles. The second factor shows a slope pattern, with decile-10 and decile-1 loadings having opposite signs but similar magnitudes. This structure resembles a ``long-short factor'', where high-return deciles are offset by low-return deciles. The third loading shows a convex pattern, forming the ``curvature'' component.

\begin{figure}[t]
    \centering
    \includegraphics[width=0.8\linewidth]{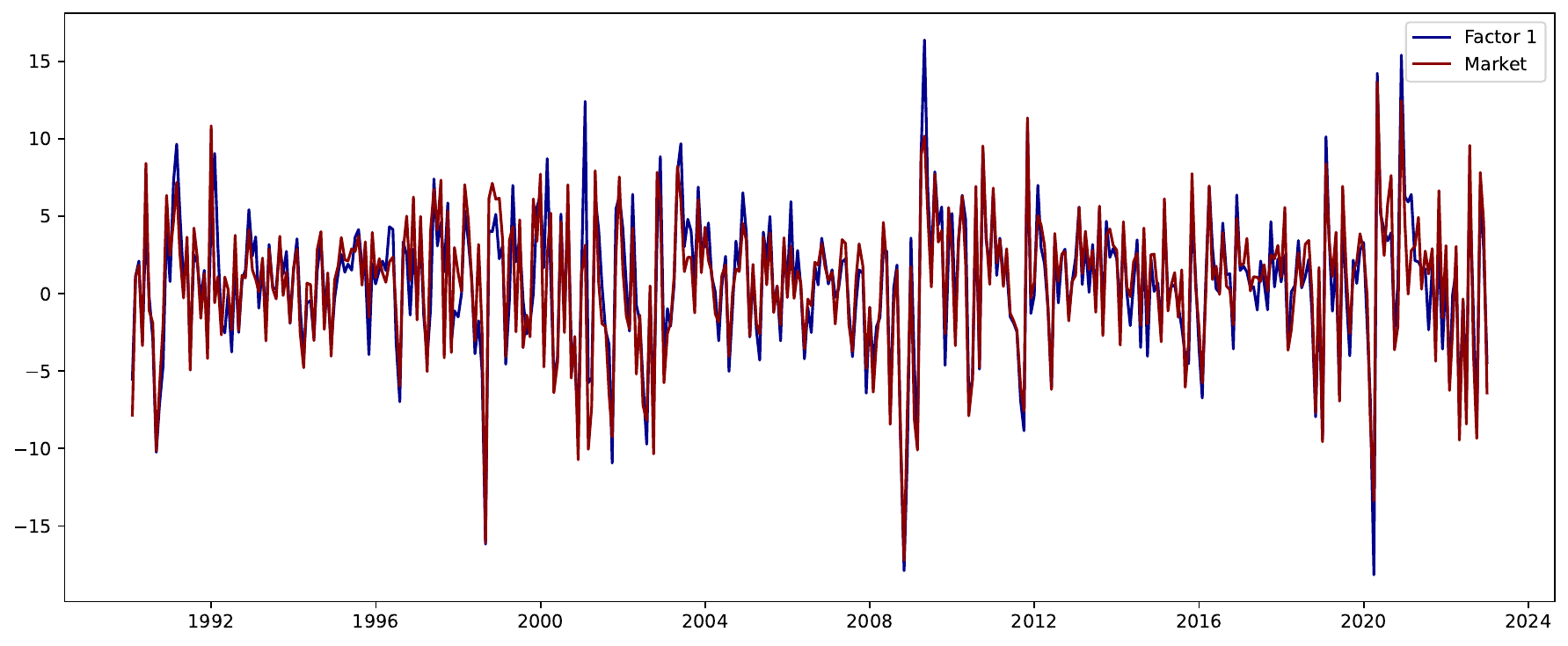}
    \caption{\small Factor 1 vs. Excess market return. Factor 1 is scaled so that its volatility is equal to the volatility of the excess market return. The correlation between Factor 1 and the market factor is 0.904.}
    \label{fig:dec_f1market}
\end{figure}

The consistent ``long-only'' pattern of the first loading across both modes, combined with the dominant eigenvalue of the first factor, suggests a strong association with the market factor. Figure \ref{fig:dec_f1market} presents the time series of the first factor along with the excess market return from the Kenneth French data library. The two series exhibit a high correlation (above 0.9), reinforcing the interpretation that this factor primarily captures systematic market risk. In summary, the three estimated factors from the CP factor model can be interpreted as the market factor, the long-short factor, and the volatility factor, respectively.



\section{Conclusion} \label{sec-conclu}
Modeling high-dimensional tensor time series has attracted growing attention, driven by the increasing availability of multidimensional datasets that go beyond the classical panel data structure. This paper considers matrix and tensor factor models with a CP low-rank structure, extending traditional vector factor models to higher-order settings. We develop ISO procedures based on contemporary covariance, preserving the tensor data structure. Theoretical properties such as the rate of convergence and limiting distributions are investigated under settings where each tensor dimension is comparable to or exceeds the number of observations, and the tensor rank may be either fixed or divergent.

Unlike auto-covariance-based estimation methods, our method explores information from contemporary covariance and can consistently estimate both loadings and factors even when observations are uncorrelated, settings where auto-covariance methods may fail. Additionally, we propose two generalized eigenvalue-ratio estimators for rank selection and justify their consistency. Extensive simulation studies highlight the merits of our method over existing methods. Furthermore, the empirical application to sorted portfolios demonstrates the practical relevance of our approach.



%
%

\bibliographystyle{apalike}
\bibliography{ref}

%
%
\newpage
\setcounter{page}{1}
\begin{appendices}
    \begin{center}
        {\LARGE
Supplementary Material of \\ ``\TITLE'' }

    \end{center}
    \section{Proofs of Main Theorem}

\begin{proof}[\bf Proof of Theorem \ref{thm:initial}]
\noindent\underline{\bf Part (i)}. Recall $\widehat\Sigma=T^{-1}\sum_{t=1}^T \cY_{t}\otimes\cY_t$, $\Theta= W (\EE F_t F_t^\top) W$, $W=\diag(w_1,...,w_r)$, $ a_i=\vec( a_{i1}\otimes a_{i2}\otimes\cdots\otimes a_{iK})$, $d=d_1d_2...d_K$. Let $e_t=\vec(\cE_t)$. Write
\begin{align*}
\widetilde\Sigma:=\mat_{[K]} \big(\widehat\Sigma \big)= A \Theta  A^\top + \Psi^* .
\end{align*}
Let $\Theta\in \R^{r\times r}$ have the eigenvalue decomposition $\Theta=\widetilde V \widetilde \Lambda \widetilde V^\top$, where the diagonal matrix $\widetilde \Lambda$ denotes its eigenvalues.
Let $U=(u_1,\ldots,u_r)$ be the orthonormal matrix corresponding to $A\widetilde V$ as in Lemma~\ref{lemma-transform-ext}. We have $\|AA^\top - UU^{\top}\|_{\rm 2}\le \delta$ and $\|A \Theta A^\top - U \widetilde \Lambda U^{\top}\|_{\rm 2}\le \lam_1\delta$ by the applications of the error bound in Lemma~\ref{lemma-transform-ext} with $\Lambda=\widetilde \Lambda$ the first time.

Let the top $r$ eigenvectors of $\widetilde\Sigma$ be $ \widehat U=(\widehat u_1, ..., \widehat u_r)\in\R^{d\times r}$.
By Wedin's perturbation theorem \citep{wedin1972} 
for any $1\le j\le r$,
\begin{align}
\|\widehat  u_j \widehat  u_j^\top - u_j  u_j^{\top} \|_{\rm 2} \le
2 \| A \Theta A^{\top} - U \widetilde \Lambda U^{\top} + \Psi^* \|_{\rm 2}/\lam_{j,\pm}
\le \big(2\lambda_1 \delta + 2 \|\Psi^*\|_{\rm 2}\big)\big/\lam_{j,\pm},
\label{eq2:thm:initial2}
\end{align}
where $\lam_{j,\pm}=\min\{\lam_{j-1}-\lam_{j},\lam_{j}-\lam_{j+1}\}$.
Combining \eqref{eq2:thm:initial2} and the inequality
$\|AA^\top - UU^{\top}\|_{\rm 2}\le \delta$,
we have
\begin{align}\label{eq3:thm:initial2}
\|\widehat u_j \widehat u_j^\top -  a_j  a_j^\top \|_{\rm 2}&\le \delta+ \big(2\lambda_1 \delta + 2 \|\Psi^*\|_{\rm 2}\big)/\lam_{j,\pm}.
\end{align}

We formulate each $\widehat u_j\in\R^d$ to be a $K$-way tensor $\widehat U_j\in \R^{d_1\times\cdots\times d_K}$. Let $\widehat U_{jk}=\mat_k(\widehat U_j)$, which is viewed as an estimate of  $ a_{jk}\vec(\otimes_{l\neq k}^K  a_{jl})^\top\in\R^{d_k\times (d/d_k)}$. Then $ a_{jk}^{\rm rcpca}$ is the top left singular vector of $\widehat U_{jk}$.
By Lemma \ref{prop-rank-1-approx},
\begin{align}
\| a_{jk}^{\rm rcpca} a_{jk}^{\rm rcpca\top} - a_{jk} a_{jk}^\top \|_{\rm 2}^2 \wedge (1/2) \le
\|\widehat u_j \widehat u_j^\top - a_j a_j^{\top} \|_{\rm 2}^2 .
\end{align}
Substituting \eqref{eq3:thm:initial2} and Lemma \ref{lem:psi} into the above equation, we have the desired results.

\noindent\underline{\bf Part (ii)}. For simplicity, consider the most extreme case where $\min\{\lam_i-\lam_{i+1},\lam_{i}-\lam_{i-1}\} \le c \lam_r$ for all $i$, with $\lambda_0=\infty, \lambda_{r+1}=0$, and $c$ is sufficiently small constant. In such cases, we need to employ Procedure \ref{alg:initialize-random} to the entire sample covariance tensor $\widehat \Sigma$. Let the eigenvalue ratio $\kappa:=\lam_1/\lam_r=O(r)$. Without loss of generality, assume $\Theta_{11}\ge \Theta_{22}\ge \cdots \ge \Theta_{rr}$ and $h=1$ in Procedure \ref{alg:initialize-random}. In general, the statement in the theorem holds for number of initialization $L\ge C d^2 \vee C dr^{2\kappa^2}$, where $a \vee b =\max\{a, b\}$.
We prove the statements through induction on factor index $i$ starting from $i=1$ proceeding to $i=r$. By the induction hypothesis, we already have estimators such that
\begin{align}\label{eq:induction_int}
\left\| \widehat a_{jk}^{\rm rcpca} \widehat a_{jk}^{\rm rcpca \top}  - a_{jk} a_{jk}^\top \right\|_2  &\le C\phi_0,  \qquad 1\le j\le i-1, 1\le k\le K,
\end{align}
in an event $\Omega$ with high probability, where
\begin{align*}
\phi_0^2 = C\left(\frac{R^{(0)}}{\lam_r} + \left(\frac{R^{(0)}}{\lam_r} \right)^{K-1} \left(\frac{\lam_1}{\lam_r} \right) + (\delta_{1}^2 + \delta /\delta_{1}) \left(\frac{\lam_1}{\lam_r} \right) + \delta_1\right) ,
\end{align*}
and $R^{(0)}=\phi^{(0)}$ is defined in \eqref{thm:initial:eq2} in Theorem \ref{thm:initial}. We shall show that the bound in \eqref{eq:induction_int} also holds for $i$ on the same event. Note that when $i=1$, the statement ``the bound in \eqref{eq:induction_int} holds for $1\le j \le i-1$ for all $1\le k\le K$'' is an empty set. Thus, we focus on proving only the inductive step.

Applying Lemma \ref{lem:rcpca}, we obtain that at the $i$-th step ($i$-th factor), we have
\begin{align*}
\left\| \widetilde a_{\ell k} \widetilde a_{\ell k}^{\top}  - a_{ik} a_{ik}^\top \right\|_2  &\le \phi_0^2,  \qquad  1\le k\le K,
\end{align*}
in the event $\Omega$ with probability at least $1-T^{-c}-d^{-c}$ for at least one $\ell\in[L]$. It follows that this estimator $\widetilde a_{\ell k}$ satisfies
\begin{align*}
\left\| \widehat\Sigma \times_{k=1}^{2K} \widetilde a_{\ell k} \right\|_2   &\ge  \left\| \sum_{j_1,j_2=1}^r \Theta_{j_1,j_2} \prod_{k=1}^K (a_{j_1 k}^\top \widetilde a_{\ell k}) (a_{j_2 k}^\top \widetilde a_{\ell k}) \right\|_2    -  \left\| \Psi \times_{k=1}^{2K} \widetilde a_{\ell k} \right\|_2 \\
&\ge \left\| \Theta_{ii} \prod_{k=1}^K (a_{i k}^\top \widetilde a_{\ell k})^2 \right\|_2  - \left\| \sum_{(j_1,j_2)\neq (i,i)}^r \Theta_{j_1,j_2} \prod_{k=1}^K (a_{j_1 k}^\top \widetilde a_{\ell k}) (a_{j_2 k}^\top \widetilde a_{\ell k}) \right\|_2   -  \left\| \Psi \times_{k=1}^{2K} \widetilde a_{\ell k} \right\|_2 ,
\end{align*}
where $\Psi$ is defined by unfolding $\Psi^*$ into a $d_1\times d_2 \times \cdots \times d_K \times d_1 \times d_2 \times \cdots \times d_K$ tensor. Let $\psi_i=CR^{(0)}/\Theta_{ii}+\delta_1^2 \kappa$. By \eqref{eq:a_decomp} and the last part of the proof of Lemma \ref{lem:rcpca}, as $\|\Psi^*\|_2/\lam_r \le \phi_0^2$ and $(1+\delta_1)\prod_{k=2}^K(\delta_k+\psi_i) w \le \phi_0^2$, it follows that
\begin{align*}
\left\| \widehat\Sigma \times_{k=1}^{2K} \widetilde a_{\ell k} \right\|_2   &\ge  (1-\phi_0^4)^K \Theta_{ii} - (1+\delta_1)\prod_{k=2}^K(\delta_k+\psi_i) \lam_1 -\phi_0^2 \Theta_{ii }    \\
&\ge (1-3\phi_0^2) \Theta_{ii}.
\end{align*}
Now consider the best initialization $\ell_*\in [L]$ by using $\ell_* =\arg\max_{s} |\widehat\Sigma \times_{k=1}^{2K} \widetilde a_{s k} |$. By the calculation above, it is immediate that
\begin{align}\label{eq:lbd}
\left\| \widehat\Sigma \times_{k=1}^{2K} \widetilde a_{\ell_* k} \right\|_2   &\ge (1-3\phi_0^2) \Theta_{ii}.
\end{align}
Let $\widetilde a_{\ell_*} =\vec( \widetilde a_{\ell_* 1} \otimes \cdots \otimes \widetilde a_{\ell_* K})$. If $\| a_i a_i^\top - \widetilde a_{\ell_*} \widetilde a_{\ell_*}^\top \|_2 \ge C \phi_0$ for a sufficiently large constant $C$, we have that
\begin{align*}
\left\| \widehat\Sigma \times_{k=1}^{2K} \widetilde a_{\ell_* k} \right\|_2   &\le  \left\| \sum_{j_1,j_2=1}^r \Theta_{j_1,j_2}  (a_{j_1 }^\top \widetilde a_{\ell_* }) (a_{j_2 }^\top \widetilde a_{\ell_* }) \right\|_2   + \left\| \Psi \times_{k=1}^{2K} \widetilde a_{\ell k} \right\|_2 \\
&\le  \left\| \sum_{j_1,j_2=i}^r \Theta_{j_1,j_2} (a_{j_1 }^\top \widetilde a_{\ell_* }) (a_{j_2 }^\top \widetilde a_{\ell_* }) \right\|_2 + \phi_0^2 \Theta_{ii }+ r \nu^{2K} \lam_1  \\
&\le  (1+\delta_1)(1-C^2 \phi_0^2) \Theta_{ii} + \phi_0^2\Theta_{ii } + r \nu^{2K} \lam_1 .
\end{align*}
If $\nu$ satisfies $r\nu^{2K} (\lam_1/\lam_r) \le c\phi_0^2$ for a small positive constant $c$, as $\delta_1 \le \phi_0^2$, we have
\begin{align*}
\left\| \widehat\Sigma \times_{k=1}^{2K} \widetilde a_{\ell_* k} \right\|_2 \le (1-C' \phi_0^2) \Theta_{ii}   ,
\end{align*}
where $C'$ is a sufficiently large constant. It contradicts \eqref{eq:lbd} above. This implies that for $\ell=\ell_*$, we have
\begin{align*}
\| a_i a_i^\top - \widetilde a_{\ell_*} \widetilde a_{\ell_*}^\top \|_2 \le C \phi_0    .
\end{align*}
By Lemma \ref{prop-rank-1-approx}, with $\widehat a_{ik}^{\rm rcpca} =\widetilde a_{\ell_* k}$, in the event $\Omega$ with probability at least $1-T^{-c}-d^{-c}$,
\begin{align*}
\left\| \widehat a_{ik}^{\rm rcpca} \widehat a_{ik}^{\rm rcpca \top}  - a_{ik} a_{ik}^\top \right\|_2  &\le C\phi_0, \quad 1\le k\le K   .
\end{align*}
This finishes the proof of part (ii) by an induction argument along with the requirements $r\nu^{2K} (\lam_1/\lam_r) \le c\phi_0^2$.
\end{proof}

\begin{lemma}\label{lem:psi}
Suppose Assumptions \ref{asmp:error}, \ref{asmp:eigenvalue}, \ref{asmp:mixing} hold and $\delta<1$. Let $\widetilde\Sigma= A \Theta  A^\top +\Psi^*$ and $1/\gamma=2/\gamma_1+1/\gamma_2$. In an event with probability at least $1-T^{-c}-d^{-c}$,
we have
\begin{align}\label{eq:lem:psi}
\|\Psi^*\|_{\rm 2} &\le C  \lambda_1  \left(\sqrt{\frac{r+ \log T}{T}} + \frac{(r+ \log T)^{1/\gamma}}{T}  \right) + C \left(\sqrt{\frac{d\log(d)}{T}}   + \frac{d \log(d)(\log T)^{\frac{2\vartheta+4}{\vartheta}}}{T}+1 \right) \notag\\
&\qquad\quad + C \lambda_1^{1/2} \left(\sqrt{\frac{d\log(d)}{T}}   + \frac{\sqrt{d r} \log(d)(\log T)^{1+\frac{2}{\vartheta}+\frac{1}{\gamma_1}}}{T} \right).
\end{align}
\end{lemma}

\begin{proof}
Let $\Upsilon_0=T^{-1} \sum_{t=1}^T\sum_{i,j=1}^r w_i w_j f_{it}f_{jt}  a_i a_j^\top$, $\overline\EE(\cdot)=\EE(\cdot|f_{it},1\le i\le r, 1\le t\le T)$. Define $e_t=\vec(\cE_t)$. Write
\begin{align*}
\widetilde\Sigma &= \frac{1}{T}\sum_{t=1}^T \vec(\cY_{t})\vec(\cY_{t})^\top \\
&= A \Theta  A^\top +  \sum_{i,j=1}^r \frac{1}{T} \sum_{t=1}^T w_i w_j\left( f_{i,t}f_{j,t} - \EE f_{i,t}f_{j,t} \right) a_i a_j^\top +   \frac{1}{T} \sum_{t=1}^T e_{t} e_t^\top\\
&\quad +   \frac{1}{T} \sum_{t=1}^T \sum_{i=1}^r w_i f_{i,t} a_i e_t^\top +   \frac{1}{T} \sum_{t=1}^T \sum_{i=1}^r w_i  f_{i,t}  e_{t} a_i^\top \\
&:= A \Theta  A^\top +\Delta_1+\Delta_2+\Delta_3+\Delta_4.
\end{align*}
That is, $\Psi^*=\Delta_1+\Delta_2+\Delta_3+\Delta_4$.

We first bound $\|\Delta_{1}\|_{\rm 2}$. Note that $\Delta_1= A\left( \widehat\Theta - \Theta \right) A^\top $.
For any unit vector $u$ in $\R^r$, there exist $u_j\in\R^r$ with $\|u_j\|_2\le 1$, $j=1,...,N_{r,\epsilon}$ such that $\max_{\|u\|_2\le 1} \min_{1\le j\le N_{r,\epsilon}} \| u -u_j\|_2\le \epsilon$. The standard volume comparison argument implies that the covering number $N_{r,\epsilon} = \lfloor(1+2/\epsilon)^r\rfloor$. Then, there exist $u_j \in\R^r$, $1\le j \le N_{r,1/3}:= 7^r$, such that $\|u_j\|_2=1$ and
\begin{align*}
\|\widehat\Theta - \Theta \|_{\rm 2} -\max_{1\le j \le N_{r,1/3}} \left|u_j^\top (\widehat\Theta - \Theta) u_j \right| \le (2/3)\|\widehat\Theta - \Theta \|_{\rm 2}.
\end{align*}
It follows that
\begin{align*}
\|\widehat\Theta - \Theta \|_{\rm 2} \le 3\max_{1\le j \le N_{r,1/3}} \left|u_j^\top (\widehat\Theta - \Theta) u_j \right| .
\end{align*}
As $1/\gamma=2/\gamma_1+1/\gamma_2$ and $W=\diag(w_1,...,w_r)$, by Theorem 1 in \citet{merlevede2011},
\begin{align}\label{eq:thm_initial:f}
\PP\left( T\left| u_j^\top W^{-1}(\widehat\Theta - \Theta) W^{-1} u_j\right|\ge x \right) &\le T\exp\left(-\frac{x^\gamma}{c_1}\right)+ \exp\left(-\frac{x^2}{c_2 T}\right) \notag\\
&\quad + \exp\left(-\frac{x^2}{c_3 T}\exp\left(\frac{x^{\gamma(1-\gamma)}}{c_4(\log x)^\gamma} \right) \right).
\end{align}
Hence,
\begin{align*}
\PP\left( T\left\| W^{-1}(\widehat\Theta - \Theta) W^{-1} \right\|_{\rm 2}/3 \ge x \right) &\le  N_{r,1/3}^2 T\exp\left(-\frac{x^\gamma}{c_1}\right)+  N_{r,1/3}^2\exp\left(-\frac{x^2}{c_2 T}\right) \notag\\
&\quad + N_{r,1/3}^2 \exp\left(-\frac{x^2}{c_3 T}\exp\left(\frac{x^{\gamma(1-\gamma)}}{c_4(\log x)^\gamma} \right) \right).
\end{align*}
Choosing $x\asymp \sqrt{T(r+\log T)} + (r+\log T )^{1/\gamma}$, in an event $\Omega_1$ with probability at least $1-T^{-c_1}/2$,
\begin{align*}
 \left\| W^{-1}(\widehat\Theta - \Theta) W^{-1} \right\|_{\rm 2} \le C\sqrt{\frac{r+\log(T)}{T}} + \frac{C(r+\log T)^{1/\gamma}}{T} .
\end{align*}
It follows that, in the event $\Omega_1$,
\begin{align*}
\|\Delta_1\|_{\rm 2} &\le \| A\|_{\rm 2}^2 \lambda_1 \cdot   \left\| W^{-1}(\widehat\Theta - \Theta) W^{-1} \right\|_{\rm 2} \\
& \le C  \lam_1 \left(\sqrt{\frac{r+ \log T}{T}} + \frac{(r+ \log T)^{1/\gamma}}{T}  \right),
\end{align*}
and,
\begin{align}\label{eq:thm_initial:upsilon}
\| \Upsilon_0 \|_{\rm 2} &\le \Big\| A \Theta  A^\top \Big\|_{\rm 2}+ \Big\| A\left( \widehat\Theta - \Theta \right) A^\top \Big\|_{\rm 2} \notag\\
&\le \| A\|_{\rm 2}^2 \lam_1  + \Big\| A\left( \widehat\Theta - \Theta \right) A^\top \Big\|_{\rm 2} \notag\\
&\le (1+\delta) \lam_1 + C  \lam_1 \left(\sqrt{\frac{r+ \log T}{T}} + \frac{(r+ \log T)^{1/\gamma}}{T}  \right) \\
&:=\Delta_{\Upsilon} .   \notag
\end{align}
Note that $\Delta_{\Upsilon} \lesssim \lam_1 .$

Next, consider $\|\Delta_2\|_{\rm 2}$. By Assumption \ref{asmp:error} and Lemma A.1 in \cite{shu2019estimation}, we have
\begin{align*}
&\PP\left(\|e_t\|_2^2 -\EE \|e_t\|_2^2 \ge x \right) = \PP\left( \xi_t^\top H^\top H \xi_t -\EE \xi_t^\top H^\top H \xi_t \ge x \right) \\
&\le 4\exp\left(-C'\left(\frac{x}{\|H^\top H\|_{\rm F}} \right)^{\frac{1}{1+2/\vartheta}} \right) \le 4\exp\left(-C'\left(\frac{x}{\sqrt{d}} \right)^{\frac{\vartheta}{\vartheta+2}} \right).
\end{align*}
Note that $\EE\|e_t\|_2^2=\EE\xi_t^\top H^\top H \xi_t=\EE \tr(H^\top H \xi_t\xi_t^\top)=\tr(H^\top H)=\tr(HH^\top )\asymp d$, and $\|H^\top H\|_{\rm F}^2=\|H H^\top \|_{\rm F}^2=\|\Sigma_e\|_{\rm F}^2\asymp d$.
Choosing $x\asymp d (\log T)^{(2\vartheta+4)/\vartheta}$, we have
\begin{align*}
\PP\left( \|e_t\|_2 \ge C\sqrt{d}(\log T)^{(\vartheta+2)/\vartheta}\right) \le 4\exp\left(-C'd^{\frac{\vartheta}{2\vartheta+4}}\log T \right).
\end{align*}
Let $N:=\|e_t e_t^\top \mathbf{1}_{\{\|e_t\|_2\le C\sqrt{d}(\log T)^{(\vartheta+2)/\vartheta}\} } \|_{ 2}$ and $\sigma_0^2:= \| \sum_{t=1}^T \EE (e_t e_t^\top \mathbf{1}_{\{\|e_t\|_2\le C\sqrt{d}(\log T)^{(\vartheta+2)/\vartheta}\} } )^2\|_{ 2}$. Then, by Assumption \ref{asmp:error}, $N\le C^2 d(\log T)^{(2\vartheta+4)/\vartheta}$ and 
\begin{align*}
\sigma_0^2&\le T \| \EE (e_t e_t^\top  e_t e_t^\top)\|_{ 2} = T \| \EE (H \xi_t \xi_t^\top H^\top H  \xi_t \xi_t^\top H^\top)\|_{ 2}  \\
&\le T \| \EE (H \xi_t \xi_t^\top H^\top H  \xi_t \xi_t^\top H^\top)\|_{\rm F} = T \| \EE (\xi_t^\top H^\top H  \xi_t \xi_t^\top H^\top H \xi_t)\|_{\rm F} \\
&= T \EE \sum_{j,l}(H^\top H)_{jl} \xi_{jt} \xi_{lt} \sum_{j' l'} (H^\top H)_{j'l'} \xi_{j't} \xi_{l't} \\
&\le C_0' T \left( \sum_{j,l}(H^\top H)_{jl}^2 +\sum_{j,l}(H^\top H)_{jj}(H^\top H)_{ll} \right) \\
&= C_0' T \left( \|H^\top H \|_{\rm F}^2 + [\tr(H^\top H)]^2 \right) \\
&\le C_0 Td.
\end{align*}

By matrix Bernstein inequality (see, e.g., Theorem 5.4.1 of \cite{vershynin2018high}),
\begin{align}\label{eq:delta2-1}
\PP\left(\left\|\sum_{t=1}^T \left[e_t e_t^\top \mathbf{1}_{\{\|e_t\|_2\le C\sqrt{d}(\log T)^{(\vartheta+2)/\vartheta}\} } -\EE e_t e_t^\top \mathbf{1}_{\{\|e_t\|_2\le C\sqrt{d}(\log T)^{(\vartheta+2)/\vartheta}\} } \right] \right\|_2 \ge x  \right)    \le 2d\exp\left( -\frac{x^2/2}{\sigma_0^2+N x/3}  \right).
\end{align}
Choosing $x\asymp \sqrt{Td\log (d)}+ d (\log d)(\log T)^{(2\vartheta+4)/\vartheta}$, with probability at least $1-d^{-c_1}$,
\begin{align}\label{eq:delta2-2}
\left\|\frac1T \sum_{t=1}^T e_t e_t^\top \mathbf{1}_{\{\|e_t\|_2\le C\sqrt{d}(\log T)^{(\vartheta+2)/\vartheta}\} } -\EE e_t e_t^\top \mathbf{1}_{\{\|e_t\|_2\le C\sqrt{d}(\log T)^{(\vartheta+2)/\vartheta}\} } \right\|_2 \le C_1 \sqrt{\frac{d\log(d)}{T}} + C_1 \cdot\frac{d\log(d)(\log T)^{\frac{2\vartheta+4}{\vartheta}} }{T}    .
\end{align}

Define $M:=\{1\le t\le T: \|e_t\|_2\ge C\sqrt{d}(\log T)^{(\vartheta+2)/\vartheta}\}$. Since $\mathbf{1}_{\{\|e_t\|_2\ge C\sqrt{d}(\log T)^{(\vartheta+2)/\vartheta}\} }$ are independent Bernoulli random variable, we have
\begin{align*}
\EE|M|=T\PP\left(\|e_t\|_2\ge C\sqrt{d}(\log T)^{(\vartheta+2)/\vartheta} \right) \le 4T\exp\left(-C' d^{\frac{\vartheta}{2\vartheta+4} } \log T \right) \le T^{-c_2}.
\end{align*}
By Chernoff bound for Bernoulli random variables,
\begin{align}\label{eq:delta2-3}
\PP(|M|\ge C)\le \exp\left(-T^{c_2} \right).
\end{align}
It follows that
\begin{align*}
\PP\left( \left\|\sum_{t=1}^T e_t e_t^\top \mathbf{1}_{\{\|e_t\|_2\ge C\sqrt{d}(\log T)^{(\vartheta+2)/\vartheta}\} } \right\|_2 \ge x \right) &\le \PP\left( |M| \max_t \| e_t\|_2^2 \ge x \right)   \\
&\le \PP(|M|\ge C) +\PP\left(|M|<C,  |M|\max_t \| e_t\|_2^2 \ge x \right) \\
&\le \exp\left(-T^{c_2} \right) + \PP\left(\max_t \| e_t\|_2^2 \ge x/C \right).
\end{align*}
Choosing $x\asymp d(\log T)^{(2\vartheta+4)/\vartheta}$, we have, with probability at least $1-\exp(-T^{c_2} )-T^{-c_2}$,
\begin{align}\label{eq:delta2-4}
 \left\|\frac1T\sum_{t=1}^T e_t e_t^\top \mathbf{1}_{\{\|e_t\|_2\ge C\sqrt{d}(\log T)^{(\vartheta+2)/\vartheta}\} } \right\|_2 \le C_2 \cdot\frac{d(\log T)^{\frac{2\vartheta+4}{\vartheta}} }{T}.
\end{align}
Similarly,
\begin{align}\label{eq:delta2-5}
\PP\left( \left\|\EE e_t e_t^\top \mathbf{1}_{\{\|e_t\|_2\ge C\sqrt{d}(\log T)^{(\vartheta+2)/\vartheta}\} } \right\|_2>0\right)=\PP \left(\| e_t\|_2 \ge C\sqrt{d}(\log T)^{(\vartheta+2)/\vartheta} \right)\le 4 \exp\left(-C' d^{\frac{\vartheta}{2\vartheta+4}} \log T \right) \le T^{-c_3}.
\end{align}
Combing \eqref{eq:delta2-2}, \eqref{eq:delta2-4}, \eqref{eq:delta2-5}, in an event $\Omega_2$ with probability at least $1-T^{-c_4}-d^{-c_1}$,
\begin{align*}
\|\Delta_2 \|_2 &\le \left\|\frac{1}{T} \sum_{t=1}^T e_{t} e_t^\top - \Sigma_e\right\|_2 + \| \Sigma_e\|_2   \\
&\le \left\|\frac1T \sum_{t=1}^T e_t e_t^\top \mathbf{1}_{\{\|e_t\|_2\le C\sqrt{d}(\log T)^{(\vartheta+2)/\vartheta}\} } -\EE e_t e_t^\top \mathbf{1}_{\{\|e_t\|_2\le C\sqrt{d}(\log T)^{(\vartheta+2)/\vartheta}\} } \right\|_2 \\
&\quad + \left\|\frac1T\sum_{t=1}^T e_t e_t^\top \mathbf{1}_{\{\|e_t\|_2\ge C\sqrt{d}(\log T)^{(\vartheta+2)/\vartheta}\} } \right\|_2
+  \left\|\EE e_t e_t^\top \mathbf{1}_{\{\|e_t\|_2\ge C\sqrt{d}(\log T)^{(\vartheta+2)/\vartheta}\} } \right\|_2 +  \| \Sigma_e\|_2 \\
&\le C_2 \sqrt{\frac{d\log(d)}{T}} + C_2 \cdot\frac{d\log(d)(\log T)^{\frac{2\vartheta+4}{\vartheta}} }{T} +C_2 .
\end{align*}

Next, consider $\|\Delta_3\|_2$. Note that $\|\Delta_4\|_2$ is the same as $\|\Delta_3\|_2$. Let $\overline\PP(\cdot)=\PP(\cdot|F_1,...,F_T)$ and $\overline\EE(\cdot)=\EE(\cdot|F_1,...,F_T)$ be the conditional probability and conditional expectation given the factor process, respectively. Similar to the derivation for $\|\Delta_2\|_2$, let
\begin{align*}
& N_1:=\left\|\sum_{i=1}^r w_i f_{it} a_i e_t^\top \mathbf{1}_{\{\|e_t\|_2\le C\sqrt{d}(\log T)^{1+\frac{2}{\vartheta}}\} } \right\|_{ 2}, \\
& \sigma_1^2:= \max\left\{\left\| \sum_{t=1}^T \overline\EE \sum_{i,j=1}^r w_i w_j f_{it} f_{jt} a_i e_t^\top e_t  a_j^\top \mathbf{1}_{\{\|e_t\|_2\le C\sqrt{d}(\log T)^{1+\frac{2}{\vartheta}}\} } \right\|_{ 2}, \right.\\
&\qquad\qquad \left.
\left\| \sum_{t=1}^T \overline\EE \sum_{i,j=1}^r w_i w_j f_{it} f_{jt} e_t  a_i^\top a_j  e_t^\top \mathbf{1}_{\{\|e_t\|_2\le C\sqrt{d}(\log T)^{1+\frac{2}{\vartheta}}\} } \right\|_{ 2} \right\}.
\end{align*}
It is easy to show
\begin{align*}
& N_1\le C\sqrt{d}(\log T)^{1+\frac{2}{\vartheta}}\left\|\sum_{i=1}^r w_i f_{it} a_i \right\|_{ 2}, \\
& \sigma_1^2 \le C_3 Td \max\left\{\left\| \frac1T\sum_{t=1}^T \sum_{i,j=1}^r w_i w_j f_{it} f_{jt} a_ia_j^\top \right\|_{ 2},
\left\| \frac{1}{Td}\sum_{t=1}^T \sum_{i,j=1}^r w_i w_j f_{it} f_{jt}  a_i^\top a_j  \right\|_{ 2} \right\}:=\sigma_2^2.
\end{align*}
By matrix Bernstein inequality,
\begin{align*}
&\overline\PP\left( \left\| \sum_{t=1}^T \left[ \sum_{i=1}^r w_i f_{i,t} a_i e_t^\top \mathbf{1}_{\{\|e_t\|_2\le C\sqrt{d}(\log T)^{(\vartheta+2)/\vartheta}\} } - \overline\EE \sum_{i=1}^r w_i f_{i,t} a_i e_t^\top \mathbf{1}_{\{\|e_t\|_2\le C\sqrt{d}(\log T)^{(\vartheta+2)/\vartheta}\} } \right] \right\|_2 \ge x \right)  \\
\le& 2d \exp\left( -\frac{x^2/2}{\sigma_1^2+N_1 x/3} \right).
\end{align*}
Choosing $x\asymp \sqrt{d} \log(d) (\log T)^{(\vartheta+2)/\vartheta} \|\sum_{i=1}^r w_i f_{it} a_i \|_{ 2} +\sqrt{\log(d)}\sigma_2$, with probability at least $1-d^{-c_4}$,
\begin{align*}
&\left\| \frac1T\sum_{t=1}^T  \sum_{i=1}^r w_i  f_{i,t} a_i e_t^\top \mathbf{1}_{\{\|e_t\|_2\le C\sqrt{d}(\log T)^{(\vartheta+2)/\vartheta}\} } - \overline\EE \sum_{i=1}^r w_i f_{i,t} a_i e_t^\top \mathbf{1}_{\{\|e_t\|_2\le C\sqrt{d}(\log T)^{(\vartheta+2)/\vartheta}\} }  \right\|_2 \\
\le& C_4\frac{\sqrt{\log(d)}}{T} \sigma_2   + C_4 \frac{\sqrt{d} \log(d)(\log T)^{1+\frac{2}{\vartheta}}\|\sum_{i=1}^r w_i f_{it} a_i \|_{ 2}}{T}.
\end{align*}
By Assumption \ref{asmp:eigenvalue}, with probability at least $1-T^{-c_5}$,
\begin{align*}
\left\|\sum_{i=1}^r w_i f_{it} a_i \right\|_{ 2} = \|F_t^\top W A^\top\|_2  \le (1+\delta) \|W F_t\|_2 \lesssim \sqrt{r} (\log(T))^{1/\gamma_1} \sqrt{\lam_1} . 
\end{align*}
By \eqref{eq:thm_initial:upsilon}, in the event $\Omega_1$, $\sigma_2^2\lesssim Td \lambda_1$. Then, with probability at least $1-T^{-c_1}/2-T^{-c_5}-d^{-c_4}$,
\begin{align*}
&\left\| \frac1T\sum_{t=1}^T  \sum_{i=1}^r w_i f_{i,t} a_i e_t^\top \mathbf{1}_{\{\|e_t\|_2\le C\sqrt{d}(\log T)^{(\vartheta+2)/\vartheta}\} } - \EE \sum_{i=1}^r w_i f_{i,t} a_i e_t^\top \mathbf{1}_{\{\|e_t\|_2\le C\sqrt{d}(\log T)^{(\vartheta+2)/\vartheta}\} }  \right\|_2 \\
\le& C_5\sqrt{\frac{d\log(d)}{T}} \sqrt{\lam_1}   + C_5 \cdot \frac{\sqrt{d r} \log(d)(\log T)^{1+\frac{2}{\vartheta}+\frac{1}{\gamma_1} }\sqrt{\lam_1}}{T}.
\end{align*}
Similar to \eqref{eq:delta2-4},
\begin{align*}
&\PP\left( \left\| \sum_{t=1}^T  \sum_{i=1}^r w_i f_{i,t} a_i e_t^\top \mathbf{1}_{\{\|e_t\|_2\ge C\sqrt{d}(\log T)^{(\vartheta+2)/\vartheta}\} }   \right\|_2 \ge x \right)  \\
\le& \PP(|M|>C)+\PP\left(|M|<C, |M|\max_t \left\|\sum_{i=1}^r w_i f_{it} a_i \right\|_{ 2} \cdot \|e_t\|_2 \ge x  \right).
\end{align*}
Choosing $x\asymp \sqrt{d r} (\log T)^{1+\frac{2}{\vartheta}+\frac{1}{\gamma_1}} \sqrt{\lam_1}$, we have with probability at least $1-\exp(-T^{c_2})-T^{-c_2}-T^{-c_5}$,
\begin{align*}
\left\| \frac1T \sum_{t=1}^T  \sum_{i=1}^r w_i f_{i,t} a_i e_t^\top \mathbf{1}_{\{\|e_t\|_2\ge C\sqrt{d}(\log T)^{(\vartheta+2)/\vartheta}\} }   \right\|_2 \le  C_6 \cdot \frac{\sqrt{d r} (\log T)^{1+\frac{2}{\vartheta}+\frac{1}{\gamma_1}} \sqrt{\lam_1}}{T} .
\end{align*}
Thus, in an event $\Omega_3$ with probability $1-T^{-c_1}/2-T^{-c_6}-d^{-c_4}$,
\begin{align*}
\|\Delta_3\|_2 \le&  \left\| \frac1T\sum_{t=1}^T  \sum_{i=1}^r w_i f_{i,t} a_i e_t^\top \mathbf{1}_{\{\|e_t\|_2\le C\sqrt{d}(\log T)^{(\vartheta+2)/\vartheta}\} } - \EE \sum_{i=1}^r w_i f_{i,t} a_i e_t^\top \mathbf{1}_{\{\|e_t\|_2\le C\sqrt{d}(\log T)^{(\vartheta+2)/\vartheta}\} }  \right\|_2 \\
&+ \left\| \frac1T \sum_{t=1}^T  \sum_{i=1}^r w_i  f_{i,t} a_i e_t^\top \mathbf{1}_{\{\|e_t\|_2\ge C\sqrt{d}(\log T)^{(\vartheta+2)/\vartheta}\} }   \right\|_2
+\left\| \EE  \sum_{i=1}^r w_i f_{i,t} a_i e_t^\top \mathbf{1}_{\{\|e_t\|_2\ge C\sqrt{d}(\log T)^{(\vartheta+2)/\vartheta}\} }   \right\|_2 \\
\le & C_7\sqrt{\frac{d\log(d)}{T}} \sqrt{\lam_1}   + C_7 \cdot \frac{\sqrt{d r} \log(d)(\log T)^{1+\frac{2}{\vartheta}+\frac{1}{\gamma_1}}\sqrt{\lam_1}}{T}.
\end{align*}

Therefore, in the event $\Omega_1\cap\Omega_2\cap\Omega_3$ with probability at least $1-T^{-c}-d^{-c}$, we have the desired bound for $\|\Psi^*\|_{\rm 2}$.

\end{proof}

\begin{lemma}\label{lem:rcpca}
Let $\lambda_1/\lam_r =\kappa $. Assume $d_1\lesssim \lam_1$ and $\delta_1^2 \kappa\le c$ for a sufficiently small positive constant $c$.
Apply random projection in Procedure \ref{alg:initialize-random} to the whole sample covariance tensor $\widehat\Sigma$ with $L\ge Cd^2 \vee Cdr^{2\kappa^2}$. Denote the estimated CP basis vectors as $\widetilde a_{\ell k}$, for $1\le \ell\le L, 1\le k\le K$. Then in an event with probability at least $1-T^{-c}-d^{-c}$, we have for any CP factor loading vectors tuple $(a_{ik}, 1\le k\le K)$, there exist $j_i\in[L]$ such that
\begin{align}
\left\| \widetilde a_{j_i,k} \widetilde a_{j_i,k}^\top  - a_{ik} a_{ik}^\top \right\|_2  &\le \psi_i,  \qquad 2\le k\le K,\\
\left\| \widetilde a_{j_i,1} \widetilde a_{j_i,1}^\top  - a_{i1} a_{i1}^\top \right\|_2  &\le \psi_i + (\delta/\delta_1) \kappa + \psi_i^{K-1} \kappa,
\end{align}
where $\psi_i=CR^{(0)}/\Theta_{ii}+\delta_1^2 \kappa$, $R^{(0)}=\phi^{(0)}$ is defined in \eqref{thm:initial:eq2} in Theorem \ref{thm:initial}, and $1\le i\le r$.
\end{lemma}
\begin{proof}
Without loss of generality, assume $\Theta_{11}\ge\Theta_{22}\ge\cdots\ge\Theta_{rr}$. Then $\Theta_{rr}\ge \lam_r, \Theta_{11}\le \lam_1$. Let $\widehat\Theta_{ij}=T^{-1}\sum_{t=1}^T w_i w_j f_{it} f_{jt}$ and $\Sigma_{\cE}=\EE \cE_t\otimes \cE_t$.
Write
\begin{align*}
\widehat\Sigma &= \frac{1}{T}\sum_{t=1}^T \cY_{t}\otimes\cY_{t}   \notag\\
&= \sum_{i,j=1}^r \Theta_{ij} \otimes_{k=1}^{K} a_{ik}\otimes_{k=K+1}^{2K} a_{jk} +  \sum_{i,j=1}^r (\widehat \Theta_{ij}- \Theta_{ij})  \otimes_{k=1}^{K} a_{ik} \otimes_{k=K+1}^{2K} a_{jk}   \notag\\
&\quad +   \frac{1}{T} \sum_{t=1}^T \sum_{i=1}^r w_i f_{it} \otimes_{k=1}^{K} a_{ik} \otimes\cE_t +   \frac{1}{T} \sum_{t=1}^T \sum_{i=1}^r w_i f_{it}  \cE_{t} \otimes_{k=K+1}^{2K} a_{ik} + \left(\frac{1}{T} \sum_{t=1}^T \cE_{t} \otimes \cE_t - \Sigma_{\cE} \right) +  \Sigma_{\cE} \notag\\
&:= \sum_{i,j=1}^r \Theta_{ij} \otimes_{k=1}^{K} a_{ik}\otimes_{k=K+1}^{2K} a_{jk} +\Delta_1+ \Delta_2+ \Delta_3+ \Delta_4+ \Delta_5,
\end{align*}
with $ a_{i,K+k}= a_{ik}$ for all $1\le k\le K$. Let $\Psi=\Delta_1+\Delta_2+\Delta_3+\Delta_4+\Delta_5$.
Let $\Xi(\theta)=\mat_{[K-1]}\widehat\Sigma\times_1\times_{K+1} \theta$. Unfold $\Psi\in \R^{d_1\times d_2\times\cdots \times d_K \times d_1\times d_2\times\cdots \times d_K}$ to be an order 4 tensor of dimension $(d/d_1)\times (d/d_1)\times d_1 \times d_1$ and denote it as $\widetilde \Psi$, and also define $\widetilde\Delta_k,k=1,...,5$ in a similar way. Then
\begin{align*}
\Xi(\theta)= \sum_{i,j=1}^r \Theta_{ij} (a_{i1}^\top \theta a_{j1}) \widetilde a_{i} \widetilde a_{j}^\top + \widetilde\Psi\times_3\times_{4} \theta    ,
\end{align*}
where $\widetilde a_i=\vec( a_{i2}\otimes \cdots  a_{iK})$. Let $\widetilde A=(\widetilde a_1,...,\widetilde a_r)\in\R^{(d/d_1)\times r}$.

First, consider the upper bound of $\| \widetilde\Psi\times_3\times_{4} \theta \|_2$. By concentration inequality for matrix Gaussian sequence (see, for example Theorem 4.1.1 in \cite{tropp2015introduction}) and employing similar arguments in the proof of Lemma \ref{lem:psi}, we have, with probability at least $1-d^{-c}$
\begin{align*}
\left\| \widetilde\Delta_4 \times_3\times_{4} \theta \right\|_2   &= \left\| \sum_{k,l} \theta_{(kl)} (\widetilde \Delta_4)_{\cdot \cdot k l} \right\|_2  \\
&\le C \max\left\{ \left\|\mat_{(1),(234)} (\widetilde \Delta_4) \right\|_2  ,   \left\|\mat_{(2),(134)} (\widetilde \Delta_4) \right\|_2 \right\} \cdot \sqrt{\log(d)} \\
&\le C \left(\sqrt{\frac{d d_1 \log(d)}{T}} + \frac{d \log(d)(\log T)^{\frac{2\vartheta+4}{\vartheta}}}{T} \right) \cdot \sqrt{\log(d)},
\end{align*}
where $\theta_{(kl)}$ is the $(k,l)$th element of $\theta$, $(\widetilde \Delta_4)_{\cdot \cdot k l}$ represents the $(k,l)$th (3,4) slice of $\widetilde \Delta_4$, and $\mat_{(1),(234)}(\cdot)$ denotes the reshaping of fourth-order tensor into a matrix by collapsing its first indices as rows, and the second, third, fourth indices as columns. In the last step, we apply the arguments in the proof of Lemma \ref{lem:psi}.
Similarly, we have, with probability at least $1-d^{-c}$,
\begin{align*}
\left\| \Sigma_{\cE} \times_3\times_{4} \theta \right\|_2 \le C\sqrt{\log(d)}.
\end{align*}
And, with probability at least $1-T^{-c}-d^{-c}$,
\begin{align*}
&\left\| (\widetilde\Delta_1+\widetilde\Delta_2+\widetilde\Delta_3+\widetilde\Delta_4) \times_3\times_{4} \theta \right\|_2   \\
\le& C  \lambda_1  \left(\sqrt{\frac{r+ \log T}{T}}  + \frac{(r+ \log T)^{1/\gamma}}{T}  \right) \cdot \sqrt{\log(d)}  + C \left(\sqrt{\frac{d d_1 \log(d)}{T}} + \frac{d \log(d)(\log T)^{\frac{2\vartheta+4}{\vartheta}}}{T} \right) \cdot \sqrt{\log(d)} \\
&+ C \lambda_1^{1/2} \left(\sqrt{\frac{d\log(d)}{T}}   + \frac{\sqrt{d r} \log(d)(\log T)^{1+\frac{2}{\vartheta}+\frac{1}{\gamma_1} }  }{T} \right) \cdot \sqrt{\log(d)} .
\end{align*}
As $d_1\lesssim \lam_1$, it follows that in an event $\Omega_0$ with probability at least $1-T^{-c}-d^{-c}$,
\begin{align}
\left\| \widetilde \Psi \times_3\times_{4} \theta \right\|_2     \le C \|\Psi^*\|_2 \sqrt{\log(d)}.
\end{align}

Consider the $i$-th factor and rewrite $\Xi(\theta)$ as follows
\begin{align}
\Xi(\theta)= \Theta_{ii} (a_{i1}^\top \theta a_{i1}) \widetilde a_{i} \widetilde a_{i}^\top + \sum_{(j_1,j_2)\neq (i,i)} \Theta_{j_1 j_2} (a_{j_1 1}^\top \theta a_{j_2 1}) \widetilde a_{j_1} \widetilde a_{j_2}^\top + \widetilde\Psi\times_3\times_{4} \theta    .
\end{align}
Suppose now we repeatedly sample $\theta_{\ell}\sim \theta$, for $\ell=1,...,L$. By the anti-concentration inequality for Gaussian random variables (see Lemma B.1 in \cite{anandkumar2014tensor}), we have
\begin{align}
\PP\left( \max_{1\le \ell \le L} ( a_{i1} \odot a_{i1})^\top \vec(\theta_{\ell}) \le \sqrt{2 \log(L)} - \frac{\log\log(L)}{4 \sqrt{\log(L)}} - \sqrt{2\log(8)} \right)\le \frac14,
\end{align}
where $\odot$ denotes Kronecker product. Let
\begin{align*}
\ell_*=\arg\max_{1\le \ell\le L}  ( a_{i1} \odot a_{i1})^\top \vec(\theta_{\ell})   .
\end{align*}
Note that $( a_{i1}\odot a_{i1})^\top \vec(\theta_{\ell})$ and $( I_{d_1^2} - (a_{i1}\odot a_{i1})(a_{i1}\odot a_{i1})^\top ) \vec(\theta_{\ell})$ are independent. Since the definition of $\ell_*$ depends only on $( a_{i1}\odot a_{i1})^\top \vec(\theta_{\ell})$, this implies that the distribution of $( I_{d_1^2} - (a_{i1}\odot a_{i1})(a_{i1}\odot a_{i1})^\top ) \vec(\theta_{\ell})$ does not depend on $\ell_*$.

By Gaussian concentration inequality of $1$-Lipschitz function, we have
\begin{align*}
\PP\left( \max_{j_1,j_2 \le r} \big( a_{j_1 1} \odot a_{j_2 1} \big)^\top \big( I_{d_1^2} - (a_{i1}\odot a_{i1})(a_{i1}\odot a_{i1})^\top \big) \vec(\theta_{\ell}) \ge \sqrt{4 \log(r)} +\sqrt{2 \log(8)} \right) \le \frac14   .
\end{align*}
Moreover, for the reminder bias term $( a_{j_1 1} \odot a_{j_2 1} )^\top  (a_{i1}\odot a_{i1})(a_{i1}\odot a_{i1})^\top \vec(\theta_{\ell})$, we have,
\begin{align*}
&\left\| \sum_{(j_1,j_2)\neq (i,i)} \Theta_{j_1,j_2}( a_{j_1 1} \odot a_{j_2 1} )^\top  (a_{i1}\odot a_{i1})(a_{i1}\odot a_{i1})^\top \vec(\theta_{\ell}) \cdot  \widetilde a_{j_1} \widetilde a_{j_2}^\top \right\|_2    \\
&\le ( a_{i1} \odot a_{i1})^\top \vec(\theta_{\ell}) \cdot \left\| \widetilde A \left( \Theta \circ \left( A_1^\top a_{i1} a_{i1}^\top A_1 - e_{ii}\right) \right)\widetilde A^\top \right\|_2 \\
&\le ( a_{i1} \odot a_{i1})^\top \vec(\theta_{\ell})  \|\widetilde A \|_2^2  \|\Theta\|_2 \left\| A_1^\top a_{i1} a_{i1}^\top A_1 - e_{ii} \right\|_2 \\
&\le (1+\delta/\delta_1) \delta_1^2 \lambda_1 ( a_{i1} \odot a_{i1})^\top \vec(\theta_{\ell}),
\end{align*}
where $\circ$ denotes Hadamard product and $e_{ii}$ is a $d_1\times d_1$ matrix with the $(i,i)$-th element be 1 and all the others be 0.

Thus, we obtain the top eigengap
\begin{align}\label{eq:lem_gap}
&( a_{i1} \odot a_{i1})^\top \vec(\theta_{\ell_*})\Theta_{ii} -  \left\| \sum_{(j_1,j_2)\neq (i,i)} \Theta_{j_1 j_2} \left(( a_{j_1 1} \odot a_{j_2 1} )^\top \vec(\theta_{\ell_*}\right) \widetilde a_{j_1} \widetilde a_{j_2}^\top \right\|_2   \notag\\
&\ge (1- 2\delta_1^2 \kappa)\left( \sqrt{2 \log(L)} - \frac{\log\log(L)}{4 \sqrt{\log(L)}} - \sqrt{2\log(8)} \right) \Theta_{ii} - \left( \sqrt{4 \log(r)} +\sqrt{2\log(8)} \right) \kappa \Theta_{ii} \notag\\
&\ge C_0 \sqrt{\log(d)} \Theta_{ii},
\end{align}
with probability at least $\frac12$, by letting $L\ge Cd \vee Cr^{2\kappa^2}$.

Since $\theta_{\ell}$ are independent samples, we instead take $L_i=L_{i1}+\cdots+L_{iM}$ for $M=\lceil C_1\log (d)/\log(2) \rceil$ and $L_{i1},...,L_{iM}\ge Cd \vee Cr^{2\kappa^2}$. We define
\begin{align*}
\ell_*^{(m)}=\arg\max_{1\le \ell\le L_{im}}  ( a_{i1} \odot a_{i1})^\top \vec(\theta_{\ell}), \quad   \ell_{*}=\arg\max_{1\le \ell\le L_{i}}  ( a_{i1} \odot a_{i1})^\top \vec(\theta_{\ell}).
\end{align*}
We then have, by independence of $\theta_{\ell}$, that the above statement \eqref{eq:lem_gap} for the $i$-th factor holds in an event $\Omega_i$ with probability at least $1-d^{-C_1}$. By Wedin's perturbation theory, we have in the event $\Omega_0\cap\Omega_i$,
\begin{align*}
\left\| \widehat a_{\ell_*} \widehat a_{\ell_*}^\top - \widetilde a_{i} \widetilde a_{i}^\top \right\|_2 \le \frac{CR^{(0)}}{\Theta_{ii}} + \delta_1^2 w,
\end{align*}
where $\widehat a_{\ell_*}$ is the top left singular vector of $\Xi(\theta_{\ell_*})$, and $R^{(0)}=\phi^{(0)}$ is defined in \eqref{thm:initial:eq2} in Theorem \ref{thm:initial}. By Lemma \ref{prop-rank-1-approx},
\begin{align}\label{eq1:lem_rcpca}
\left\| \widetilde a_{\ell_*,k} \widetilde a_{\ell_*,k}^\top - a_{ik} a_{ik}^\top \right\|_2 \le \frac{CR^{(0)}}{\Theta_{ii}}+ \delta_1^2 \kappa, \qquad 2\le k\le K.
\end{align}

Now consider to obtain $\widetilde a_{\ell_*,1}$. Write $\psi_i=CR^{(0)}/\Theta_{ii}+\delta_1^2 \kappa$. Note that
\begin{align*}
\widehat\Sigma \times_{k=2}^K \widetilde a_{\ell_*,k} \times_{k=K+2}^{2K} \widetilde a_{\ell_*,k} =& \prod_{k=2}^K \left(\widetilde a_{\ell_*,k}^\top a_{ik} \right)^2 \Theta_{ii}  a_{i1} a_{i1}^\top +\Psi  \times_{k=2}^K \widetilde a_{\ell_*,k} \times_{k=K+2}^{2K} \widetilde a_{\ell_*,k} \\
&+  \sum_{(j_1,j_2)\neq (i,i)} \prod_{k=2}^K \left(\widetilde a_{\ell_*,k}^\top a_{j_1 k} \right)\left(\widetilde a_{\ell_*,k}^\top a_{j_2 k} \right) \Theta_{j_1 j_2}  a_{j_1 1} a_{j_2 1}^\top   .
\end{align*}
By Lemma \ref{lem:psi} and \eqref{eq1:lem_rcpca}, in the event $\Omega_0\cap\Omega_1$,
\begin{align*}
&\left\| \Psi  \times_{k=2}^K \widetilde a_{\ell_*,k} \times_{k=K+2}^{2K} \widetilde a_{\ell_*,k} \right\|_2    \le \| \Psi^*\|_2,\\
&\prod_{k=2}^K \left(\widetilde a_{\ell_*,k}^\top a_{ik} \right)^2  \ge (1-\psi_i^2)^{K-1}  .
\end{align*}
Since
\begin{align}\label{eq:a_decomp}
\max_{j_1\neq i}\big| a_{j_1 k}^\top \widetilde a_{\ell_*,k}  \big| &=\max_{j_1\neq i}\big| \widetilde a_{\ell_*,k}^\top a_{ik} a_{ik}^\top a_{j_1 k} + \widetilde a_{\ell_*,k}^\top (I - a_{ik} a_{ik}^\top ) a_{j_1 k}   \big|    \notag\\
&\le \max_{j_1\neq i}\big| \widetilde a_{\ell_*,k}^\top a_{ik} \big| \big| a_{ik}^\top a_{j_1 k} \big| + \max_{j_1\neq i}\big\|\widetilde a_{\ell_*,k}^\top (I - a_{ik} a_{ik}^\top ) \big\|_2 \big\| (I - a_{ik} a_{ik}^\top )  a_{j_1 k}   \big\|_2 \notag\\
&\le \sqrt{1-\psi_i^2} \delta_k + \psi_i \sqrt{1-\delta_k^2} \le \delta_k +\psi_i,
\end{align}
we have
\begin{align*}
\left\| \sum_{(j_1,j_2)\neq (i,i)} \prod_{k=2}^K \left(\widetilde a_{\ell_*,k}^\top a_{j_1 k} \right)\left(\widetilde a_{\ell_*,k}^\top a_{j_2 k} \right) \Theta_{j_1 j_2}  a_{j_1 1} a_{j_2 1}^\top \right\|_2 &\le (1+\delta_1)\prod_{k=2}^K(\delta_k+\psi_i) \lambda_1 \\
& \le C_K (\delta/\delta_1+\psi_i^{K-1}) \kappa \Theta_{ii}.
\end{align*}
By Wedin's perturbation theory,
\begin{align}\label{eq2:lem_rcpca}
\left\| \widetilde a_{\ell_*,1} \widetilde a_{\ell_*,1}^\top - a_{i1} a_{i1}^\top \right\|_2 \le \frac{CR^{(0)}}{\Theta_{ii}}+ (\delta/\delta_1) \kappa + \psi_i^{K-1} \kappa
\end{align}

Repeat the same argument again for all $1\le i\le r$ factors, and let $L=\sum_{i} L_i\ge Cd^2 \vee Cdr^{2\kappa^2} \ge Cdr\log(d) \vee Cr^{2\kappa^2+1}\log(d)$. We have, in the event $\Omega_0\cap\Omega_1\cap\cdots\cap\Omega_r$ with probability at least $1-T^{-c}-d^{-c}$, \eqref{eq1:lem_rcpca} and \eqref{eq2:lem_rcpca} hold for all $i$.

\end{proof}

\begin{proof}[\bf Proof of Theorem \ref{thm:projection}]
Recall $\widehat A_k^{(m)}=(\widehat a_{1k}^{(m)},\ldots,\widehat a_{rk}^{(m)})\in\R^{d_k\times r}$, $\widehat\Sigma_k^{(m)}=\widehat A_k^{(m)\top}\widehat A_k^{(m)}$, and $\widehat B_k^{(m)} = \widehat A_k^{(m)}(\widehat\Sigma_k^{(m)})^{-1} = (\widehat b_{1k}^{(m)},...,\widehat b_{rk}^{(m)}) \in\R^{d_k\times r}$. Let $\overline\EE(\cdot)=\EE(\cdot|F_{t}, 1\le t\le T)$ and $\overline\PP(\cdot)=\PP(\cdot|F_{t},1\le t\le T)$. Also let
\begin{align*}
\overline\lambda_{i}= \frac{1}{T} \sum_{t=1}^T w_i^2 f_{it}^2    .
\end{align*}
Without loss of generality, assume $\Theta_{11}\ge\Theta_{22}\ge\cdots\ge\Theta_{rr}$. Then $\EE \overline\lam_r=\Theta_{rr}\ge \lam_r, \EE \overline\lam_1=\Theta_{11}\le \lam_1$. Write
\begin{align}\label{eq:sigmahat-dcomp}
\widehat\Sigma &= \frac{1}{T}\sum_{t=1}^T \cY_{t}\otimes\cY_{t}   \notag\\
&= \sum_{i=1}^r \overline\lambda_i \otimes_{k=1}^{2K} a_{ik} +  \sum_{i\neq j}^r \frac{1}{T} \sum_{t=1}^T w_i w_j f_{it}f_{jt}  \otimes_{k=1}^{K} a_{ik} \otimes_{k=K+1}^{2K} a_{jk} + \frac{1}{T} \sum_{t=1}^T \cE_{t} \otimes \cE_t \notag\\
&\quad +   \frac{1}{T} \sum_{t=1}^T \sum_{i=1}^r w_i f_{it} \otimes_{k=1}^{K} a_{ik} \otimes\cE_t +   \frac{1}{T} \sum_{t=1}^T \sum_{i=1}^r w_i f_{it}  \cE_{t} \otimes_{k=K+1}^{2K} a_{ik}    \notag\\
&:= \sum_{i=1}^r \overline\lambda_i \otimes_{k=1}^{2K} a_{ik} +\Delta_1+ \Delta_2+ \Delta_3+ \Delta_4,
\end{align}
with $ a_{i,K+k}= a_{ik}$ for all $1\le k\le K$. Let $\Psi=\Delta_1+\Delta_2+\Delta_3+\Delta_4$.

By Theorem \ref{thm:initial}, in an event $\Omega_0$ with probability at least $1-T^{-C_1}-d^{-C_1}$,
\begin{align*}
\| \widehat a_{ik}^{(0)}\widehat a_{ik}^{(0)\top}  - a_{ik}  a_{ik}^\top \|_{\rm 2} &\le \psi_0.
\end{align*}
At $m$-th step, let
\begin{align}
\psi_{m,i,k} :=\| \widehat a_{ik}^{(m)}\widehat a_{ik}^{(m)\top}  - a_{ik} a_{ik}^\top \|_{\rm 2},\qquad \psi_{m,k} :=\max_i\psi_{m,i,k}, \qquad \psi_m =\max_k\psi_{m,k}.
\end{align}
Let $ g_{i\ell}= b_{i\ell}/\| b_{i\ell}\|_2$ and $\widehat g_{i\ell}^{(m)}=\widehat b_{i\ell}^{(m)}/\|\widehat b_{i\ell}^{(m)}\|_2$.
Given $\widehat a_{i\ell}^{(m)}$ ($1\le i\le r, 1\le \ell \le K$), the $(m+1)$th iteration produces estimates $\widehat a_{ik}^{(m+1)}$, which is the top left singular vector of $\widehat\Sigma\times_{\ell\in [ 2K] \backslash\{k,K+k\} } \widehat b_{i\ell}^{(m)\top}$, or equivalently $\widehat\Sigma\times_{\ell\in [ 2K] \backslash\{k,K+k\} } \widehat g_{i\ell}^{(m)\top}$. Note that $\widehat\Sigma = \sum_{j=1}^r  \overline\lambda_j \otimes_{\ell=1}^{2K}  a_{j\ell} +  \Psi$, with $ a_{j,\ell+K}= a_{j\ell}$. The ``noiseless'' version of this update is given by
\begin{equation}\label{error-free-update}
\widehat\Sigma\times_{\ell\in [ 2K] \backslash\{k,K+k\} }  g_{i\ell}^{\top}=  \overline\lambda_{i} a_{ik} a_{ik}^\top + \Psi \times_{\ell\in [ 2K] \backslash\{k,K+k\} }  g_{i\ell}^{\top}.
\end{equation}
At $(m+1)$-th iteration, for any $1\le i\le r$, we have
\begin{align*}
\widehat\Sigma\times_{\ell\in [ 2K] \backslash\{k,K+k\} } \widehat g_{i\ell}^{(m)\top} = \sum_{j=1}^r   \widetilde\lambda_{j,i}  a_{jk} a_{jk}^\top +  \Psi \times_{\ell\in [ 2K] \backslash\{k,K+k\} } \widehat g_{i\ell}^{(m)\top},
\end{align*}
where
\begin{align}\label{eq:lambdaji}
\widetilde\lambda_{j,i}=\overline\lambda_j\prod_{\ell\in [ 2K] \backslash\{k,K+k\} }  a_{j\ell}^\top\widehat g_{i\ell}^{(m)}   .
\end{align}
Let
\begin{align*}
\lambda_{j,i}&=\Theta_{jj}\prod_{\ell\in [ 2K] \backslash\{k,K+k\} }   a_{j\ell}^\top\widehat  g_{i\ell}^{(m)}, \\
\alpha&=\sqrt{1-\delta_{\max}}-(r^{1/2}+1)\psi_0/\sqrt{1-1/(4r)},\\
\phi_{m,\ell}&=1 \wedge \frac{\psi_{m,\ell}\sqrt{2r}}{\alpha \sqrt{1-1/(4r)}}  ,  \\
\phi_{m}&=\max_{\ell} \phi_{m,\ell}.
\end{align*}
We may assume without loss of generality $ a_{j\ell}^\top\widehat a_{j\ell}^{(m)}\ge 0$ for all $(j,\ell)$.
Similar to the proofs of Theorem 3 in \cite{han2023tensor}, we can show
\begin{align}\label{a-bd}
&\max_{j\le r}\|\widehat a_{j\ell}^{(m)} -  a_{j\ell} \|_2 \le \psi_{m,\ell}/\sqrt{1-1/(4r)}, \ \
\displaystyle \big\|\widehat b_{j\ell}^{(m)}\big\|_2 \le \|\widehat B_\ell^{(m)}\|_{\rm 2}
\le \bigg(\sqrt{1-\delta_\ell}-\frac{r^{1/2}\psi_0}{\sqrt{1-1/(4r)}}\bigg)^{-1}, \\
&\big\|\widehat g_{j\ell}^{(m)} -  b_{j\ell}/\| b_{j\ell}\|_2\big\|_2
\le (\psi_{m,\ell}/\alpha)\sqrt{2r/(1-1/(4r))}.  \label{b-bd}
\end{align}
Moreover,
\eqref{a-bd} provides
\begin{align}\label{g-bd}
\max_{i\neq j}\big| a_{i\ell}^\top\widehat g_{j\ell}^{(m)}\big| \le \psi_{m,\ell}/\sqrt{1-1/(4r)},\ \
\big| a_{j\ell}^\top\widehat g_{j\ell}^{(m)} \big| \ge \alpha,
\end{align}
as $\widehat a_{i\ell}^{(m)\top}\widehat g_{j\ell}^{(m)}=I\{i=j\}/\|\widehat b_{j\ell}^{(m)}\|_2$.
Then, for $j\neq i$,
\begin{align*}
\lambda_{j,i}/\lambda_{i,i} \le \big(\lam_{1}/\Theta_{ii}\big)   \prod_{\ell\neq k}^K \left(\frac{\psi_{m,\ell}/\sqrt{1-1/(4r)} }{1- \psi_{m,\ell}/\sqrt{1-1/(4r)} } \right)^2.
\end{align*}
Employing similar arguments in the proof of Lemma \ref{lem:psi}, in an event $\Omega_1$ with probability at least $1-T^{-c_1}$, we have
\begin{align}\label{eq:thetah}
\left\|\widehat\Theta - \Theta\right\|_{\rm 2} & \le C_1 \lam_1 \left(\sqrt{\frac{r+\log T}{T}} + \frac{(r+\log T)^{1/\gamma}}{T}  \right),
\end{align}
In the event $\Omega_{1}$, we also have
\begin{align*}
\max_{1\le j_1,j_2\le r}\left| \frac{1}{T}\sum_{t=1}^T w_{j_1} w_{j_2} f_{j_1,t} f_{j_2,t} -\EE w_{j_1} w_{j_2} f_{j_1,t} f_{j_2,t} \right|   &\le C_1  \sqrt{\Theta_{j_1,j_1}\Theta_{j_2,j_2}}\left(\sqrt{\frac{r+\log T}{T}} + \frac{(r+\log T)^{1/\gamma}}{T}  \right) .
\end{align*}
It follows that in the event $\Omega_{1}$, for any $1\le j\le r$,
\begin{align*}
|\overline \lambda_j -\Theta_{jj}| \le C_1 \Theta_{jj} \left(\sqrt{\frac{r+\log T}{T}} + \frac{(r+\log T)^{1/\gamma}}{T}  \right)  .
\end{align*}
By Wedin's theorem \citep{wedin1972}, in the event $\Omega_1$,
\begin{align}\label{wedin72}
\|\widehat a_{ik}^{(m+1)} \widehat a_{ik}^{(m+1)\top} - a_{ik}  a_{ik}^\top \|_{\rm 2} &\le \frac{2\left\| \sum_{j\neq i}^r \widetilde\lambda_{j,i}  a_{jk} a_{jk}^\top  \right\|_{\rm 2} + 2\| \Psi \times_{\ell\in [ 2K] \backslash\{k,K+k\} } \widehat g_{i\ell}^{(m)\top} \|_{\rm 2} }{\widetilde\lambda_{i,i}} \notag\\
&\le \frac{4\| A_k\|_{\rm 2}^2 \max_{j\neq i}|\lambda_{j,i}  | + 4  \| \Psi \times_{\ell\in [ 2K] \backslash\{k,K+k\} } \widehat  g_{i\ell}^{(m)\top} \|_{\rm 2} }{\alpha^{2k-2} \Theta_{ii} } .
\end{align}

To bound the numerator of \eqref{wedin72}, we write
\begin{align*}
&\Delta_{1,1}=\sum_{j_2\neq i}^r \frac{1}{T} \sum_{t=1}^T w_i w_{j_2} f_{it}f_{j_2,t}   \otimes_{\ell=1}^{K}  a_{i\ell}\otimes_{\ell=K+1}^{2K}  a_{j_2\ell} \times_{\ell\in [ 2K] \backslash\{k,K+k\} } \widehat  g_{i\ell}^{(m)\top} , \\
&\Delta_{1,2}=\sum_{j_1\neq i}^r \frac{1}{T} \sum_{t=1}^T w_i w_{j_1} f_{j_1,t}f_{it}   \otimes_{\ell=1}^{K}  a_{j_1\ell}\otimes_{\ell=K+1}^{2K}  a_{i\ell} \times_{\ell\in [ 2K] \backslash\{k,K+k\} } \widehat  g_{i\ell}^{(m)\top} , \\
&\Delta_{1,3}=\sum_{j_1\neq j_2\neq i}^r \frac{1}{T} \sum_{t=1}^T w_{j_1} w_{j_2} f_{j_1,t}f_{j_2,t}   \otimes_{\ell=1}^{K}  a_{j_1\ell}\otimes_{\ell=K+1}^{2K}  a_{j_2\ell} \times_{\ell\in [ 2K] \backslash\{k,K+k\} } \widehat  g_{i\ell}^{(m)\top} .
\end{align*}
For any vectors $\widetilde   g_{i\ell},\widecheck   g_{i\ell}\in\R^{d_{\ell}}$, define
\begin{align*}
&\Delta_{2,k}(\widetilde  g_{i\ell},\widecheck   g_{i\ell}, \ell\neq k) = \frac{1}{T} \sum_{t=1}^T \cE_{t} \otimes \cE_t \times_{\ell=1,\ell\neq k}^K \widetilde  g_{i\ell}^\top \times_{\ell=K+1,\ell\neq K+k}^{2K} \widecheck  g_{i,\ell-K}^\top \in\R^{d_k \times d_k}, \\
&\Delta_{3,k}(\widetilde  g_{i\ell}, \ell\neq k) =\frac{1}{T} \sum_{t=1}^T w_i f_{it} \cE_{t}  \times_{\ell=1,\ell\neq k}^K \widetilde  g_{i\ell}^\top\in\R^{d_k} , \\
&\Delta_{4,k}(\widetilde  g_{i\ell}, \ell\neq k) =\frac{1}{T} \sum_{t=1}^T  \cE_{t} \otimes ( w_j f_{jt}, \ j\neq i)^\top \times_{\ell=1,\ell\neq k}^K \widetilde  g_{i\ell}^\top  \in\R^{d_k\times (r-1)} . 
\end{align*}
As $\Delta_{q,k}(\widetilde  g_{i\ell},\widecheck   g_{i\ell}, \ell\neq k)$ is linear in $\widetilde   g_{i\ell},\widecheck   g_{i\ell}$, by \eqref{g-bd}, the numerator on the right hand side of \eqref{wedin72} can be bounded by
\begin{align}\label{norm-bd}
&\| \Psi \times_{\ell\in [ 2K] \backslash\{k,K+k\} } \widehat  g_{i\ell}^{(m)\top} \|_{\rm 2} \notag\\
\le& \| \Delta_1 \times_{\ell\in [ 2K] \backslash\{k,K+k\} } \widehat  g_{i\ell}^{(m)\top} \|_{\rm 2}    + \| \Delta_{2,k}(\widehat  g_{i\ell}^{(m)},\widehat  g_{i\ell}^{(m)}, \ell\neq k) \|_{\rm 2} + 2\| \Delta_{3,k}(\widehat  g_{i\ell}^{(m)}, \ell\neq k) \|_{\rm 2}   \notag\\
&\quad  + 2\| A_k\|_{\rm 2}\| \Delta_{4,k}(\widehat  g_{i\ell}^{(m)}, \ell\neq k) \|_{\rm 2} \max_{j\neq i}\prod_{\ell \neq k}^K \left| a_{j \ell}^\top \widehat g_{i\ell}^{(m)} \right| \notag\\
\le&  \sum_{q=1,2,3} \| \Delta_{1,q} \|_{\rm 2}  + \| \Delta_{2,k}(  g_{i\ell},   g_{i\ell}, \ell\neq k) \|_{\rm 2}   +(2K-2) \phi_{m,k}  \| \Delta_{2,k} \|_{\rm S} \notag\\
&\quad + 2\| \Delta_{3,k}(  g_{i\ell}, \ell\neq k) \|_{\rm 2} +  (4K-4) \phi_{m,k} \| \Delta_{3,k} \|_{\rm S} \notag\\
&\quad +  2\| A_k\|_{\rm 2}\prod_{\ell\neq k}^K \big(\psi_{m,\ell}/\sqrt{1-1/(4r)}\big)  \| \Delta_{4,k}(  g_{i\ell}, \ell\neq k) \|_{\rm 2} \notag\\
&\quad + 2\| A_k\|_{\rm 2}(2K-2) \phi_{m,k} \prod_{\ell\neq k}^K \big(\psi_{m,\ell}/\sqrt{1-1/(4r)}\big) \| \Delta_{4,k} \|_{\rm S},
\end{align}
where
\begin{align*}
&\| \Delta_{2,k}\|_{\rm S}= \max_{\substack{ \|\widetilde  g_{i\ell}\|_2=\|\widecheck  g_{i\ell}\|_2=1, \\ \widetilde  g_{i\ell},\widecheck  g_{i\ell}\in \R^{d_{\ell}} } } \| \Delta_{2,k}(\widetilde  g_{i\ell},\widecheck   g_{i\ell}, \ell\neq k) \|_{\rm 2} ,\\
&\| \Delta_{q,k}\|_{\rm S}= \max_{\substack{ \|\widetilde  g_{i\ell}\|_2=1, \\ \widetilde  g_{i\ell} \in \R^{d_{\ell}} } } \| \Delta_{q,k}(\widetilde  g_{i\ell}, \ell\neq k) \|_{\rm 2}, \quad q=3,4 .
\end{align*}
Note that
\begin{align*}
\Delta_{1,1}= a_{ik}\left(\prod_{\ell\neq k}^K  a_{i\ell}^\top \widehat  g_{i\ell}^{(m)} \right) \frac{1}{T}\sum_{t=1}^T w_i f_{i,t} \cdot (w_{j_2} f_{j_2,t},\ j_2\neq i) \  {\rm diag}\left(\prod_{\ell\neq k}^K  a_{j_2 \ell}^\top \widehat  g_{i\ell}^{(m)},j_2\neq i  \right) ( a_{j_2 k}, j_2\neq i)^\top.
\end{align*}
By \eqref{eq:thetah}, in the event $\Omega_1$,
\begin{align}\label{Delta-bd-11}
\|\Delta_{1,1}\|_{\rm 2} &\lesssim \sqrt{\lam_1\Theta_{ii}}\left(\sqrt{\frac{r+\log T }{T}} + \frac{(r+\log T)^{1/\gamma}}{T}  \right) \prod_{\ell \neq k}^K  \psi_{m,\ell}.
\end{align}
Similarly, in the event $\Omega_1$,
\begin{align}
\|\Delta_{1,2}\|_{\rm 2} &\lesssim \sqrt{\lam_1\Theta_{ii}} \left(\sqrt{\frac{r+\log T }{T}} + \frac{(r+\log T)^{1/\gamma}}{T}  \right) \prod_{\ell \neq k}^K \psi_{m,\ell} , \label{Delta-bd-12} \\
\|\Delta_{1,3}\|_{\rm 2} &\lesssim \lam_1 \left(\sqrt{\frac{r+\log T }{T}} + \frac{(r+\log T)^{1/\gamma}}{T}  \right) \prod_{\ell \neq k}^K \psi_{m,\ell}^2 .  \label{Delta-bd-13}
\end{align}
Let $\Upsilon_{0,i,k}=T^{-1} \sum_{t=1}^T w_i^2 f_{it}^2 a_{ik}  a_{ik}^\top$ and $\Upsilon_{0,-i,k}=T^{-1} \sum_{t=1}^T \sum_{j_1,j_2\neq i}^r w_{j_1} w_{j_2} f_{j_1t}f_{j_2t}  a_{j_1k} a_{j_2k}^\top$. Then, in the event $\Omega_1$,
\begin{align}
\| \Upsilon_{0,i,k} \|_{\rm 2} &\le \Theta_{ii} + C_1 \Theta_{ii} \left(\sqrt{\frac{r+\log T}{T}} + \frac{(r+\log T)^{1/\gamma}}{T}  \right) :=\Delta_{\Upsilon_i} \asymp \Theta_{ii}, \label{eq:thm_projection:upsilon} \\
\| \Upsilon_{0,-i,k} \|_{\rm 2} &\le \lam_1 + C_1 \lam_1 \left(\sqrt{\frac{r+\log T }{T}} + \frac{(r+\log T)^{1/\gamma}}{T}  \right) :=\Delta_{\Upsilon_{-i}} \asymp \lambda_1. \label{eq:thm_projection:upsilon2}
\end{align}
Recall $e_t=\vec(\cE_t)$. Similar to the proof of Lemma \ref{lem:psi}, we can show, in an event $\Omega_2$ with probability at least $1-T^{-c_2}-d^{-c_2}$,
\begin{align}\label{Delta-bd-2}
&\| \Delta_{2,k} \|_{\rm S} \le \left\|\frac1T \sum_{t=1}^T e_t e_t^\top \right\|_2 \lesssim \sqrt{\frac{d\log (d)}{T}} + \frac{d\log(d) (\log T)^{\frac{2\vartheta+4}{\vartheta}} }{T}+1, \notag \\
&\| \Delta_{3,k} \|_{\rm S} \le \left\| \frac1T \sum_{t=1}^T w_i f_{it} e_t^\top \right\|_2 \lesssim \sqrt{\frac{d\log (d)}{T}} \sqrt{\Theta_{ii}} + \frac{\sqrt{d r} \log(d)(\log T)^{1+\frac{2}{\vartheta}+\frac{1}{\gamma_1} }  }{T} \sqrt{\Theta_{ii}}, \\
& \| \Delta_{4,k} \|_{\rm S} \le  \left\| \frac1T \sum_{t=1}^T ( w_j f_{jt}, \ j\neq i)^\top e_t^\top \right\|_2 \lesssim \sqrt{\frac{d\log (d)}{T}} \sqrt{\lam_1} + \frac{\sqrt{d r} \log(d)(\log T)^{1+\frac{2}{\vartheta}+\frac{1}{\gamma_1} }  }{T} \sqrt{\lam_1} . \notag
\end{align}
We claim that in certain events $\Omega_3$, with probability at least $1-T^{-c_3}-d^{-c_3}$, for any $1\le \ell\le K$, the following bounds hold,
\begin{align}\label{Delta-bd-2n}
&\| \Delta_{2,k}(  g_{i\ell},   g_{i\ell}, \ell\neq k) \|_{\rm 2} \le  C_1\sqrt{\frac{d_k \log(d)}{T}} + \frac{C_1d_k\log(d)(\log T)^{\frac{2\vartheta+4}{\vartheta}}}{T} + C_1, \notag \\
&\| \Delta_{3,k}(  g_{i\ell},  \ell\neq k) \|_{\rm 2} \le C_1 \sqrt{\frac{d_k\log(d)}{T} }  \sqrt{\Theta_{ii}}+ \frac{C_1 \sqrt{d_k r} \log(d)(\log T)^{1+\frac{2}{\vartheta}+\frac{1}{\gamma_1} } }{T}   \sqrt{\Theta_{ii}},\\
&\| \Delta_{4,k}(  g_{i\ell},  \ell\neq k) \|_{\rm 2} \le C_1 \sqrt{\frac{d_k\log(d)}{T} }  \sqrt{\lam_1} +  \frac{C_1 \sqrt{d_k r} \log(d)(\log T)^{1+\frac{2}{\vartheta}+\frac{1}{\gamma_1} }  }{T}   \sqrt{\lam_1}. \notag
\end{align}

Define
\begin{align} 
R_{k,i} &= \sqrt{\frac{d_k\log (d)}{T}} + \frac{d_k\log(d) (\log T)^{\frac{2\vartheta+4}{\vartheta}} }{T}+1 + \sqrt{\frac{\Theta_{ii}d_k \log(d)  }{T}} + \frac{\sqrt{\Theta_{ii}} \sqrt{d_k r} \log(d)(\log T)^{1+\frac{2}{\vartheta}+\frac{1}{\gamma_1} }  }{T},\\
R^* &= \max_i\max_k R_{k,i}/\Theta_{ii}  .
\end{align}
As $ g_{i\ell}$ is true and deterministic, it follows from \eqref{norm-bd}, \eqref{Delta-bd-11}, \eqref{Delta-bd-12}, \eqref{Delta-bd-13}, \eqref{Delta-bd-2}, \eqref{Delta-bd-2n}, in the event $\cap_{q=0}^3\Omega_{q}$, for some numeric constant $C_2>0$
\begin{align}\label{norm-bd-1}
&\| \Psi \times_{\ell\in [ 2K] \backslash\{k,K+k\} } \widehat  g_{i\ell}^{(m)\top} \|_{\rm 2} \notag \\
\le& C_2 R_{k,i}  + C_{1,K} R^{(0)}   \phi_{m,k}    + C_{1,K} \sqrt{\lam_1\Theta_{ii}}\prod_{\ell\neq k}\psi_{m,\ell} \left(\sqrt{\frac{r+\log T }{T}} + \frac{(r+\log T)^{1/\gamma}}{T}  \right) \notag \\
&\quad  + C_{1,K}\lambda_1\prod_{\ell\neq k}\psi_{m,\ell}^2\left(\sqrt{\frac{r+\log T }{T}} + \frac{(r+\log T)^{1/\gamma}}{T}  \right) + C_{1,K} \sqrt{\frac{d_k\log(d)}{T} }  \sqrt{\lam_1} \prod_{\ell\neq k}\psi_{m,\ell},
\end{align}
where $R^{(0)}=\phi^{(0)}$ is defined in \eqref{thm:initial:eq2} in Theorem \ref{thm:initial}.
Substituting \eqref{norm-bd-1} into \eqref{wedin72}, by the definition of $\phi_{m,k}$, we have, in the event $\cap_{q=0}^3\Omega_{q}$,
\begin{align}\label{bdd1:thm-projection}
&\| \widehat  a_{ik}^{(m+1)}\widehat  a_{ik}^{(m+1)\top} -  a_{ik}  a_{ik}^\top \|_{\rm 2} \notag\\
&\le \frac{4(1+\delta_{\max}) \lambda_1 \prod_{\ell\neq k}\psi_{m,\ell}^2 }{\alpha^{2K-2}\Theta_{ii} [\sqrt{(1-\delta_{\max})(1-1/(4r))}\alpha]^{2K-2}  } +    \frac{4C_2R_{k,i}   }{\alpha^{2K-2}\Theta_{ii} }
+ \frac{4C_{1,K} R^{(0)}   \phi_{m,k}}{\alpha^{2K-2}\Theta_{ii} }   \notag\\
&\quad + \frac{4C_{1,K} \sqrt{\lambda_1}}{\alpha^{2K-2}\Theta_{ii}} \prod_{\ell\neq k}\psi_{m,\ell}  \sqrt{\frac{d_k\log(d)}{T} }  \notag\\
&\quad + \frac{4C_{1,K}}{\alpha^{2K-2}}\sqrt{\lambda_1/\Theta_{ii}}\prod_{\ell\neq k}\psi_{m,\ell} \left(\sqrt{\frac{r+\log T }{T}} + \frac{(r+\log T)^{1/\gamma}}{T}  \right)     \notag\\
&\quad + \frac{4C_{1,K}}{\alpha^{2K-2}}(\lambda_1/\Theta_{ii})\prod_{\ell\neq k}\psi_{m,\ell}^2\left(\sqrt{\frac{r+\log T }{T}} + \frac{(r+\log T)^{1/\gamma}}{T}  \right)  \notag \\
&\le  C_{\alpha,K} R_{k,i} /\Theta_{ii} + C_{\alpha,K} (\sqrt{r} \psi_0) \psi_{m,k}   + C_{\alpha,K}\sqrt{\lambda_1/\lam_r} \prod_{\ell\neq k}\psi_{m,\ell} R_{k,i} /\Theta_{ii} \notag\\
&\quad + C_{\alpha,K}(\lambda_1/\lam_r)\prod_{\ell\neq k}\psi_{m,\ell}^2 + C_{\alpha,K}\sqrt{\lambda_1/\lam_r}\prod_{\ell\neq k}\psi_{m,\ell} \left(\sqrt{\frac{r+\log T }{T}} + \frac{(r+\log T)^{1/\gamma}}{T}  \right) \notag\\
&\le C_{\alpha,K} R^* +\rho \psi_m,
\end{align}
where the last inequality comes from condition \eqref{thm-projection:eq1a} with $\rho<1$.
As $R^*\lesssim 1$ and $\psi_0\lesssim 1$, we have $R^*\lesssim \psi^{\ideal}$. 
It follows that, after $O(\log(\psi_0/\psi^{\ideal} ))$ iterations,
\begin{align}\label{bdd2:thm-projection}
\psi_{m,i,k} \lesssim \psi^{\ideal}.
\end{align}

In the end, we divide the rest of the proof into 3 steps to prove \eqref{Delta-bd-2n}.

\smallskip
\noindent\underline{\bf Step 1.}
We prove \eqref{Delta-bd-2n} for the $\|\Delta_{2,k}(  g_{i\ell},   g_{i\ell}, \ell\neq k)\|_{\rm 2}$.
Let $P_{g_{ik}}=g_{iK}^\top \odot \cdots \odot g_{i,k+1}^\top \odot I_{d_k} \odot g_{i,k-1}^\top \odot \cdots \odot g_{i1}^\top\in\R^{d_k\times d}$, where $\odot$ represents Kronecker product. Also let $e_{t,ik}=\cE_t\times_{\ell\neq k}^K g_{i\ell}$. Then $e_{t,ik}=P_{g_{ik}} H \xi_t \in \R^{d_k}$.

By Assumption \ref{asmp:error} and Lemma A.1 in \cite{shu2019estimation}, we have
\begin{align*}
&\PP\left(\|e_{t,ik}\|_2^2 -\EE \|e_{t,ik}\|_2^2 \ge x \right) = \PP\left( \xi_t^\top H^\top P_{g_{ik}}^\top P_{g_{ik}} H \xi_t -\EE \xi_t^\top H^\top P_{g_{ik}}^\top P_{g_{ik}} H \xi_t \ge x \right) \\
&\le 4\exp\left(-C'\left(\frac{x}{\|H^\top P_{g_{ik}}^\top P_{g_{ik}} H\|_{\rm F}} \right)^{\frac{1}{1+2/\vartheta}} \right) 
\end{align*}
Note that $\EE\|e_{t,ik}\|_2^2=\EE\xi_t^\top H^\top P_{g_{ik}}^\top P_{g_{ik}} H \xi_t=\tr(HH^\top P_{g_{ik}}^\top P_{g_{ik}})\asymp d_k$, and $\|H^\top P_{g_{ik}}^\top P_{g_{ik}} H\|_{\rm F}^2=\|H H^\top P_{g_{ik}}^\top P_{g_{ik}} \|_{\rm F}^2\asymp d_k$.
Choosing $x\asymp d_k(\log T)^{(2\vartheta+4)/\vartheta}$, we have
\begin{align*}
\PP\left( \|e_{t,ik}\|_2 \ge C\sqrt{d_k} (\log T)^{(\vartheta+2)/\vartheta}\right) \le 4\exp\left(-C'd_k^{\frac{\vartheta}{2\vartheta+4}} \log T \right).
\end{align*}
Let $N:=\|e_{t,ik} e_{t,ik}^\top \mathbf{1}_{\{\|e_{t,ik}\|_2\le C\sqrt{d_k} (\log T)^{(\vartheta+2)/\vartheta}\} } \|_{ 2}$ and $\sigma_0^2:= \| \sum_{t=1}^T \EE (e_{t,ik} e_{t,ik}^\top \mathbf{1}_{\{\|e_{t,ik}\|_2\le C\sqrt{d_k} (\log T)^{(\vartheta+2)/\vartheta}\} } )^2\|_{ 2}$. Then, by Assumption \ref{asmp:error}, $N\le C^2 d_k(\log T)^{(2\vartheta+4)/\vartheta}$ and $\sigma_0^2\le C_0 Td_k$.
By matrix Bernstein inequality (see, e.g., Theorem 5.4.1 of \cite{vershynin2018high}),
\begin{align*}
&\PP\left(\left\|\sum_{t=1}^T \left[e_{t,ik} e_{t,ik}^\top \mathbf{1}_{\{\|e_{t,ik}\|_2\le C\sqrt{d_k} (\log T)^{(\vartheta+2)/\vartheta}\} } -\EE e_{t,ik} e_{t,ik}^\top \mathbf{1}_{\{\|e_{t,ik}\|_2\le C\sqrt{d_k} (\log T)^{(\vartheta+2)/\vartheta}\} } \right] \right\|_2 \ge x  \right)    \\
\le& 2d_k \exp\left( -\frac{x^2/2}{\sigma_0^2+N x/3}  \right).
\end{align*}
Choosing $x\asymp \sqrt{Td_k\log (d)}+ d_k \log(d)(\log T)^{(2\vartheta+4)/\vartheta}$, with probability at least $1-d^{-c_1}$,
\begin{align}\label{eq:delta2-2n}
&\left\|\frac1T \sum_{t=1}^T e_{t,ik} e_{t,ik}^\top \mathbf{1}_{\{\|e_{t,ik}\|_2\le C\sqrt{d_k} (\log T)^{(\vartheta+2)/\vartheta}\} } -\EE e_{t,ik} e_{t,ik}^\top \mathbf{1}_{\{\|e_{t,ik}\|_2\le C\sqrt{d_k} (\log T)^{(\vartheta+2)/\vartheta}\} } \right\|_2 \notag\\
\le& C_1 \sqrt{\frac{d_k\log(d)}{T}} + C_1 \cdot\frac{d_k\log(d)(\log T)^{\frac{2\vartheta+4}{\vartheta}} }{T}    .
\end{align}

Define $M:=\{1\le t\le T: \|e_{t,ik}\|_2\ge C\sqrt{d_k} (\log T)^{(\vartheta+2)/\vartheta}\}$. Since $\mathbf{1}_{\{\|e_{t,ik}\|_2\ge C\sqrt{d_k} (\log T)^{(\vartheta+2)/\vartheta}\} }$ are independent Bernoulli random variable, we have
\begin{align*}
\EE|M|=T\PP\left(\|e_{t,ik}\|_2\ge C\sqrt{d_k} (\log T)^{1+\frac{2}{\vartheta}} \right) \le 4T\exp\left(-C' d_k^{\frac{\vartheta}{2\vartheta+4}} \log T \right) \le T^{-c_2}.
\end{align*}
By Chernoff bound for Bernoulli random variables,
\begin{align*}
\PP(|M|\ge C)\le \exp\left(-T^{c_2} \right).
\end{align*}
It follows that
\begin{align*}
\PP\left( \left\|\sum_{t=1}^T e_{t,ik} e_{t,ik}^\top \mathbf{1}_{\{\|e_{t,ik}\|_2\ge C\sqrt{d_k} (\log T)^{(\vartheta+2)/\vartheta}\} } \right\|_2 \ge x \right) &\le \PP\left( |M| \max_t \| e_{t,ik}\|_2^2 \ge x \right)   \\
&\le \PP(|M|\ge C) +\PP\left(|M|<C,  |M|\max_t \| e_{t,ik}\|_2^2 \ge x \right) \\
&\le \exp\left(-T^{c_2} \right) + \PP\left(\max_t \| e_{t,ik}\|_2^2 \ge x/C \right) .
\end{align*}
Choosing $x\asymp d_k(\log T)^{(2\vartheta+4)/\vartheta}$, we have, with probability at least $1-\exp(-T^{c_2} )-T^{-c_2}$,
\begin{align}\label{eq:delta2-4n}
 \left\|\frac1T\sum_{t=1}^T e_{t,ik} e_{t,ik}^\top \mathbf{1}_{\{\|e_{t,ik}\|_2\ge C\sqrt{d_k} (\log T)^{(\vartheta+2)/\vartheta}\} } \right\|_2 \le C_2 \cdot\frac{d_k (\log T)^{\frac{2\vartheta+4}{\vartheta}} }{T}.
\end{align}
Similarly,
\begin{align}\label{eq:delta2-5n}
&\PP\left( \left\|\EE e_{t,ik} e_{t,ik}^\top \mathbf{1}_{\{\|e_{t,ik}\|_2\ge C\sqrt{d_k} (\log T)^{(\vartheta+2)/\vartheta}\} } \right\|_2>0\right)=\PP \left(\| e_{t,ik}\|_2 \ge C\sqrt{d_k} (\log T)^{\frac{\vartheta+2}{\vartheta}} \right) \notag\\
\le& 4 \exp\left(-C' d_k^{\frac{\vartheta}{2\vartheta+4}} \log T \right) \le T^{-c_3}.
\end{align}
Combing \eqref{eq:delta2-2n}, \eqref{eq:delta2-4n}, \eqref{eq:delta2-5n}, in an event with probability at least $1-T^{-c_4}-d^{-c_1}$,
\begin{align*}
&\|\Delta_{2,k}(  g_{i\ell},   g_{i\ell}, \ell\neq k) \|_2 \le \left\|\frac{1}{T} \sum_{t=1}^T e_{t,ik} e_{t,ik}^\top - \EE e_{t,ik} e_{t,ik}^\top \right\|_2 + \| \EE e_{t,ik} e_{t,ik}^\top \|_2   \\
&\le \left\|\frac1T \sum_{t=1}^T e_{t,ik} e_{t,ik}^\top \mathbf{1}_{\{\|e_{t,ik}\|_2\le C\sqrt{d_k} (\log T)^{(\vartheta+2)/\vartheta}\} } -\EE e_{t,ik} e_{t,ik}^\top \mathbf{1}_{\{\|e_{t,ik}\|_2\le C\sqrt{d_k} (\log T)^{(\vartheta+2)/\vartheta}\} } \right\|_2  + \| P_{g_{ik}} H H^\top P_{g_{ik}}^\top \|_2 \\
&\quad + \left\|\frac1T\sum_{t=1}^T e_{t,ik} e_{t,ik}^\top \mathbf{1}_{\{\|e_{t,ik}\|_2\ge C\sqrt{d_k} (\log T)^{(\vartheta+2)/\vartheta}\} } \right\|_2 +  \left\|\EE e_{t,ik} e_{t,ik}^\top \mathbf{1}_{\{\|e_{t,ik}\|_2\ge C\sqrt{d_k} (\log T)^{(\vartheta+2)/\vartheta}\} } \right\|_2  \\
&\le C_2 \sqrt{\frac{d_k\log(d)}{T}} + C_2 \cdot\frac{d_k\log(d)(\log T)^{\frac{2\vartheta+4}{\vartheta}}  }{T} +C_2 .
\end{align*}

\smallskip
\noindent\underline{\bf Step 2.}
Now we prove \eqref{Delta-bd-2n} for $\|\Delta_{3,k}(  g_{i\ell},  \ell\neq k)\|_{\rm 2}$.
Let
\begin{align*}
& N_1:=\left\| w_i f_{it} e_{t,ik}^\top \mathbf{1}_{\{\|e_{t,ik}\|_2\le C\sqrt{d_k} (\log T)^{(\vartheta+2)/\vartheta}\} } \right\|_{ 2}, \\
& \sigma_1^2:= \max\left\{\left\| \sum_{t=1}^T \overline\EE w_i^2 f_{it}^2  e_{t,ik}^\top e_{t,ik}   \mathbf{1}_{\{\|e_{t,ik}\|_2\le C\sqrt{d_k} (\log T)^{(\vartheta+2)/\vartheta}\} } \right\|_{ 2},
\left\| \sum_{t=1}^T \overline\EE w_i^2 f_{it}^2 e_{t,ik}  e_{t,ik}^\top \mathbf{1}_{\{\|e_{t,ik}\|_2\le C\sqrt{d_k} (\log T)^{(\vartheta+2)/\vartheta}\} } \right\|_{ 2} \right\}.
\end{align*}
It is easy to show
\begin{align*}
& N_1\le C\sqrt{d_k}(\log T)^{1+\frac{2}{\vartheta}}\left\| w_i f_{it} a_i \right\|_{ 2}, \\
& \sigma_1^2 \le C_3 Td_k \max\left\{\left\| \frac1T\sum_{t=1}^T w_i^2 f_{it}^2  \right\|_{ 2},
\left\| \frac{1}{Td_k}\sum_{t=1}^T w_i^2 f_{it}^2   \right\|_{ 2} \right\}:=\sigma_2^2.
\end{align*}
By matrix Bernstein inequality,
\begin{align*}
&\overline\PP\left( \left\| \sum_{t=1}^T \left[ w_i f_{i,t} e_{t,ik}^\top \mathbf{1}_{\{\|e_{t,ik}\|_2\le C\sqrt{d_k} (\log T)^{(\vartheta+2)/\vartheta}\} } - \overline\EE w_i f_{i,t} e_{t,ik}^\top \mathbf{1}_{\{\|e_{t,ik}\|_2\le C\sqrt{d_k} (\log T)^{(\vartheta+2)/\vartheta}\} } \right] \right\|_2 \ge x \right) \\
\le& 2d_k \exp\left( -\frac{x^2/2}{\sigma_1^2+N_1 x/3} \right).
\end{align*}
Choosing $x\asymp \sqrt{d_k} \log(d)(\log T)^{(\vartheta+2)/\vartheta}\| w_i f_{it} \|_{ 2} +\sqrt{\log(d)}\sigma_2$, with probability at least $1-d^{-c_4}$,
\begin{align*}
&\left\| \frac1T\sum_{t=1}^T  w_i f_{i,t} e_{t,ik}^\top \mathbf{1}_{\{\|e_{t,ik}\|_2\le C\sqrt{d_k} (\log T)^{(\vartheta+2)/\vartheta}\} } - \overline\EE w_i f_{i,t} e_{t,ik}^\top \mathbf{1}_{\{\|e_{t,ik}\|_2\le C\sqrt{d_k} (\log T)^{(\vartheta+2)/\vartheta}\} }  \right\|_2 \\\le& C_4\frac{\sqrt{\log(d)}\sigma_2 + \sqrt{d_k} \log(d)(\log T)^{(\vartheta+2)/\vartheta}\| w_i f_{it} a_i \|_{ 2}}{T}.
\end{align*}
By Assumption \ref{asmp:eigenvalue}, with probability at least $1-T^{-c_5}$,
\begin{equation*}
\left\| w_i f_{it} \right\|_{ 2}  \lesssim \sqrt{r} (\log(T))^{1/\gamma_1} \sqrt{\Theta_{ii}} . 
\end{equation*}
Similarly, in the event $\Omega_1$, $\sigma_2^2\lesssim Td_k \Theta_{ii}$. Then, with probability at least $1-T^{-c_1}/2-T^{-c_5}-d^{-c_4}$,
\begin{align*}
&\left\| \frac1T\sum_{t=1}^T w_i f_{i,t} e_{t,ik}^\top \mathbf{1}_{\{\|e_{t,ik}\|_2\le C\sqrt{d_k} (\log T)^{(\vartheta+2)/\vartheta}\} } - \EE w_i f_{i,t} e_{t,ik}^\top \mathbf{1}_{\{\|e_{t,ik}\|_2\le C\sqrt{d_k} (\log T)^{(\vartheta+2)/\vartheta}\} }  \right\|_2 \\
\le& C_5\sqrt{\frac{d_k\log(d)}{T}} \sqrt{\Theta_{ii}}   + C_5 \cdot \frac{\sqrt{d_k r} \log(d)(\log T)^{1+\frac{2}{\vartheta}+\frac{1}{\gamma_1} }\sqrt{\Theta_{ii}}}{T}.
\end{align*}
Similar to \eqref{eq:delta2-4n},
\begin{align*}
&\PP\left( \left\| \sum_{t=1}^T w_i f_{i,t} e_{t,ik}^\top \mathbf{1}_{\{\|e_{t,ik}\|_2\ge C\sqrt{d_k} (\log T)^{(\vartheta+2)/\vartheta}\} }   \right\|_2 \ge x \right)  \\
\le& \PP(|M|>C)+\PP\left(|M|<C, |M|\max_t \left\| w_i f_{it} \right\|_{ 2} \cdot \|e_{t,ik}\|_2 \ge x  \right).
\end{align*}
Choosing $x\asymp \sqrt{d_k r} (\log T)^{1+\frac{2}{\vartheta}+\frac{1}{\gamma_1} } \sqrt{\Theta_{ii}}$, we have with probability at least $1-\exp(-T^{c_2})-T^{-c_2}-T^{-c_5}$,
\begin{align*}
\left\| \frac1T \sum_{t=1}^T w_i f_{i,t} e_{t,ik}^\top \mathbf{1}_{\{\|e_{t,ik}\|_2\ge C\sqrt{d_k} (\log T)^{(\vartheta+2)/\vartheta}\} }   \right\|_2 \le  C_6 \cdot \frac{\sqrt{d_k r} (\log T)^{1+\frac{2}{\vartheta}+\frac{1}{\gamma_1} }\sqrt{\Theta_{ii}}}{T}  .
\end{align*}
Thus, in an event with probability $1-T^{-c_1}/2-T^{-c_6}-d^{-c_4}$,
\begin{align*}
\|\Delta_3(  g_{i\ell},  \ell\neq k)\|_2 \le&  \left\| \frac1T\sum_{t=1}^T w_i f_{i,t} e_{t,ik}^\top \mathbf{1}_{\{\|e_{t,ik}\|_2\le C\sqrt{d_k} (\log T)^{(\vartheta+2)/\vartheta}\} } - \EE w_i f_{i,t} e_{t,ik}^\top \mathbf{1}_{\{\|e_{t,ik}\|_2\le C\sqrt{d_k} (\log T)^{(\vartheta+2)/\vartheta}\} }  \right\|_2 \\
&+ \left\| \frac1T \sum_{t=1}^T w_i  f_{i,t} e_{t,ik}^\top \mathbf{1}_{\{\|e_{t,ik}\|_2\ge C\sqrt{d_k} (\log T)^{(\vartheta+2)/\vartheta}\} }   \right\|_2
+\left\| \EE w_i f_{i,t} e_{t,ik}^\top \mathbf{1}_{\{\|e_{t,ik}\|_2\ge C\sqrt{d_k} (\log T)^{(\vartheta+2)/\vartheta}\} }   \right\|_2 \\
\le & C_7\sqrt{\frac{d_k\log(d)}{T}} \sqrt{\Theta_{ii}}   + C_7 \cdot \frac{\sqrt{d_k r} \log(d)(\log T)^{1+\frac{2}{\vartheta}+\frac{1}{\gamma_1} }\sqrt{\Theta_{ii}}}{T} .
\end{align*}

\smallskip
\noindent\underline{\bf Step 3.}
Inequality \eqref{Delta-bd-2n} for $\|\Delta_{4,k}(  g_{i\ell},  \ell\neq k)\|_{\rm 2}$ follow from the same argument as the above step.

\end{proof}

\begin{proof}[\bf Proof of Theorem \ref{thm:factors}]
Let
\begin{align*}
\psi_0 &= \max_{1\le i\le r}\max_{1\le k\le K} \| \widehat a_{ik}^{\rm rcpca} \widehat  a_{ik}^{{\rm rcpca}\top} -  a_{ik}  a_{ik}^\top \|_{2}  , \\
\psi &=\max_{1\le i\le r}\max_{1\le k\le K} \|\widehat a_{ik}^{\iso} \widehat a_{ik}^{\iso\top} - a_{ik} a_{ik}^\top \|_{2}  .
\end{align*}
As $1 \lesssim \lambda_r$, Theorem \ref{thm:projection} implies that, in an event $\Omega$ with probability at least $1-T^{-C}- d^{-C}$,
\begin{align*}
\psi\le C_1 \left(  \sqrt{\frac{d_{\max}\log d}{\lambda_r T}} + \frac{1}{\lambda_r} \right) . 
\end{align*}
Note that
\begin{align*}
\widehat w_i \widehat f_{it} - w_i f_{it} &= \sum_{j=1}^r w_j f_{jt} \otimes_{k=1}^K a_{jk} \times_{k=1}^K \widehat b_{ik}^{\iso\top}  + \cE_t \times_{k=1}^K \widehat b_{ik}^{\iso\top} - w_i f_{it}\\
&= w_i f_{it} \left(\prod_{k=1}^K a_{ik}^\top \widehat b_{ik}^{\iso} -1 \right) + \sum_{j\neq i} w_j f_{jt}  \prod_{k=1}^K a_{ik}^\top \widehat b_{jk}^{\iso} + \cE_t \times_{k=1}^K \widehat b_{ik}^{\iso\top} \\
&:= {\rm I}_1 + {\rm I}_2 + {\rm I}_3.
\end{align*}
By \eqref{g-bd} in the proof of Theorem \ref{thm:projection}, $\max_{1\le i,j\le r}\max_{1\le k\le K} | a_{ik}^\top \widehat  b_{jk}^{\iso} - \mathbf{1}_{\{i=j\}}| \lesssim \psi$ in the event $\Omega$. By \eqref{Delta-bd-2n}, it follows that ${\rm I}_1\le C_1(w_i\psi)$, ${\rm I}_2\le C_1(\sqrt{\lambda_1} r \psi^K)$ and ${\rm I}_3\le C_1$, in the same event $\Omega$. Since $d_{\max} r\lesssim d$, condition \eqref{thm-projection:eq1a} leads to $(\lambda_1/\lambda_r)^{1/2} r \psi^{K-1} \lesssim (\lambda_1/\lambda_r)^{1/2} \psi_0^{K-1} \lesssim 1$. Then ${\rm I}_2\le C_2(\lambda_r^{1/2}\psi)$ in the event $\Omega$. Hence, in the event $\Omega$, $w_i^{-1}|\widehat w_i \widehat f_{it} - w_i f_{it}|\le C_0(\psi+\sigma w_i^{-1})$, which completes the proof.
\end{proof}

\begin{proof}[\bf Proof of Theorem \ref{thm:clt}]
By the definition of the iterative algorithm, after convergence to a stationary point, $\widehat a_{ik}$ is the top eigenvector of the matrix
\begin{align}
\widehat \Sigma_{ik}  &= \widehat\Sigma\times_{\ell\in [ 2K] \backslash\{k,K+k\} } \widehat g_{i\ell}^{\top} \notag \\
=& \widetilde\lambda_{i,i}  a_{ik} a_{ik}^\top  +   \frac{1}{T} \sum_{t=1}^T w_i f_{it}\left( \prod_{\ell\neq k}^K a_{i\ell}^\top \widehat g_{i\ell}^{(m)\top}  \right) a_{ik} \otimes\left( \cE_t \times_{\ell\in [ K] \backslash\{k\} } \widehat g_{i\ell}^{\top} \right) \notag \\
&+ \frac{1}{T} \sum_{t=1}^T w_i f_{it} \left( \prod_{\ell\neq k}^K a_{i\ell}^\top \widehat g_{i\ell}^{(m)\top}  \right) \left( \cE_{t} \times_{\ell\in [ K] \backslash\{k\} } \widehat g_{i\ell}^{\top}\right) \otimes a_{ik} +
\frac{1}{T} \sum_{t=1}^T \cE_{t} \otimes \cE_t \times_{\ell\in [ 2K] \backslash\{k,K+k\} } \widehat g_{i\ell}^{\top} \notag \\
&+ \sum_{j\neq i}^r   \widetilde\lambda_{j,i}  a_{jk} a_{jk}^\top   +   \sum_{j_1\neq j_2}^r \frac{1}{T} \sum_{t=1}^T w_{j_1} w_{j_2} f_{j_1t}f_{j_2t}  \otimes_{k=1}^{K} a_{j_1k} \otimes_{k=K+1}^{2K} a_{j_2k} \times_{\ell\in [ 2K] \backslash\{k,K+k\} } \widehat g_{i\ell}^{\top}  \notag\\
&+   \frac{1}{T} \sum_{t=1}^T \sum_{j\neq i}^r w_j f_{jt} \otimes_{k=1}^{K} a_{jk} \otimes\cE_t \times_{\ell\in [ 2K] \backslash\{k,K+k\} } \widehat g_{i\ell}^{\top}  +   \frac{1}{T} \sum_{t=1}^T \sum_{j\neq i}^r w_j f_{jt}  \cE_{t} \otimes_{k=K+1}^{2K} a_{jk} \times_{\ell\in [ 2K] \backslash\{k,K+k\} } \widehat g_{i\ell}^{\top}   \notag\\
&:= \widetilde\lambda_{i,i}  a_{ik} a_{ik}^\top +\Psi_1 +\Psi_2 +\Psi_3 +\Psi_ 4+\Psi_5 +\Psi_6  +\Psi_7  \notag \\
&:= \widetilde\lambda_{i,i}  a_{ik} a_{ik}^\top + \Psi,
\end{align}
where $\widetilde \lambda_{j,i}$ is defined in \eqref{eq:lambdaji}, $\widehat g_{i\ell}=\widehat b_{i\ell}/\|\widehat b_{i\ell}\|_2$, and $\Psi=\sum_{j=1}^7 \Psi_j$.

Let $P_{ a_{ik},\perp}=I_{d_k}- a_{ik} a_{ik}^\top=  a_{ik, \perp}  a_{ik,\perp}^\top $. By Theorem \ref{thm:projection}, the final estimates of $\widehat a_{ik}$ satisfies, in an event $\Omega$ with probability at least $1-T^{-C}-d^{-C}$,
\begin{align}\label{thmclt:eq1}
\|\widehat a_{ik}\widehat a_{ik}^{\top}  - a_{ik} a_{ik}^\top \|_{2} &\le C_{0} \psi^{\ideal},
\end{align}
where $\psi^{\ideal}$ is defined in \eqref{eq:final_rate}.
Using resolvent based series expansion of projection matrices (e.g., Theorem 1 in \cite{xia2021normal}), we have the following expansion,
\begin{align}
\widehat a_{ik}\widehat a_{ik}^{\top}  - a_{ik} a_{ik}^\top =& \frac{1}{\widetilde \lambda_{i,i}} P_{ a_{ik},\perp} \Psi P_{ a_{ik}} + \frac{1}{\widetilde \lambda_{i,i}}   P_{ a_{ik}} \Psi P_{ a_{ik},\perp} \notag \\
&+ \frac{1}{\widetilde \lambda_{i,i}^2} \left( P_{ a_{ik}} \Psi P_{ a_{ik},\perp} \Psi P_{ a_{ik},\perp} + P_{ a_{ik},\perp} \Psi P_{ a_{ik}} \Psi P_{ a_{ik},\perp}  + P_{ a_{ik},\perp} \Psi P_{ a_{ik},\perp} \Psi P_{ a_{ik}}  \right) \notag \\
&- \frac{1}{\widetilde \lambda_{i,i}^2} \left( P_{ a_{ik},\perp} \Psi P_{ a_{ik}} \Psi P_{ a_{ik}} + P_{ a_{ik}} \Psi P_{ a_{ik},\perp} \Psi P_{ a_{ik}}  + P_{ a_{ik}} \Psi P_{ a_{ik}} \Psi P_{ a_{ik},\perp}  \right)  \notag \\
&+\cR_3(\Psi).
\end{align}
Moreover, $\|\cR_3(\Psi)\|_2\le C_1 \|\Psi\|_2^3/\widetilde \lambda_{i,i}^3 \le C_2 (\psi^{\ideal})^3$ under the event $\Omega$.

\noindent\underline{\bf Case (i)}. Let $u= a_{ik}$. Then
\begin{align*}
u^\top \left( \widehat a_{ik}\widehat a_{ik}^{\top}  - a_{ik} a_{ik}^\top \right) u &=   \left( \widehat a_{ik}^{\top}  a_{ik} \right)^2 -1  = - \frac{1}{\widetilde \lambda_{i,i}^2} a_{ik}^\top \Psi P_{ a_{ik},\perp} \Psi  a_{ik} +  a_{ik}^{\top}  \cR_3(\Psi) a_{ik} \\
&= - \frac{1}{\widetilde \lambda_{i,i}^2} \left( a_{ik,\perp}^\top \Psi a_{ik}  \right)^\top \left( a_{ik,\perp}^\top \Psi a_{ik}  \right)  +  a_{ik}^{\top}  \cR_3(\Psi) a_{ik}.
\end{align*}
By Theorem \ref{thm:projection} and \eqref{thmclt:eq1}, in the event $\Omega$, $\| \widetilde \lambda_{i,i}^{-1} a_{ik,\perp}^\top \Psi a_{ik} \|_2 \le C_0 \psi^{\ideal}$. It follows that, in the event $\Omega$,
\begin{align}\label{thmclt:eq2}
\left( \widehat a_{ik}^{\top}  a_{ik} \right)^2 -1 \le C_3 (\psi^{\ideal})^2 .
\end{align}
From the condition \eqref{thm-projection:eq1a} and \eqref{bdd1:thm-projection}, we have $\psi^{\ideal} \lesssim \psi_0^2$, where $\psi_0$ is the error bound for the initialization.
By the proofs of Theorem \ref{thm:projection}, i.e. the derivation of \eqref{bdd1:thm-projection}, we can show, in the event $\Omega$,
\begin{align}
&\left\| \frac{1}{\widetilde \lambda_{i,i}} \left( \Psi_4+\Psi_5+\Psi_6+\Psi_7 \right)  \right\|_2 \le C_4 (\psi^{\ideal})^2 , \label{clt:eq1}\\
&\left\| \frac{1}{\widetilde \lambda_{i,i}} \Psi_3   \right\|_2 \le C_4 (\sqrt{r} \psi_0) \psi^{\ideal} + C_4 \frac{1}{\Theta_{ii}}\left( \frac{d_k\log(d)}{T} + \sqrt{\frac{d_k \log(d)}{T}} + 1\right) , \label{clt:eq2}  \\
&\left\| \frac{1}{\widetilde \lambda_{i,i}}  \Psi_1 - \frac{1}{\Theta_{ii}\prod_{\ell\neq k} ( a_{i\ell}^\top g_{il}) T} \sum_{t=1}^T w_i f_{it} a_{ik} \otimes\left( \cE_t \times_{\ell\in [ K] \backslash\{k\} } g_{i\ell}^{\top} \right)  \right\|_2 \le C_4 (\psi^{\ideal})^2  + C_4 (\sqrt{r} \psi_0) \psi^{\ideal} ,  \label{clt:eq3} \\
&\left\| \frac{1}{\widetilde \lambda_{i,i}}   \Psi_2 - \frac{1}{\Theta_{ii} \prod_{\ell\neq k} ( a_{i\ell}^\top g_{il})  T} \sum_{t=1}^T w_i f_{it} \left( \cE_t \times_{\ell\in [ K] \backslash\{k\} } g_{i\ell}^{\top} \right) \otimes a_{ik}    \right\|_2 \le C_4 (\psi^{\ideal})^2  + C_4 (\sqrt{r} \psi_0) \psi^{\ideal} .  \label{clt:eq4}
\end{align}
As $a_{ik,\perp}^\top\Psi_1 a_{ik}=0$, in the event $\Omega$,
\begin{align}\label{thmclt:eq3}
&\left\| \frac{1}{\widetilde \lambda_{i,i}} a_{ik,\perp} \Psi a_{ik} -  \frac{1}{\Theta_{ii}\prod_{\ell\neq k} ( a_{i\ell}^\top g_{il})} a_{ik,\perp}^\top \left[\frac{1}{ T} \sum_{t=1}^T w_i f_{it} \left( \cE_t \times_{\ell\in [ K] \backslash\{k\} } g_{i\ell}^{\top} \right) \otimes a_{ik} \right] a_{ik}   \right\|_2 \notag\\
\le& C_4 (\psi^{\ideal})^2 + C_4 (\sqrt{r} \psi_0) \psi^{\ideal} + C_4 \frac{1}{\Theta_{ii}}\left( \frac{d_k\log(d)}{T} + \sqrt{\frac{d_k \log(d)}{T}} + 1\right) .
\end{align}

\noindent\underline{\bf Case (ii)}.  Let $u\perp a_{ik}$. Define $v= u P_{ a_{ik},\perp} $. Then
\begin{align*}
&u^\top \left( \widehat a_{ik}\widehat a_{ik}^{\top}  - a_{ik} a_{ik}^\top \right) a_{ik} = (u^\top \widehat a_{ik}) (\widehat a_{ik}^\top a_{ik} ) \\
=& \frac{1}{\widetilde \lambda_{i,i}} u^\top P_{ a_{ik},\perp} \Psi a_{ik} +
 \frac{1}{\widetilde \lambda_{i,i}^2} \left( u^\top P_{ a_{ik},\perp} \Psi P_{ a_{ik},\perp} \Psi a_{ik}  -  u^\top P_{ a_{ik},\perp} \Psi P_{ a_{ik}} \Psi a_{ik}   \right)  +\cR_3(\Psi)  \\
=& \frac{1}{\widetilde \lambda_{i,i}} v^\top \Psi a_{ik} +
 \frac{1}{\widetilde \lambda_{i,i}^2} \left( v^\top  \Psi a_{ik,\perp} a_{ik,\perp}^\top \Psi a_{ik}  -   v^\top \Psi a_{ik}a_{ik}^\top \Psi a_{ik}   \right)  +\cR_3(\Psi).
\end{align*}
Note that, in the event $\Omega$, $\| \widetilde \lambda_{i,i}^{-1} \Psi \|_2 \le C_0 \psi^{\ideal}$. By \eqref{clt:eq1}, \eqref{clt:eq2}, \eqref{clt:eq3}, \eqref{clt:eq4}, we have,
\begin{align}\label{thmclt:eq4}
&\sup_{u \perp a_{ik}} \left| u^\top \left( \widehat a_{ik}\widehat a_{ik}^{\top}  - a_{ik} a_{ik}^\top \right) a_{ik} - \frac{1}{\Theta_{ii}\prod_{\ell\neq k} ( a_{i\ell}^\top g_{il})} u^\top P_{ a_{ik},\perp} \left[\frac{1}{ T} \sum_{t=1}^T w_i f_{it} \left( \cE_t \times_{\ell\in [ K] \backslash\{k\} } g_{i\ell}^{\top} \right) \otimes a_{ik} \right] a_{ik}  \right|   \notag\\
&\le C_4 (\psi^{\ideal})^2 + C_4 (\sqrt{r} \psi_0) \psi^{\ideal} + C_4 \frac{1}{\Theta_{ii}}\left( \frac{d_k\log(d)}{T} + \sqrt{\frac{d_k \log(d)}{T}} + 1\right) .
\end{align}

Now, let's move to the proof of Theorem \ref{thm:clt}. Without loss of generality, assume $\widehat a_{ik}^\top a_{ik} >0$. For $u$ such that $\lim\inf_{d_k\to\infty} \|P_{ a_{ik},\perp} u\|_2>0$, we have
\begin{align*}
u^\top (\widehat a_{ik} - a_{ik})  = u^\top P_{ a_{ik},\perp} \widehat a_{ik} + (u^\top a_{ik}) ( a_{ik}^\top \widehat a_{ik} -1).
\end{align*}
By \eqref{thmclt:eq2},
\begin{align*}
\left| (u^\top a_{ik}) ( a_{ik}^\top \widehat a_{ik} -1)  \right|   \le C \| u^\top a_{ik} \|_2 (\psi^{\ideal})^2.
\end{align*}
In addition,
\begin{align*}
u^\top P_{ a_{ik},\perp} \widehat a_{ik} = u^\top P_{ a_{ik},\perp} \widehat a_{ik} (\widehat a_{ik}^\top a_{ik} ) +    u^\top P_{ a_{ik},\perp} \widehat a_{ik} (1-\widehat a_{ik}^\top a_{ik} ).
\end{align*}
By \eqref{thmclt:eq4} and \eqref{thmclt:eq1},
\begin{align*}
&\sup_{u } \left| u^\top P_{ a_{ik},\perp} \widehat a_{ik} - \frac{1}{\Theta_{ii}\prod_{\ell\neq k} ( a_{i\ell}^\top g_{il})} u^\top P_{ a_{ik},\perp} \left[\frac{1}{ T} \sum_{t=1}^T w_i f_{it} \left( \cE_t \times_{\ell\in [ K] \backslash\{k\} } g_{i\ell}^{\top} \right) \otimes a_{ik} \right] a_{ik}  \right|   \notag\\
\le& C \|P_{ a_{ik},\perp} u\|_2 \left[ (\psi^{\ideal})^2 + (\sqrt{r} \psi_0) \psi^{\ideal} + \frac{1}{\Theta_{ii}}\left( \frac{d_k\log(d)}{T} + \sqrt{\frac{d_k \log(d)}{T}} + 1\right) \right].
\end{align*}
Combing the bound above, we have
\begin{align*}
&\sup_{u } \left| u^\top (\widehat a_{ik} - a_{ik})  - \frac{1}{\Theta_{ii}\prod_{\ell\neq k} ( a_{i\ell}^\top g_{il})} u^\top P_{ a_{ik},\perp} \left[\frac{1}{ T} \sum_{t=1}^T w_i f_{it} \left( \cE_t \times_{\ell\in [ K] \backslash\{k\} } g_{i\ell}^{\top} \right)  \right]  \right|   \notag\\
\le& C \|P_{ a_{ik},\perp} u\|_2 \left[ (\psi^{\ideal})^2 + (\sqrt{r} \psi_0) \psi^{\ideal} + \frac{1}{\Theta_{ii}}\left( \frac{d_k\log(d)}{T} + \sqrt{\frac{d_k \log(d)}{T}} + 1\right) \right] + C \| u^\top a_{ik} \|_2 (\psi^{\ideal})^2.
\end{align*}
By \eqref{Delta-bd-2n} and $\widehat a_{ik}^\top a_{ik} >0$, if $\sqrt{d_k/T} \gg 1/ \sqrt{\Theta_{ii}}$, i.e. $\Theta_{ii} \gg T/d_k$, then the leading term in $u^\top (\widehat a_{ik} - a_{ik})$ is
\begin{align*}
\frac{1}{\Theta_{ii}\prod_{\ell\neq k} ( a_{i\ell}^\top g_{il})} u^\top P_{ a_{ik},\perp} \left[\frac{1}{ T} \sum_{t=1}^T w_if_{it} \left( \cE_t \times_{\ell\in [ K] \backslash\{k\} } g_{i\ell}^{\top} \right) \right] =\frac{1}{\Theta_{ii} } u^\top P_{ a_{ik},\perp} \left[\frac{1}{ T} \sum_{t=1}^T w_i f_{it} \left( \cE_t \times_{\ell\in [ K] \backslash\{k\} } b_{i\ell}^{\top} \right) \right].
\end{align*}
Thus, \eqref{eqn:clt1} follows from the central limit theorem of the above leading term.

Otherwise, $1/\Theta_{ii}$ is the leading order term of $u^\top (\widehat a_{ik} - a_{ik})$. Then we have \eqref{eqn:clt2}.

\end{proof}

\begin{proof}[\bf Proof of Theorem \ref{thm:clt2}]
First, by \eqref{thmclt:eq2}, we have \eqref{eqn2:clt1}. Moreover, by \eqref{thmclt:eq3} and case (i) in the proof of Theorem \ref{thm:clt}, we have
\begin{align*}
&\left| \left( \widehat a_{ik}^{\top}  a_{ik} \right)^2 -1 -  \frac{1}{\Theta_{ii}^2 \prod_{\ell\neq k} ( a_{i\ell}^\top g_{il})^2} \left[\frac{1}{ T} \sum_{t=1}^T w_i f_{it} \left( \cE_t \times_{\ell\in [ K] \backslash\{k\} } g_{i\ell}^{\top} \right) \right]^\top  P_{ a_{ik},\perp}  \left[\frac{1}{ T} \sum_{t=1}^T w_i f_{it} \left( \cE_t \times_{\ell\in [ K] \backslash\{k\} } g_{i\ell}^{\top} \right) \right]   \right| \notag\\
&\le C_4 (\psi^{\ideal})^3 + C_4 (\sqrt{r} \psi_0) (\psi^{\ideal})^2 + C_4 \frac{1}{\Theta_{ii}^2}\left( \frac{d_k\log(d)}{T} + \sqrt{\frac{d_k \log(d)}{T}} + 1\right)^2 .
\end{align*}
Then \eqref{eqn2:clt2} and \eqref{eqn2:clt3} can be derived by applying similar arguments in the proof of Theorem \ref{thm:clt}.

\end{proof}

\begin{proof}[\bf Proof of Theorem \ref{thm:rank}]
The proof of the consistency of $\hat r^{\rm uer}$ draws on methods similar to those in \cite{ahn2013} and \cite{han2022rank}, given that our CP tensor factor model can be equated to a vector factor model. Moreover, the consistency of $\hat r^{\rm ip}$ aligns with the proofs in \cite{han2022rank}, as our CP tensor factor model can also be regarded as a Tucker factor model with a uniform Tucker rank of $(r,...,r)$. Specifically, lemmas akin to Lemmas 11 and 12 (or Lemmas 14 and 15) in \cite{han2022rank} can be derived under our assumptions. It leads to $\PP(\hat r_k=r, 1\le k\le K)\to1$. We omit the detailed proofs as they are laborious, albeit straightforward, adaptations for a specialized case of the Tucker factor model.
\end{proof}

\section{Techinical Lemmas}

\begin{lemma}\label{lemma-transform-ext}
Let  $A \in \R^{d_1\times r}$ and $B \in \R^{d_2\times r}$ with
$\|A^\top A - I_r\|_{\rm 2}\vee \|B^\top B - I_r\|_{\rm 2} \le\delta$ and $d_1\wedge d_2\ge r$.
Let $A=\widetilde U_1 \widetilde D_1 \widetilde U_2^\top$ be the SVD of $A$,
$U = \widetilde U_1\widetilde U_2^\top$, $B=\widetilde V_1 \widetilde D_2 \widetilde V_2^\top$
the SVD of $B$, and $V = \widetilde V_1\widetilde V_2^\top$.
Then, $\|A \Lambda A^\top - U \Lambda U^{\top}\|_{\rm 2}\le \delta \|\Lambda\|_{\rm 2}$
for all nonnegative-definite matrices $\Lambda$ in $\R^{r\times r}$, and
$\|A Q B^\top - U Q V^{\top}\|_{\rm 2}\le \sqrt{2}\delta \|Q\|_{\rm 2}$
for all $r\times r$ matrices $Q$.
\end{lemma}

\begin{lemma}\label{prop-rank-1-approx}
Let $M\in \R^{d_1\times d_2}$ be a matrix with $\|M\|_{\rm F}=1$ and
${a}$ and ${b}$ be unit vectors respectively in $\R^{d_1}$ and $\R^{d_2}$.
Let $\widehat a$ be the top left singular vector of $M$.
Then,
\begin{equation}\label{prop-rank-1-approx-1}
\big(\|{\widehat a} {\widehat a}^\top - {a} {a}^\top\|_{\rm 2}^2\big) \wedge (1/2)
\le \|\vec(M)\vec(M)^\top - \vec({a} {b}^\top)\vec({a} {b}^\top)^\top\|_{\rm 2}^2.
\end{equation}
\end{lemma}

Lemmas \ref{lemma-transform-ext} and \ref{prop-rank-1-approx} are Propositions 5 and 3 in \cite{han2023tensor}, respectively.

\section{More Simulation Results} \label{appendix:simulation}
\begin{enumerate}
    \item In this subsection we show the simulation results of Configuration I, II and III with AR coefficient on $g_{it}$, $\phi$, equal to $0.5$. We can see that AC-ISO algorithm has a better performance since the signal strength in the auto-covariance grow with $\phi$. CC-ISO algorithm, however, outperforms AC-ISO algorithm even with stronger serial correlation in the factor process.

\begin{figure}[htbp!]
    \centering
    \includegraphics[width = 0.8\columnwidth]{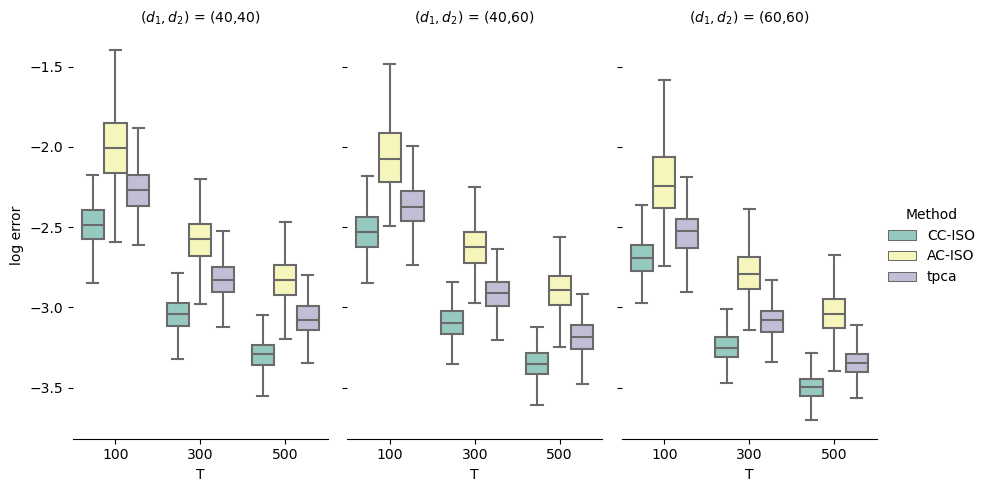}
    \caption{Boxplots of estimation errors over 500 replications under Configuration I with $\phi = 0.5$}
    \label{fig:config1_ar.5}
\end{figure}

\begin{figure}[htbp!]
    \centering
    \includegraphics[width = 0.8\columnwidth]{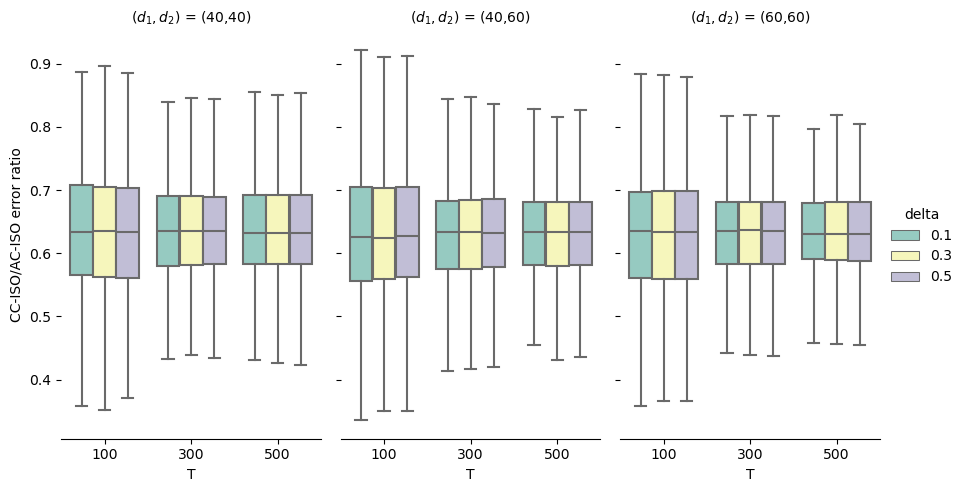}
    \includegraphics[width = 0.8\columnwidth]{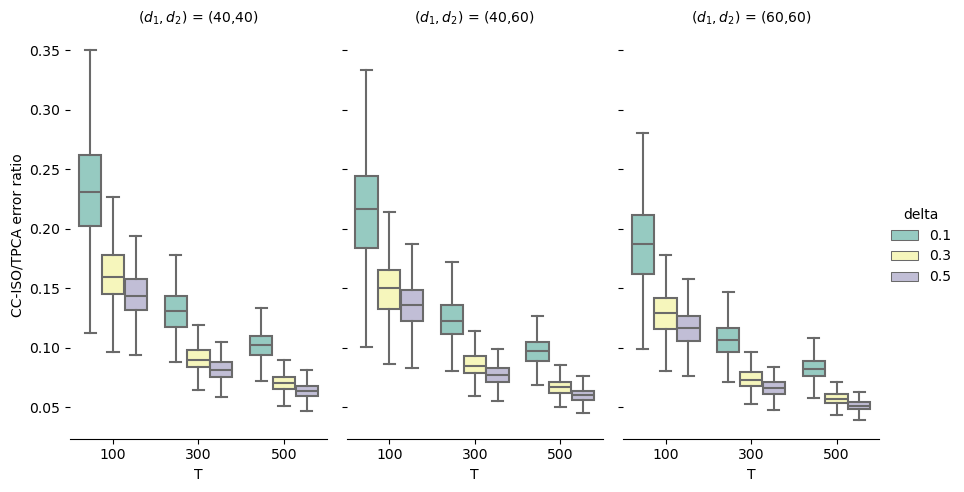}
    \caption{Boxplots of the estimation error over 500 replications under configuration II with $\phi = 0.5$. Note: The first panel shows the ratio of the estimation error of CC-ISO on AC-ISO. The second panel shows the ratio of the estimation error of CC-ISO on TPCA.}
    \label{fig:config2_ar.5}
\end{figure}

\begin{figure}[htbp!]
    \centering
    \includegraphics[width = 0.8\columnwidth]{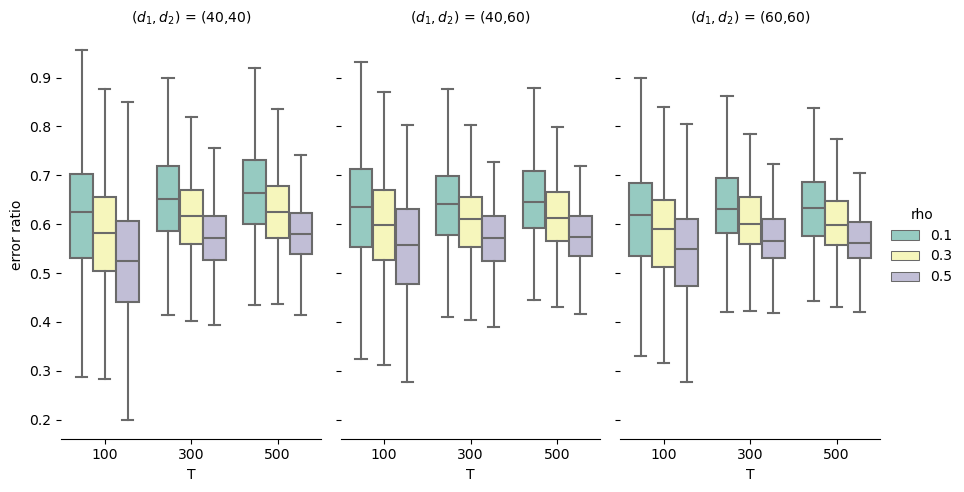}
    \caption{Boxplots of estimation errors over 500 replications under Configuration III with $\phi = 0.5$}
    \label{fig:config3_ar.5}
\end{figure}
\item We conduct the simulation to evaluate the performance of CC-ISO under Pareto-distributed errors. We set $d_1 = d_2$ and let $d_k$ vary over $\{20,40,60,80 \}$ and T over $\{100,200,300,400,500\}$. The signal strength is set to $(r-i+1)\sqrt{d}$. Each element in $\calE_t$ is generated i.i.d. from Pareto distribution with shape parameter $\alpha=2.5$. Figure \ref{fig:sim_pareto} shows the boxplots of estimation errors for the factor loadings over 200 repetitions. While Pareto-distributed errors lead to a higher occurrence of outliers compared to normally distributed errors whose results are shown in the main text, CC-ISO remains consistent, demonstrating its robustness in heavy-tailed settings.
\begin{figure}
          \centering
          \includegraphics[width=0.5\linewidth]{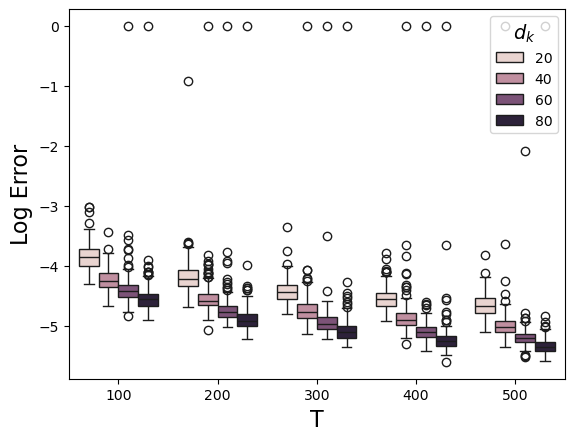}
          \caption{Boxplots of estimation errors with pareto errors over 200 repetitions}
          \label{fig:sim_pareto}
      \end{figure}
      \item We conduct simulation for correlated factors. Specifically, we generate data following DGP Setting II from the main text, with $\eta=0.1$. In this setting, $f_{it}$ are initially generated independently across $i$. To introduce correlation among the factors, we pre-multiply $f_t$ by a Toeplitz matrix with parameter $\zeta$, defined as
    $$
    \begin{aligned}
        f_{t}^{(corr)} &= \Sigma_{f} f_t, \\
        \Sigma_{f,ij} &= \zeta^{|i-j|},
    \end{aligned}
    $$
    where $\Sigma_{f,ij}$ denotes the $(i,j)$ entry of $\Sigma_f$. By varying $\zeta$, we control the degree of correlation among factors. Figure \ref{fig:sim_corrfactor} presents the simulation results, which show that our estimator remains consistent in the presence of correlated factors.
    \begin{figure}
        \centering
        \includegraphics[width=\linewidth]{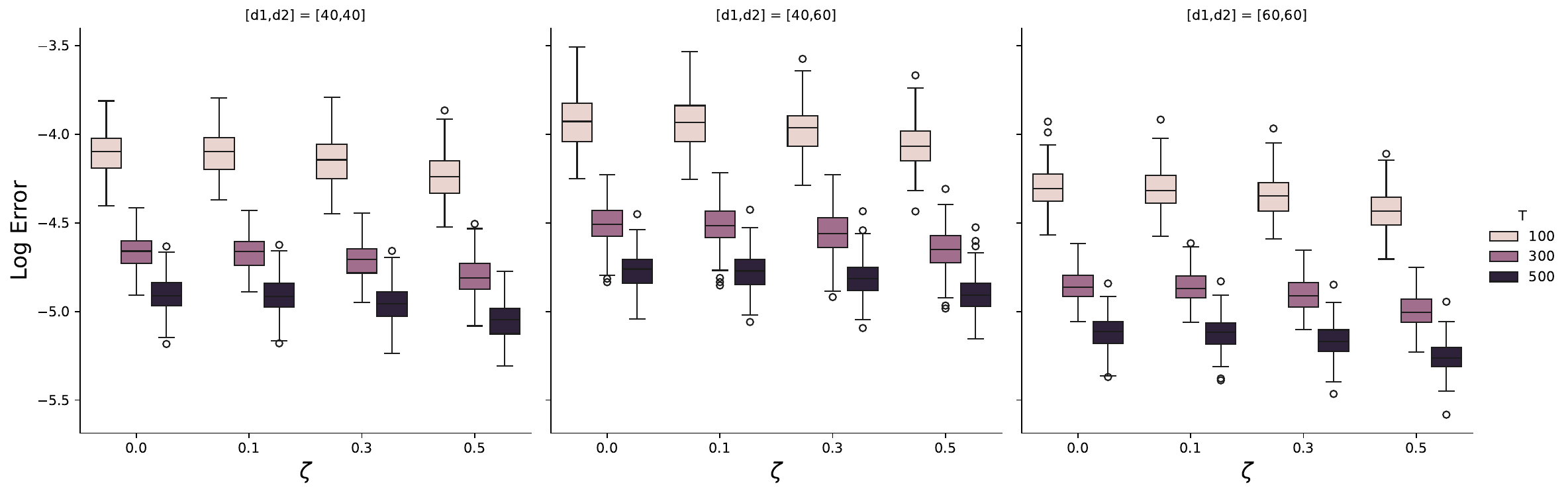}
        \caption{Boxplots of estimation errors with correlated factors}
        \label{fig:sim_corrfactor}
    \end{figure}
    \item We conduct a simulation study to compare the performance of the initial estimator (RC-PCA) and the iterative algorithm (CC-ISO) under both distinct and non-distinct eigenvalues of $\Sigma_0$ (as defined in Algorithm 1). The data-generating process (DGP) is set with dimensions $(d_1, d_2, T) = (\bar{d},\bar{d},500)$. In the distinct-eigenvalue setting, we set $\eta = 0.1$ so that $\delta \leq 0.2$ and $w_i = (r-i+1) \sqrt{d}$, where $d = d_1d_2$. For the non-distinct-eigenvalue setting, following Setting V in the main text, we set $w_i = d^{\alpha/2}$ and fix $\delta = 0$. Figures \ref{fig:sim_cpcaiso} and \ref{fig:sim_rcpcaiso} present the results for these two settings, respectively. In both settings, the CC-ISO algorithm outperforms RC-PCA in terms of estimation error and variance. In particular, due to non-zero $\delta$ in the distinct-eigenvalue setting, RC-PCA estimates loadings with a non-vanishing bias, but this bias is removed by CC-ISO.
    \begin{figure}[htbp!]
        \centering
        \includegraphics[width=0.5\linewidth]{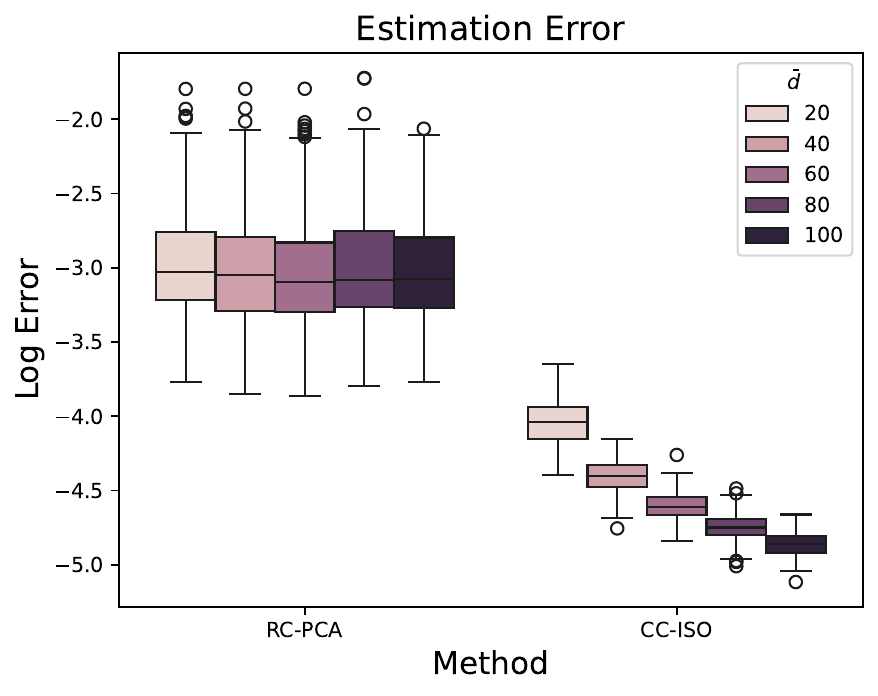}
        \caption{Boxplots of estimation errors of RC-PCA and ISO with distinct eigenvalues of $\Sigma_0$}
        \label{fig:sim_cpcaiso}
    \end{figure}

    \begin{figure}[htbp!]
        \centering
        \includegraphics[width=0.5\linewidth]{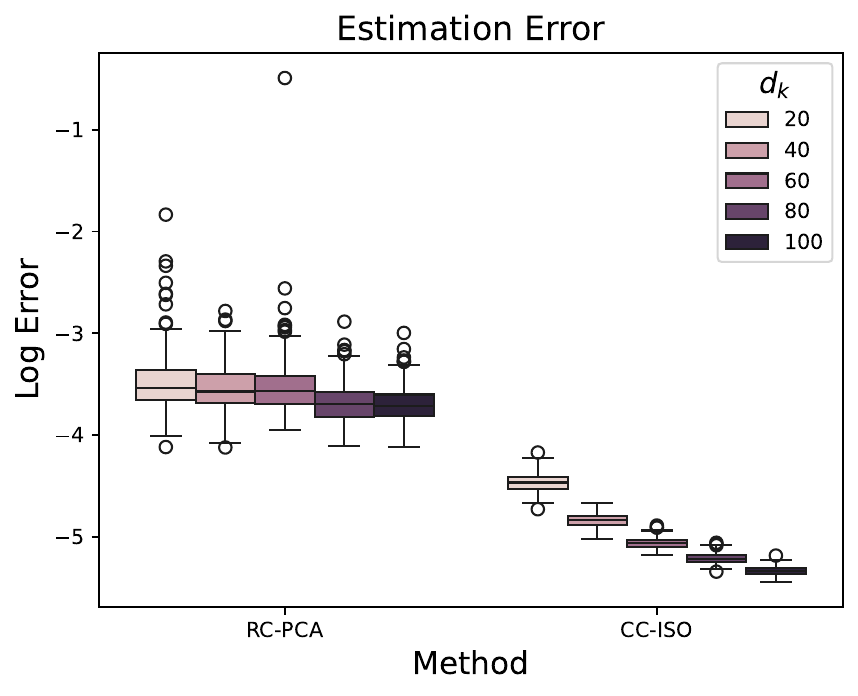}
        \caption{Boxplots of estimation errors of RC-PCA and ISO with non-distinct eigenvalues of $\Sigma_0$}
        \label{fig:sim_rcpcaiso}
    \end{figure}
    \item We conduct simulation study to evaluate Theorem 4.4(i) with estimated variance $\hat \sigma_{u,ik}$ and $\hat \Theta$, which are proposed in \cite{chy2025}, our accompanying paper on the forecast with tensor data. Specifically, we propose a thresholding estimator $\hat \Sigma_e^{\calT}$:
$$
\hat \Sigma_e^{\calT} = \calT_\lambda \left(\frac{1}{T} \sum_{t=1}^T \hat e_t \hat e_t^\top \right),
$$
where $\left(\hat \Sigma_e^\calT\right)_{(j,l)} = \calT_\lambda \left(\frac{1}{T} \sum_{t=1}^T \hat e_{jt} \hat e_{lt} \right)$, $\calT_\lambda (\cdot)$ is a thresholding operator and $\hat e_t$ is the vectorized estimated error using the CC-ISO algothrithm. Then we can estimate the variance $\sigma_{u,ik}^2$ via $\hat{\sigma}_{u,ik}^2= \hat h_{ik}^\top \hat \Sigma_e^{\calT} \hat h_{ik}$, where $\hat h_{ik}= \hat b_{iK}\odot \cdots \odot \hat b_{i,k+1} \odot \hat P_{ a_{ik},\perp}u \odot \hat b_{i,k-1}\odot \cdots \odot \hat b_{i1}$, and $\hat \Theta= \hat W (\frac{1}{T}\sum_{t=1}^T \hat F_t \hat F_t^\top) \hat W$. All settings are the same as Configuration (VII) in the main text but we let $T = \lceil 800 + d^{3/4} \rceil$ for the consistency of $\hat \sigma_{u,ik}$. The results are shown in Figure \ref{fig:sim_clt_wcs}.
\begin{figure}[htbp!]
    \centering
    \includegraphics[width=0.8\linewidth]{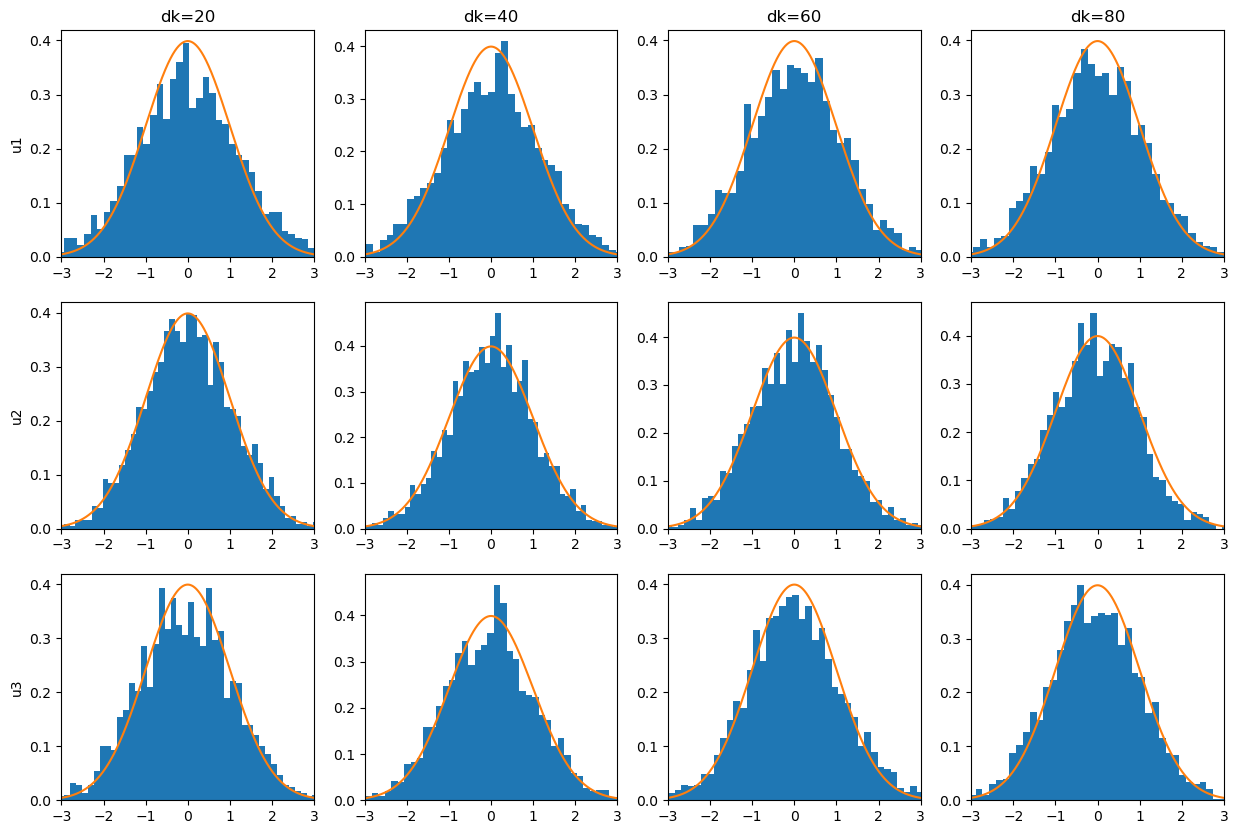}
    \caption{Histogram of $\sqrt{\hat \Theta_{ii}T} \hat \sigma_{u,ik}^{-1/2} u_l^\top\left( \hat a_{ik} - \operatorname{sign}\left( \hat a_{ik}^\top a_{ik} \right) \cdot a_{ik} \right)$ with weak cross-sectionally correlated errors.}
    \label{fig:sim_clt_wcs}
\end{figure}
\item We conduct a simulation study to evaluate the performance of the CC-ISO algorithm under unbalanced dimensions. The data is generated according to DGP Setting II in the main text, with $\eta = 0.1$. We fix $d_2 = 40$ while varying $d_1$ from 20 to 200 and $T$ from 100 to 500. In this setup, since the noise tensor does not exhibit cross-sectional dependence, the rate from Remark \ref{rmk:thm2} can be simplified to $\max_i\|\widehat a_{ik}^{\iso} \widehat a_{ik}^{\iso\top}  - a_{ik} a_{ik}^\top \|_{2}=O_{\PP}(\sqrt{d_{k}/(dT)})$, with $d=d_1d_2$.
    As $T$ increases, the estimation errors for both $\widehat a_{i1} (1\le i\le r)$ and $\widehat a_{i2} (1\le i\le r)$ decrease. Furthermore, with $d_2$ fixed, an increase in $d_1$ improves the convergence rate for the second-mode loading vectors $\widehat a_{i2}^{\iso}$, while the convergence rate for the first-mode loading vectors $\widehat a_{i1}^{\iso}$ remains unchanged. The simulation results, presented in Figure \ref{fig:sim_unbalanaceddim}, are consistent with this theoretical analysis.


    \begin{figure}
        \centering
        \includegraphics[width=\linewidth]{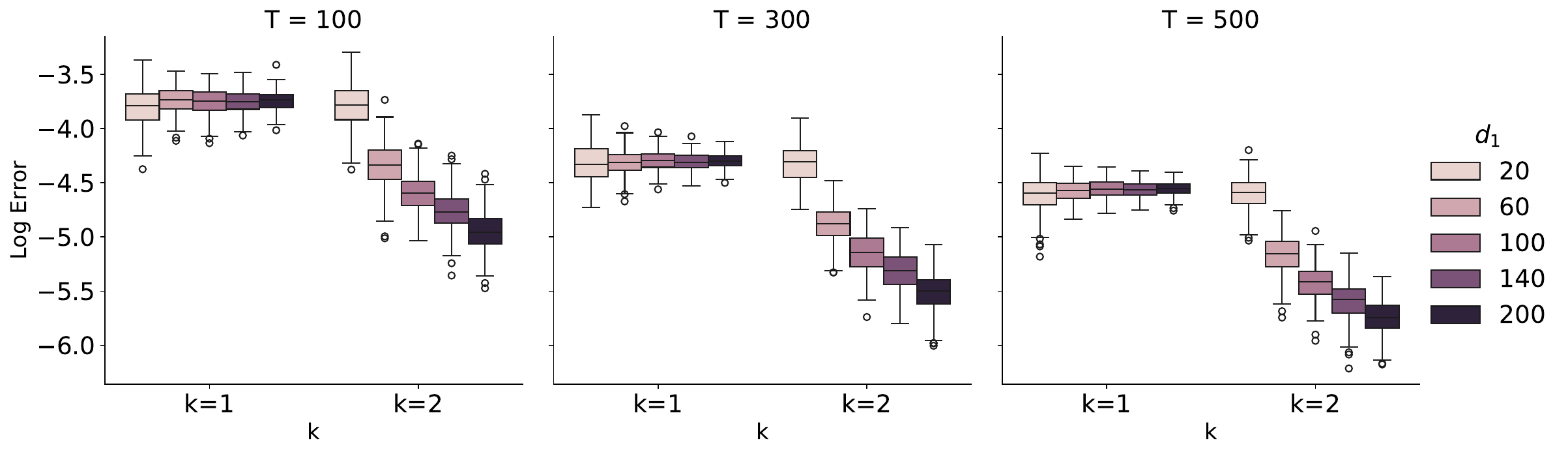}
        \caption{Boxplots of estimation errors under unbalanced dimensions. The two cases, $k=1$ and $k=1$, represent the estimation error $\max_i\|\widehat a_{ik}^{\iso} \widehat a_{ik}^{\iso\top}  - a_{ik} a_{ik}^\top \|_{2}$ for each respective mode.}
        \label{fig:sim_unbalanaceddim}
    \end{figure}
\item We measure the computation time of the proposed algorithms, RC-PCA and CC-ISO, and compare them against TPCA \cite{babii2022tensor} and the GE method proposed by \cite{chang2023modelling} with $K = 3$. Simulations are conducted with data dimensions $(80,80,500)$ and rank $r = 3$. To evaluate the computation cost of Randomized Projection, we consider two settings: one with well-separated eigenvalues and another with closely spaced eigenvalues. The tuning parameters are set to $L = 2r^2$, $c_0 = 0.1$ and $\nu = 0.8$. Table \ref{tab:time_cost} reports the median computation time and the median number of iterations (of CC-ISO) over 200 repetitions.

In the well-separated eigenvalues setting, RC-PCA requires more time than TPCA and GE due to the computation of $\Sigma_0$, which has a dimension of $6400 \times 6400$ in our DGP. However, this additional cost yields a benefit: RC-PCA attains lower estimation errors, especially when factor loadings are not orthogonal. In the close-eigenvalue setting, the computational cost of RC-PCA increases further because the Randomized Projection step involves performing SVD at least $2 \times L$ times. Nevertheless, this added complexity leads to improved estimation performance in difficult scenarios with nearly equal eigenvalues, highlighting a trade-off between computational efficiency and accuracy. For CC-ISO, each iteration takes approximately 2.7 seconds, which is comparable to the computation time of TPCA.

\begin{table}[htbp]
  \centering
    \begin{tabular}{crrrrr}
    \toprule
    DGP   & \multicolumn{1}{c}{RC-PCA} & \multicolumn{1}{c}{CC-ISO} & \multicolumn{1}{c}{\# Iteration} & \multicolumn{1}{c}{TPCA} & \multicolumn{1}{c}{GE, K=3} \\
    \midrule
    Distinct Eigenvalues & 16.94 & 8.39 & 3     & 2.26 & 6.48 \\
    Close Eigenvalues & 131.19 & 8.19 & 3     & 2.13 & 6.41 \\
    \bottomrule
    \end{tabular}%
    \caption{ \small Median of computation time (in second) of RC-PCA (column 1), CC-ISO(column 2), Tensor PCA(column 4), and generalized eigen-analysis based estimation (column 5) , along with the median number of iterations (column 3) for CC-ISO. The first row presents results for the distinct-eigenvalue setting. The second row corresponds to the setting with close eigenvalues, where Randomized Projection is applied.}
    \label{tab:time_cost}%
\end{table}
\item We conducted a robustness check for $c_0$ and $\nu$. Using Setting V in the paper with $\bar{d} = 20$ and $T = 500$, we vary $c_0 \in [0.05,0.5]$ and $\nu \in [0.5,0.9]$. Figure \ref{fig:rp_err} reports the results, with the first box in each group showing the error of the initial estimators without randomized projection (C-PCA). RC-PCA performs consistently across all values of $c_0$ and $\nu$, and clearly outperforms C-PCA.
 \begin{figure}
        \centering
        \includegraphics[width=\linewidth]{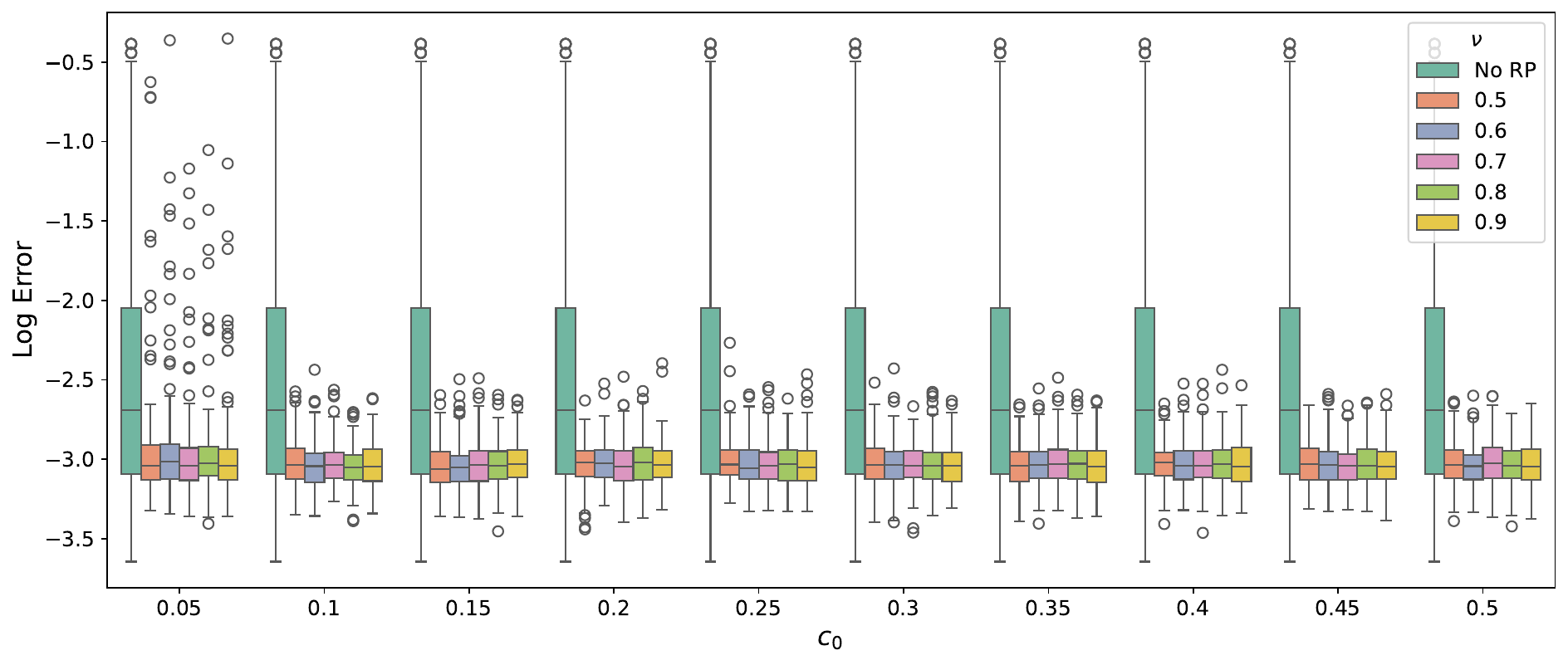}
        \caption{Boxplots of estimation errors, C-PCA vs. RC-PCA, across different $c_0$ and $\nu$. }
        \label{fig:rp_err}
    \end{figure}

\end{enumerate}
\clearpage

\section{More Empirical Results}\label{appendix:empirical}
\begin{figure}[htbp!]
    \centering
    \includegraphics[width=\linewidth]{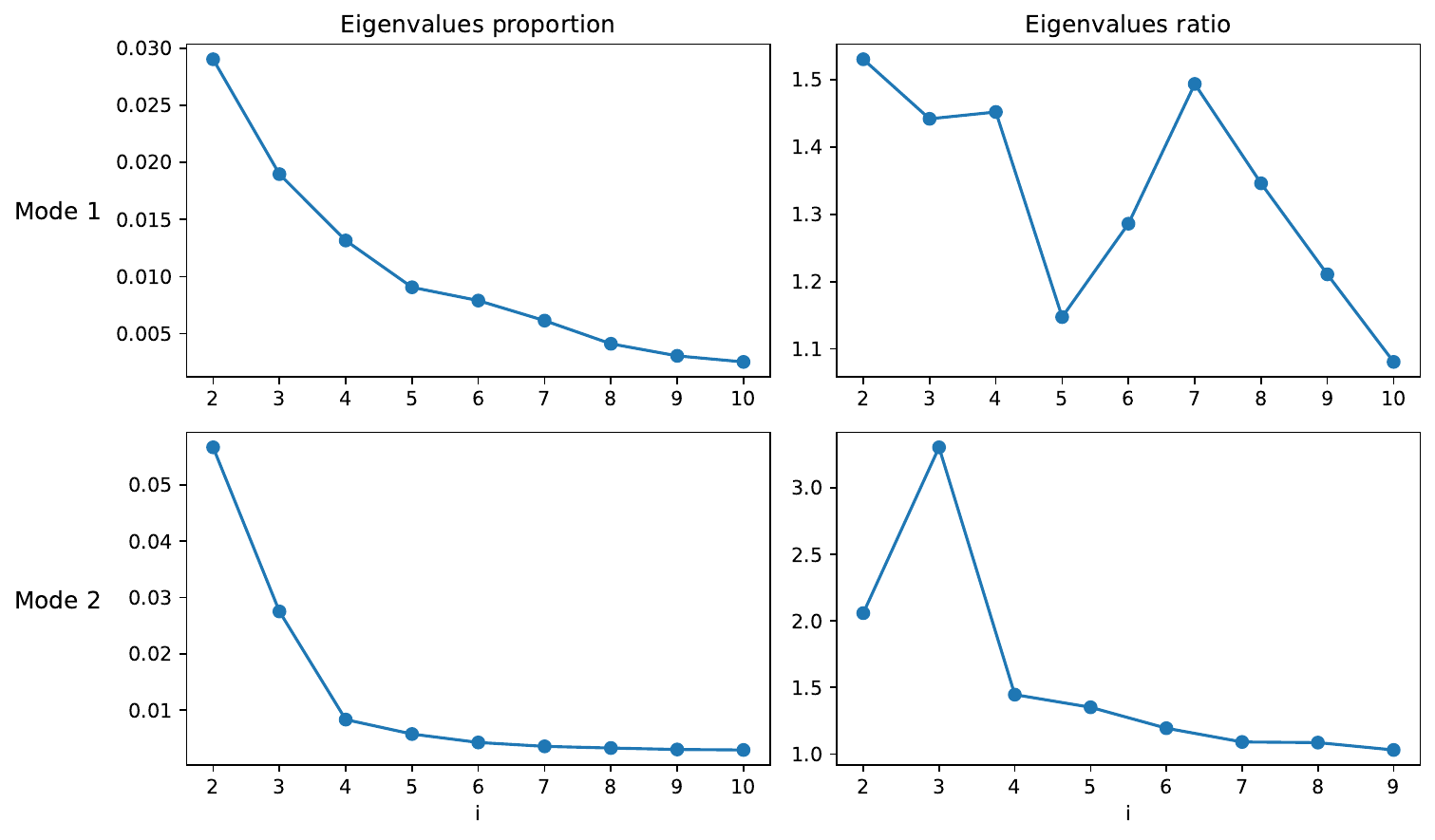}
    \caption{\small The mode-wide eigenvalue ratio plot (the first column) and the mode-wide scree plot (the second column) of the excess return of characteristic decile portfolios, starting from the second largest eigenvalue. The first row shows the plot of the first mode, i.e., the characteristic mode, and the second row shows the second mode, i.e., the decile mode. In the scree plot, the y-axis indicates the proportion of the $i^{th}$ largest eigenvalue to the sum of eigenvalues. In the eigenvalue ratio plot, the y-axis indicates the ratio of the $i^{th}$ eigenvalue to the $(i+1)^{th}$ eigenvalue. The plots suggest the rank pair $(2,3)$ for the Tucker factor model.}
    \label{fig:decile_mode_screeplot}
\end{figure}
\section{More Discussion on Initial Factor Estimates}\label{appendix:initial}
Theorem \ref{thm:projection} relies on initial factor estimates being reasonably close to the ground truth. This requirement is, in fact, standard and typically unavoidable from a theoretical standpoint. Least-squares estimation for both Tucker and CP factor models involves non-convex optimization problems with many local optima, most of which are not close to the ground truth; see, for example, \cite{chen2023statistical,han2024cp}. This challenge is well recognized and is a main reason why estimation of tensor factor models is notoriously difficult.

    In the optimization field, there are two general strategies to handle this issue. One approach is to incorporate occasional jumps into the iterations to enable escape from poor local optima (see, e.g., \cite{belkin2018eigenvectors}). The other is to ensure that the initialization is close to a good local optimum. Our iterative algorithm 3 is related to the second strategy. Similarly, the HOSVD used in \cite{lettau20243d} provides a high-quality initialization close to the ground truth, which can be refined by the iterative HOOI algorithm. Since HOOI also falls into the second category, its performance crucially depends on good initialization \citep{zhang2018tensor}. As highlighted in \cite{KB2009review} and other related literature, theoretical convergence guarantees for alternating least squares in CP decomposition are still lacking. In this sense, our RC-PCA (Algorithm 1) plays a role analogous to HOSVD, while our ISO procedure (Algorithm 3) serves as the CP analogue of HOOI in the Tucker factor model analyzed in \cite{lettau20243d}.

\end{appendices}

\end{document}